\documentclass{emulateapj}

\shorttitle{[$\alpha$/Fe] Trends in Dwarf Galaxies}
\shortauthors{Kirby et al.}
\slugcomment{Accepted to ApJ, 2010 November 16}

\begin{document}
\newcommand{\teff}{$T_{\rm{eff}}$}
\newcommand{\mathteff}{T_{\rm eff}}
\newcommand{\logg}{$\log g$}
\newcommand{\mathlogg}{\log g}
\newcommand{\feh}{[Fe/H]}
\newcommand{\mathfeh}{{\rm [Fe/H]}}
\newcommand{\afe}{[$\alpha$/Fe]}
\newcommand{\mathafe}{{\rm [\alpha/Fe]}}
\newcommand{\ah}{[$\alpha$/H]}
\newcommand{\mathah}{{\rm [\alpha/H]}}
\newcommand{\vt}{$v_t$}
\newcommand{\mathvt}{v_t}

\newcommand{\agefor}{1.3}
\newcommand{\aoutfor}{1.5}
\newcommand{\alphafor}{0.98}
\newcommand{\tauinfor}{0.31}
\newcommand{\ageleoi}{1.4}
\newcommand{\aoutleoi}{3.9}
\newcommand{\alphaleoi}{0.71}
\newcommand{\tauinleoi}{0.35}
\newcommand{\agescl}{1.1}
\newcommand{\aoutscl}{5.4}
\newcommand{\alphascl}{0.83}
\newcommand{\tauinscl}{0.27}
\newcommand{\ageleoii}{1.6}
\newcommand{\aoutleoii}{6.6}
\newcommand{\alphaleoii}{0.66}
\newcommand{\tauinleoii}{0.42}
\newcommand{\agesex}{0.8}
\newcommand{\aoutsex}{9.6}
\newcommand{\alphasex}{0.50}
\newcommand{\tauinsex}{0.22}
\newcommand{\agedra}{0.7}
\newcommand{\aoutdra}{9.5}
\newcommand{\alphadra}{0.34}
\newcommand{\tauindra}{0.22}
\newcommand{\agecvni}{0.9}
\newcommand{\aoutcvni}{8.8}
\newcommand{\alphacvni}{0.36}
\newcommand{\tauincvni}{0.21}
\newcommand{\ageumi}{0.4}
\newcommand{\aoutumi}{11.0}
\newcommand{\alphaumi}{0.26}
\newcommand{\tauinumi}{0.17}
\newcommand{\tauindgc}{0.69}
\newcommand{\tauinage}{0.96}
\newcommand{\tIaagescl}{0.82}
\newcommand{\epshnagescl}{0.82}
\newcommand{\Zwindagescl}{0.82}
\newcommand{\sclgaslostorig}{$1.8 \times 10^8~M_{\sun}$}
\newcommand{\sclgaslostZwind}{$4.5 \times 10^6~M_{\sun}$}

\title{Multi-Element Abundance Measurements from Medium-Resolution
  Spectra. \\ IV. Alpha Element Distributions in Milky Way Dwarf
  Satellite Galaxies\altaffilmark{1}}

\author{Evan~N.~Kirby\altaffilmark{2,3},
  Judith~G.~Cohen\altaffilmark{3},
  Graeme~H.~Smith\altaffilmark{4},
  Steven~R.~Majewski\altaffilmark{5},
  Sangmo~Tony~Sohn\altaffilmark{6},
  Puragra~Guhathakurta\altaffilmark{4}}

\altaffiltext{1}{Data herein were obtained at the W.~M. Keck
  Observatory, which is operated as a scientific partnership among the
  California Institute of Technology, the University of California,
  and NASA.  The Observatory was made possible by the generous
  financial support of the W.~M. Keck Foundation.}
\altaffiltext{2}{Hubble Fellow.}
\altaffiltext{3}{California Institute of Technology, 1200
  E.\ California Blvd., MC 249-17, Pasadena, CA 91125}
\altaffiltext{4}{University of California Observatories/Lick
  Observatory, University of California, 1156 High St., Santa Cruz, CA
  95064}
\altaffiltext{5}{Department of Astronomy, University of Virginia,
  P.~O.\ Box 400325, Charlottesville, VA, 22904-4325}
\altaffiltext{6}{Space Telescope Science Institute, 3700 San Martin
  Dr., Baltimore, MD 21218, USA}

\keywords{galaxies: dwarf --- galaxies: abundances --- galaxies:
  evolution --- Local Group}

%%%%%%%%%%%%%%%%%%%%%%%%%%%%%%%%%
%%%%%%%%%    ABSTRACT    %%%%%%%%
%%%%%%%%%%%%%%%%%%%%%%%%%%%%%%%%%

\begin{abstract}

We derive the star formation histories of eight dwarf spheroidal
(dSph) Milky Way satellite galaxies from their alpha element abundance
patterns.  Nearly 3000 stars from our previously published catalog
(\citeauthor*{kir10b}) comprise our data set.  The average
[$\alpha$/Fe] ratios for all dSphs follow roughly the same path with
increasing [Fe/H].  We do not observe the predicted knees in the
[$\alpha$/Fe] vs.\ [Fe/H] diagram, corresponding to the metallicity at
which Type~Ia supernovae begin to explode.  Instead, we find that
Type~Ia supernova ejecta contribute to the abundances of all but the
most metal-poor ($\mathfeh < -2.5$) stars.  We have also developed a
chemical evolution model that tracks the star formation rate, Types~II
and Ia supernova explosions, and supernova feedback.  Without metal
enhancement in the supernova blowout, massive amounts of gas loss
define the history of all dSphs except Fornax, the most luminous in
our sample.  All six of the best-fit model parameters correlate with
dSph luminosity but not with velocity dispersion, half-light radius,
or Galactocentric distance.

\end{abstract}

%%%%%%%%%%%%%%%%%%%%%%%%%%%%%%%%%
%%%%%%%%%   SECTION 1   %%%%%%%%%
%%%%%%%%%%%%%%%%%%%%%%%%%%%%%%%%%

\section{Introduction}
\label{sec:intro}

Understanding the origins of galaxies requires understanding the
histories of their dark matter growth, gas flows, and star formation.
Of these, the dark matter growth is the most straightforward to model
\citep[e.g.,][]{die07,spr08}.  The gas flow history presents more
difficult obstacles, such as collisional dissipation, gas cooling,
stellar feedback, and conversion into stars.  Despite the challenges,
some models---built on top of dark matter simulations---track all of
these processes over cosmic time \citep[e.g.,][]{gov07}.  The results
of these models have observational consequences for the properties of
the present stellar populations of galaxies.

\subsection{Methods for Determining Star Formation Histories}

The star formation histories (SFHs) of galaxies may be deduced from
the colors and magnitudes of the population and from the spectra of
the stars and gas, if present.  Distant, unresolved galaxies display
only a single, composite spectral energy distribution, which may be
examined through calibrations of spectrophotometric indices
\citep[e.g.,][]{gra08} or, in some cases, spectral synthesis
\citep{mcw08,col09}.  Nearer stellar systems may be resolved both
photometrically and spectroscopically.  The {\it Hubble Space
  Telescope} (HST) has enabled the characterization of the SFHs of
many nearby galaxies \citep{wei08,dal09,ber09}, including most of the
dwarf galaxies in the Local Group \citep{hol06,orb08}.

Photometrically derived SFHs are most sensitive to young stars and
metal-rich stars because the separation between isochrones increases
with decreasing age and increasing metallicity.  Elemental abundances
obtained from spectroscopy do not give absolute ages, but they can
provide finer relative time resolution for old, metal-poor
populations.  \citet{gil91} showed that star formation bursts of
varying duration and frequency in dwarf galaxies engrave signatures on
the ratio of oxygen to iron as a function of metallicity.  Because
oxygen-rich Type~II supernovae (SNe) explode within tens of Myr of a
starburst, the oxygen content of stars forming soon after the burst
will be high.  Within hundreds of Myr, iron-rich Type~Ia SNe begin to
explode.  The injection of iron into the interstellar medium (ISM)
depresses the oxygen-to-iron ratio of subsequently forming stars.
These processes are generalizable to other elements.  The abundances
of the next several elements with even atomic number beyond
oxygen---the alpha elements (Ne, Mg, Si, S, Ar, Ca, and Ti)---roughly
scale with oxygen abundance.  The abundances of iron-peak elements (V,
Cr, Mn, Co, and Ni) roughly scale with iron abundance.  The trend of
the alpha-to-iron-peak ratio with iron-peak abundance, a proxy for
elapsed time or integrated star formation, reveals the relative star
formation history with a resolution of about 10~Myr, the approximate
timescale for a Type~II SN.

\subsection{Chemical Evolution Models}

A glance at a diagram of [Mg/Fe] vs.\ [Fe/H] gives a qualitative sense
of a galaxy's star formation history.  Converting quantitative
abundances into a quantitative SFH requires a chemical evolution
model.  \citet{pag97} described in detail how to create such a model,
and \citet{tol09} reviewed recent progress on modeling the SFHs of
Local Group dwarf galaxies.  \citet{mat08} described the levels of
approximation that the models assume.  In general, more sophisticated
and presumably more accurate models reduce the number of
approximations.  The most basic assumptions are instantaneous
recycling and instantaneous mixing.  Consideration of stellar
lifetimes and SN delay times removes the first approximation.
Three-dimensional hydrodynamical simulations remove the second
approximation.

A chemical evolution model reflects the history not only of star
formation but also of gas flow.  A complete explanation of metallicity
and alpha element distributions requires both inflows and outflows.
The metallicity distribution functions (MDFs) of nearby Galactic G
dwarfs cannot be explained with a closed box model
\citep{van62,sch63}.  \citet{pag97} discussed some of the proposed
solutions to the G dwarf problem, including variable nucleosynthesis
yields, bimodal star formation, and pre-enrichment.  One of the most
promising solutions is infalling matter \citep{lar72}.  Gases
undoubtedly flow out of the galaxy, either from SN winds
\citep{mat71,lar74} or stripping from the influence of external or
host galaxies \citep{tin79,lin83}.  For example, interactions with the
Milky Way could remove gas from the satellite galaxies discussed here.
Both inflows and outflows affect the star formation rate (SFR)
throughout the history of the galaxy.  Therefore, they shape the MDF
and the trend of [$\alpha$/Fe] with [Fe/H].

Chemical evolution models suffer from uncertainties in the initial
mass function of stars and stellar lifetimes \citep{rom05},
nucleosynthesis yields \citep{rom10}, and the delay time distribution
(DTD) for Type~Ia SNe \citep{mat09}.  However, these limitations have
not prevented the models from providing good fits to abundance data.
Even models with some of the first theoretical SN yields \citep{woo93}
successfully reproduced the observed metallicity distribution and
abundance patterns in the Galaxy \citep{pag95}.  Models with newer SN
yields also match the solar neighborhood abundance distributions very
well \citep[e.g.,][]{rom10}.  Nonetheless, uncertainties in the model
assumptions do complicate the interpretation of the model results.
For example, changing the Type~Ia DTD, particularly the turn-on time,
affects the derived timescale for star formation.  The best way to
circumvent these uncertainties is to apply the same model consistently
to several systems and compare them differentially.  Although the
absolute ages or SFRs may be affected by systematic errors in the
model, the relative quantities between different galaxies will be
meaningful.

Local Group dwarf galaxies make good subjects for chemical evolution
models.  First, the Local Group contains many resolved dwarf galaxies
\citep{mat98,tol09} with stars bright enough for medium- or
high-resolution spectroscopy.  Second, dwarf galaxies span a wide
range of properties, including velocity dispersion and luminosity.
The populations of the lowest luminosity galaxies enable the study of
star formation on small scales \citep{martin08a,nor08}.  The changes
in populations for more luminous or more massive galaxies show how
star formation responds to galaxy size \citep{mat98,kir10a}.  Third,
dwarf galaxies host some of the most metal-poor stars known
\citep{kir08b,kir09,geh09,coh09,coh10,fre10a,fre10b,sim10,nor10a,nor10b,sta10,taf10}. %sim10a
and sim10b These stars retain the chemical imprint of the ISM when the
Universe was less than 1~Gyr old.  Therefore, dwarf galaxies permit
the study of star formation not only on small scales but also at early
times.  Finally, dwarf galaxies may be the primary building blocks for
the Milky Way (MW) halo \citep{sea78,whi78}.  The stellar populations
of the surviving dwarf galaxies may reflect the stellar populations of
the dissolved building blocks, and they may show how the surviving
satellites evolved since the time of rapid accretion onto the MW.

In a series of articles, \citet{lan03,lan04,lan07,lan10} and
\citet*{lan06,lan08} presented numerical models that tracked the
evolution of several elements in dSphs.  The models plausibly
explained the MDFs and the available multi-element abundance
measurements in dSphs.  However, large samples of published abundance
measurements in any individual dSph have been sparse until recently
\citep{she09,kir09,kir10b,let10}.  Other chemical evolution models of
dSphs have examined the effects of reionization \citep{fen06} and star
formation stochasticity \citep{car08}.  \citet{rec01} constructed one
of the first hydrodynamical models of dwarf galaxy evolution.  In
particular, they simulated a galaxy similar to IZw18.
\citet{mar06,mar08} published hydrodynamical simulations of an
isolated, Draco-like dSph.  Their models relaxed the assumption of
instantaneous mixing and allowed inhomogeneous chemical enrichment.
Some of the newest hydrodynamical models \citep{rev09,saw10} tracked
both the kinematics and abundances of the stars as they form.  They
attempted to explain not only chemical abundance patterns but also
dynamical properties of dSphs, such as the seemingly universal
dynamical mass measured within their optical radii \citep{mat98,str08}
and out to the edge of their light distributions \citep{gil07}.

\subsection{History of Chemical Analysis of Milky Way Satellites}

%GHS
The earliest indications of heavy element abundance spreads among red
giants of the dSph systems in Draco, Ursa Minor, Sculptor, and Fornax
were first obtained by the multichannel scanner observations of
\citet{zin78,zin81}, initial efforts at spectroscopy
\citep{nor78,kin80,kin81,ste84,smi84,leh92}, and both broad and narrow
band photometry \citep*{dem79,smi83}.  The globular clusters of the
Fornax system proved to differ in their metallicities
\citep{zinper81}.  The presence of carbon stars
\citep{aar80,aar82,aarhod83,azz85} and so-called anomalous Cepheids
\citep{dem75,nor75,hir80,smi86} further indicated the potential
complexity of the stellar populations in dSphs.  Carbon stars are
exceedingly rare in globular clusters, while the period-luminosity
relations of the anomalous Cepheids implied that they are more massive
than typical cluster Cepheids \citep{zin76}.  As a consequence, by the
mid-1980s, circumstantial evidence was building to suggest that dSphs
had more complex and possibly more extensive star formation and
chemical evolution histories than globular clusters.

Since that time, the application of ground-based CCD and HST imaging
has lead to greatly improved color-magnitude diagrams (CMDs) that have
clearly shown the presence of significant internal age spreads within
{\it some} of the Milky Way's retinue of dSphs, such as Carina,
Fornax, Leo~I, and Sextans
\citep*[e.g.,][]{mig90,mig97,sme96,hur98,buo99,gal99a,gal99b,sav00,lee09}.
Spectroscopy with large ground-based telescopes has demonstrated the
presence of abundance inhomogeneities in the majority of these systems
\citep[e.g.,][]{sun93,sme99,she01b,she03,tol01,tol03,tol04,win03,pon04,gei05,mcw05a,mcw05b,bat06,koc06,bos07,sbo07,gul09,coh09,coh10,kir09}.

%Another major difference between dSphs and Galactic globular clusters
%has been revealed by radial velocity observations of individual red
%giants.  Velocity dispersion studies have led to the recognition of
%substantial dark matter contents in the local dSph systems
%\citep*{aar83,fab83,lin83,gal94,arm95,vog95,mat97,mat98,kle01,kle02,mas06,koc07,wal07,wal09,pen08}.
%The dark matter halos of dSphs have likely played a role in governing
%their dynamical and metallicity evolution
%\citep*{dek86,cari02,fra03,mas05}.
%GHS

\subsection{Chemical Evolution Models for the New Catalog}

In this article, we interpret the multi-element abundance
distributions in eight dSphs with our own chemical evolution model.
The data set is our catalog of abundances based on spectral synthesis
of medium-resolution spectra from the DEIMOS spectrograph on the
Keck~II telescope \citep[][Paper~II]{kir10b}.  The catalog contains
2961 stars with abundance measurements.  The number of stars in each
dSph ranges from 141 (Sextans) to 827 (Leo~I).  It is the largest
homogeneous chemical abundance data set in dwarf galaxies.  The
typical areal coverage is about $300~{\rm arcmin}^2$ at or near the
center of each dSph.  The median uncertainty on [Fe/H] is 0.12~dex.
The fraction of the sample with [Mg/Fe] uncertainties less than 0.2
(0.3)~dex is 42\% (53\%).  That fraction increases to 54\% (69\%) for
[Ti/Fe], which is easier to measure than [Mg/Fe].  For
$\langle[\alpha/\rm{Fe}]\rangle$ (the average of [Mg/Fe], [Si/Fe],
[Ca/Fe], and [Ti/Fe]), the fraction increases to 71\% (88\%).

Our one-zone model is simple, but it incorporates some of the newest
SN yields and the most recently measured DTD for Type~Ia SNe.  The
biggest advantage of our data set is that it is homogeneous.  All of
the spectra were obtained with the same spectrograph configuration,
and all of the abundances were measured with the same spectral
synthesis code.  Thus, the derived star formation and gas flow
histories from our model---despite its simplicity---will be easy to
interpret differentially.  In other words, the absolute ages and star
formation rates may be affected by model uncertainties, but the trends
with galaxy properties, such as luminosity, should reflect the true
SFHs.

We begin by describing our model (Sec.~\ref{sec:model}).  Then, we
apply the model to the eight dSphs by finding the solution that best
matches the abundances.  We discuss how our results compare to
previous photometric and spectroscopic studies (Sec.~\ref{sec:dsphs}).
Next, we change some of the model variables to estimate the systematic
errors in the derived SFHs (Sec.~\ref{sec:exploration}).  Then, we
explore how the abundance distributions, SFHs, and gas flow histories
change with galaxy properties such as luminosity and velocity
dispersion (Sec.~\ref{sec:trends}).  Finally, we enumerate our
conclusions (Sec.~\ref{sec:conclusions}).

%%%%%%%%%%%%%%%%%%%%%%%%%%%%%%%%%
%%%%%%%%%   SECTION 2   %%%%%%%%%
%%%%%%%%%%%%%%%%%%%%%%%%%%%%%%%%%

\section{Chemical Evolution Model}
\label{sec:model}

\begin{deluxetable*}{lll}
\tablecolumns{3}
\tablewidth{0pt}
\tablecaption{Chemical Evolution Model Variables\label{tab:gcevars}}
\tablehead{\colhead{Variable} & \colhead{Description} & \colhead{Units}}
\startdata
$t$ & Time since start of simulation & Gyr \\
%$j$ & Chemical element & \nodata \\
$M$ & Mass of a single star & $M_{\sun}$ \\
$\xi_j(t)$ & Gas mass in element $j$ & $M_{\sun}$ \\
$X_j(t)$ & Mass fraction in element $j$ & dimensionless \\
$Y$ & Primordial helium mass fraction ($X_{\rm He}(0)$) & dimensionless \\
$M_{\rm gas}(t)$ & Total gas mass & $M_{\sun}$ \\
$Z(t)$ & Metal fraction (all elements heavier than He) & dimensionless \\
$\dot{\xi}_j(t)$ & Time derivative of $\xi_j$ & $M_{\sun}~{\rm Gyr}^{-1}$ \\
$\dot{\xi}_{j,*}(t)$ & Star formation rate, or rate of gas loss in element $j$ due to star formation &  $M_{\sun}~{\rm Gyr}^{-1}$ \\
$\dot{\xi}_{j,{\rm II}}(t)$ & Type~II SN or HN yield rate for element $j$ &  $M_{\sun}~{\rm Gyr}^{-1}$ \\
$\epsilon_{\rm HN}$ & Fraction of HNe among stars with $M \ge 20~M_{\sun}$ & dimensionless \\
$\zeta_{j,{\rm II}}(M,Z)$ & Mass of element $j$ ejected by one Type~II SN & $M_{\sun}$ \\
$\dot{\xi}_{j,{\rm Ia}}(t)$ & Type~Ia SN yield rate for element $j$ &  $M_{\sun}~{\rm Gyr}^{-1}$ \\
$t_{\rm delay}$ & Type~Ia SN delay time & Gyr \\
$\Psi_{\rm Ia}(t_{\rm delay})$ & Type~Ia SN delay time distribution & ${\rm SN}~{\rm Gyr}^{-1}~{M_{\sun}}^{-1}$ \\
$\zeta_{j,{\rm Ia}}$ & Mass of element $j$ ejected by one Type~Ia SN & $M_{\sun}$ \\
$\dot{\xi}_{j,{\rm AGB}}(t)$ & AGB yield rate for element $j$ &  $M_{\sun}~{\rm Gyr}^{-1}$ \\
$\zeta_{j,{\rm AGB}}(M,Z)$ & Mass of element $j$ ejected by one AGB star & $M_{\sun}$ \\
$A_*$ & Normalization of star formation rate law (free parameter) & $M_{\sun}~{\rm Gyr}^{-1}$ \\
$\alpha$ & SFR exponent of $M_{\rm gas}$ (free parameter) & dimensionless \\
$A_{\rm in}$ & Normalization of gas infall rate (free parameter) & $M_{\sun}~{\rm Gyr}^{-1}$ \\
$\tau_{\rm in}$ & Gas infall time constant (free parameter) & Gyr \\
$A_{\rm out}$ & Gas lost per SN (free parameter) & $M_{\sun}~{\rm SN}^{-1}$ \\
$M_{\rm gas}(0)$ & Initial gas mass (free parameter) & $M_{\sun}$ \\
\enddata
\end{deluxetable*}

In order to provide a rough interpretation of the abundance trends in
\citeauthor*{kir10b}'s catalog, we have developed a rudimentary model of
chemical evolution.  Table~\ref{tab:gcevars} defines the symbol for
each variable or constant in the model.  The model supposes that a
dwarf galaxy at any instant is a chemically homogeneous system that
can accrete or lose gas.  The ejecta of Type~II SNe enrich the gas
according to the total lifetime of massive ($10 < M/M_{\sun} < 100$)
stars, while the Type~Ia SNe follow the observed DTD \citep[][see
  below]{mao10}.  Stars form according to the \citet{kro93} initial
mass function (IMF, $dN/dM = 0.31 M^{-2.2}$ for $0.5 < M/M_{\sun} < 1$
and $dN/dM = 0.31 M^{-2.7}$ for $M > 1~M_{\sun}$).

The calculation tracks the mass of H, He, Mg, Si, Ca, Ti, and Fe at
each time step ($\Delta t = 1$~Myr).  The calculation is terminated
when the system reaches zero gas mass.

We define $\xi_j(t)$ as the galaxy's gas mass of element $j$ at time
$t$.  The galaxy's total gas mass at time $t$ is

\begin{eqnarray}
M_{\rm gas}(t) &=& \sum_j \xi_j(t) \label{eq:mgasexact} \\
               &\approx& \xi_{\rm H}(t) + \xi_{\rm He}(t) + 20.4[\xi_{\rm Mg}(t) + \xi_{\rm Si}(t) + \nonumber \\
               & & \xi_{\rm Ca}(t) + \xi_{\rm Ti}(t)] + 1.07\xi_{\rm Fe}(t) \label{eq:mgasapprox}
\end{eqnarray}

\noindent
The summation in Equation~\ref{eq:mgasexact} is over all elements in
the periodic table.  However, our model tracks only seven elements.
Therefore, we assume the ratio of the sum of all elements from Li to
Ti, inclusive, to the sum of Mg, Si, Ca, and Ti is the same as in the
Sun.  This ratio is 20.4 \citep{and89}.  Similarly, we assume the
solar ratio for the sum of all elements V through Ge compared to Fe:
1.07.  Elements beyond Ge are neglected.  Equation~\ref{eq:mgasapprox}
reflects these approximations.  For convenience, we define the
metallicity of the gas as follows:

\begin{equation}
Z = \frac{M_{\rm gas}(t) - \xi_{\rm H}(t) - \xi_{\rm He}(t)}{M_{\rm gas}(t)} \label{eq:z}
\end{equation}

\noindent
We also define the gas-phase mass fraction in an element $j$:

\begin{equation}
X_j(t) = \frac{\xi_j(t)}{M_{\rm gas}(t)}
\end{equation}

The following subsections explain the components of the models.  Each
component is expressed as the time change in $\xi_j(t)$, where
$\dot{\xi}_j \equiv d\xi_j(t)/dt$.

\subsection{Star Formation Rate}
\label{sec:sfr}

For simplicity, we assume that the star formation rate is a power law
in the gas mass of the galaxy.  With this assumption,

\begin{equation}
\dot{\xi}_{j,*} = A_* X_j(t) \left(\frac{M_{\rm gas}(t)}{10^6~M_{\sun}}\right)^{\alpha} \label{eq:sfr}
\end{equation}

\noindent
The variables $A_*$ and $\alpha$ are free parameters in the model.  In
the complete chemical evolution equation (Eq.~\ref{eq:gce}), the sign
of $\dot{\xi}_{j,*}$ is negative because $\xi_j$ represents the gas
mass, which is depleted due to star formation.

Equation \ref{eq:sfr} is a generalization of a Kennicutt-Schmidt law
\citep{sch59,ken98}, which connects the SFR to the gas surface
density, $\Sigma_{\rm gas}$.  Surface density is perhaps more
appropriate for disks than spheroids.  Desiring a more
three-dimensional property, we have used the gas mass, $M_{\rm gas}$,
instead of $\Sigma_{\rm gas}$.  The volume density, $\rho_{\rm gas}$,
would be a better description, but the difference between $M_{\rm
  gas}$ and $\rho_{\rm gas}$ is simply a constant because our model is
one-zoned.

\subsection{Type~II Supernovae}
\label{sec:II}

In our model, stars more massive than $10~M_{\sun}$ and less massive
than $100~M_{\sun}$ explode according to their total lifetimes
\citep{pad93,kod97}:

\begin{equation}
\tau_*(M) = \left(1.2 \left(M/M_{\sun}\right)^{-1.85} + 0.003\right)~{\rm Gyr} \label{eq:lifetime_massive}
\end{equation}

\noindent
This formula is valid for stars more massive than $6.6~M_{\sun}$,
(inclusive of our entire mass range for Type~II SNe).  \citet{mae89}
give slightly different formulas for stars less massive than
$60~M_{\sun}$, but the differences do not affect the chemical
evolution model appreciably.

Stars more massive than $100~M_{\sun}$ do not form in this model.  The
Type~II SN ejecta are mixed homogeneously and instantaneously into the
interstellar medium (ISM) of the entire dSph.

We adopt the Type~II SN nucleosynthetic yields of \citet{nom06}.  The
symbol $\zeta_{j,{\rm II}}(M,Z)$ represents the mass in element $j$
ejected from the Type~II SN explosion of a star with an initial mass
$M$.  It is a function of both initial stellar mass and metallicity.
\citeauthor{nom06}\ tabulated the yields for seven initial masses
ranging from $13~M_{\sun}$ to $40~M_{\sun}$ and four metallicities
from $Z=0$ to $Z=0.02$.  The total mass of the ejecta is always less
than the birth mass of the star because the star loses some mass
during its lifetime and because some mass is locked up forever in a SN
remnant.

\citeauthor{nom06}\ modeled both normal core-collapse SNe and very
energetic hypernovae (HNe).  The lowest mass HN they model is
$20~M_{\sun}$.  The fraction of stars at least this massive that
explode as HNe is $\epsilon_{\rm HN}$.  \citeauthor{nom06}\ adopted
$\epsilon_{\rm HN} = 0.5$ for their own model of the solar
neighborhood.  \citet{rom10} explored the cases of $\epsilon_{\rm HN}
= 0$ and 1.  In our own experimentation, we have found that
$\epsilon_{\rm HN} = 0$ produces good matches to the dSph abundance
patterns at the lowest values of [Fe/H], and we adopt this value for
the model.  In Sec.~\ref{sec:epshn05}, we explore the effect of
increasing $\epsilon_{\rm HN}$ on the model.

The following integral gives the instantaneous change in gas mass from
the ejecta of Type~II SNe ($M_{\sun}~{\rm Gyr}^{-1}$):

\begin{eqnarray}
\dot{\xi}_{j,\rm{II}} &=& 0.31~M_{\sun}^{0.7}\:\int_{10~M_{\sun}}^{100~M_{\sun}} \zeta_{j,{\rm II}}(M,Z(t-\tau_*(M)))\, \nonumber \\
 & & \; \times \; \dot{\xi}_*(t-\tau_*(M)) \, M^{-2.7} \, dM \label{eq:SNII}
\end{eqnarray}

\noindent
The coefficient $0.31~M_{\sun}^{0.7}$ is the normalization from the
IMF.  This integral depends on the SN yields ($\zeta_{j,{\rm II}}$),
the recent star formation history ($\dot{\xi}_*$), and the high-mass
IMF slope ($M^{-2.7}$).  In practice, this integral is performed
numerically with Newton-Cotes integration over an array of 100
logarithmically spaced masses between $10~M_{\sun}$ and
$100~M_{\sun}$.  The values of $\zeta_{j,{\rm II}}$ and $\dot{\xi}_*$
are interpolated onto this array.  The metallicity used to look up the
appropriate SN yields is consistent with the metallicity of the gas at
the time the exploding star formed.  (In other words, at any given
time step, the metallicities of the lower mass SNe are less than the
metallicities of higher mass SNe from more recently formed stars.)

The instantaneous Type~II SN rate (SN~Gyr$^{-1}$) is given by a
related integral:

\begin{equation}
\dot{N}_{\rm{II}} = 0.31~M_{\sun}^{0.7}\:\int_{10~M_{\sun}}^{100~M_{\sun}} \dot{\xi}_*(t-\tau_*(M))\,M^{-2.7}\,dM \label{eq:N_SNII}
\end{equation}

\noindent
This integral is performed over the same array of massive star
lifetimes as a function of mass as for Eq.~\ref{eq:SNII}.  The value
will be used to determine the mass lost from SN winds
(Sec.~\ref{sec:winds}).

\subsection{Type~Ia Supernovae}
\label{sec:Ia}

\begin{figure}[t!]
\includegraphics[width=\linewidth]{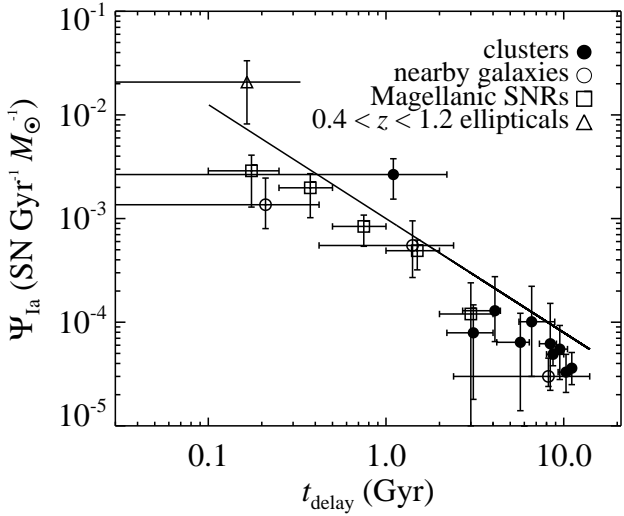}
\caption{Type~Ia supernova delay time distribution, as measured by
  \citet{mao10}.  The data come from a variety of star formation
  environments, given in the figure legend.  Equation \ref{eq:dtd}
  gives the expression for this function.  Compare this figure to
  \citeauthor{mao10}'s Fig.~2.\label{fig:dtd}}
\end{figure}

We adopt the Type~Ia SN yields of \citet{iwa99}.  The mass of element
$j$ ejected per Type~Ia SN is $\zeta_{j,{\rm Ia}}$.  The SNe explode
according to a function that approximates the delay time distribution
observed by \citet[][see Fig.~\ref{fig:dtd}]{mao10}.  The following
equation describes the adopted delay time distribution.

\begin{eqnarray}
\Psi_{\rm Ia} &=& \left\{\begin{array}{lcr}
     0 &~~~& t_{\rm delay} < 0.1~{\rm Gyr} \\
     \begin{array}{l} (1 \times 10^{-3}~{\rm SN~Gyr}^{-1}~M_{\sun}^{-1}) \\ \:\:\:\:\: \times \; \left(\frac{t_{\rm delay}} {\rm{Gyr}}\right)^{-1.1}\end{array} &~~~& t_{\rm delay} \ge 0.1~{\rm Gyr} \\
                              \end{array} \right. \label{eq:dtd}
\end{eqnarray}

\noindent
The variable $t_{\rm delay}$ is used instead of $t$ to indicate that
the DTD will be integrated from time $t$ into the past.

Unfortunately, the abundance distributions derived from the chemical
evolution model depend sensitively on the normalization and turn-on
time of $\Psi_{\rm Ia}$.  Both of these quantities---particularly the
turn-on time---have large uncertainties.  The normalization affects
[Fe/H] and the slope of [$\alpha$/Fe] with [Fe/H].  We have chosen $1
\times 10^{-3}~{\rm SN~Gyr}^{-1}~M_{\sun}^{-1}$ for the normalization
because that is the value that \citet{mao10} reported.  Even though
the data in Fig.~\ref{fig:dtd} are easily consistent with half that
value, the larger value better reproduces the slope of [$\alpha$/Fe]
with [Fe/H] for many of the dSphs.  The turn-on time determines the
time or [Fe/H] at which [$\alpha$/Fe] begins to drop.  We have chosen
0.1~Gyr because that is approximately the maximum value acceptable for
the DTD data (Fig.~\ref{fig:dtd}).  See Sec.~\ref{sec:tIa3} for a
discussion of the effect of increasing this minimum delay time to
0.3~Gyr.

The instantaneous Type~Ia SN rate is given by combining $\Psi_{\rm
  Ia}$ with the past star formation history:

\begin{equation}
\dot{N}_{\rm{Ia}} = \int_t^0 \dot{\xi}_{*}(t_{\rm delay}) \, \Psi_{\rm Ia}(t-t_{\rm delay}) \, dt_{\rm delay} \: . \label{eq:SNIa}
\end{equation}

\noindent
The mass returned to the ISM is the product of the SN~Ia yields
($\zeta_{j,{\rm Ia}}$) and the Ia rate:

\begin{equation}
\dot{\xi}_{j,\rm{Ia}} = \zeta_{j,{\rm Ia}} \dot{N}_{\rm{Ia}} \label{eq:N_SNIa}
\end{equation}

\subsection{Asymptotic Giant Branch Stars}

Winds from low- and intermediate-mass stars return a small but
significant amount of mass to the ISM.  The stars lose less than 1\%
of this mass before reaching the asymptotic giant branch
\citep[AGB,][]{van97}.  Therefore, we consider mass loss on the AGB
only.

We adopt the AGB yields of \citet{kar10}, who tracked all of the
elements we consider here except Ca and Ti.  (We assume that the
fraction of Ca and Ti in AGB ejecta is the same as in the material
that formed the star.)  We assume all of the mass is ejected in the
final time step of the star's lifetime.  This assumption is
appropriate because an AGB star's thermal pulsation period, during
which it loses most of its mass, lasts on the order of 1~Myr
\citep{mar07}, which is the length of one time step in our model.
Equation~\ref{eq:lifetime_massive} gives the lifetimes of stars more
massive than $6.6~M_{\sun}$.  Less massive stars obey
\citeauthor{pad93}'s (\citeyear{pad93}) and \citeauthor{kod97}'s
(\citeyear{kod97}) equation:

\begin{equation}
\tau_*(M) = 10^{\frac{0.334-\sqrt{1.790-0.2232[7.764-\log (M/M_{\sun})]}}{0.1116}}~{\rm Gyr} \label{eq:lifetime_lowmass}
\end{equation}

Each AGB star ejects $\zeta_{j,{\rm AGB}}$ solar masses of element
$j$.  Stars lighter than $10~M_{\sun}$ participate in AGB mass loss
whereas stars heavier than $10~M_{\sun}$ explode as Type~II SNe
(Sec.~\ref{sec:II}).  The lower mass limit we consider for AGB stars
is $0.865~M_{\sun}$, which is the stellar lifetime corresponding to
the age of the Universe, 13.6~Gyr, according to
Eq.~\ref{eq:lifetime_lowmass}.  The AGB mass return rate in
$M_{\sun}~{\rm Gyr}^{-1}$ is given by

\begin{eqnarray}
\dot{\xi}_{j,\rm{AGB}} &=& 0.31~M_{\sun}^{0.2}\:\int_{0.865~M_{\sun}}^{1~M_{\sun}} \zeta_{j,{\rm AGB}}(M,Z(t-\tau_*(M)))\, \nonumber \\
 & & \; \times \; \dot{\xi}_*(t-\tau_*(M)) \, M^{-2.2} \, dM \nonumber \\
 & & \; + \; 0.31~M_{\sun}^{0.7}\:\int_{1~M_{\sun}}^{10~M_{\sun}} \zeta_{j,{\rm AGB}}(M,Z(t-\tau_*(M)))\, \nonumber \\
 & & \; \times \; \dot{\xi}_*(t-\tau_*(M)) \, M^{-2.7} \, dM  \label{eq:AGB}
\end{eqnarray}

Compared to SN ejecta, AGB ejecta affect the chemical evolution of the
elements considered here to a small degree.  AGB ejecta are more
important for other elements, such as C, N, and O.

\subsection{Gas Infall}
\label{sec:infall}

Infall of gas during the star formation lifetime of a dSph is required
to explain its MDF \citep[][Paper~III]{kir10a}.  Therefore, our model
allows pristine gas to fall into the dSph.  The gas has a helium
fraction of $Y = X_{\rm He}(0) = 0.2486$, which is the value obtained
when the WMAP7 \citep{lar10} baryon-to-photon ratio is applied to the
formula of \citet{ste07}.  The rest of the infalling gas is hydrogen.

The MDFs of the dSphs are generally more peaked than a closed box
model predicts.  One scenario that explain such a distribution is gas
infall that first increases and then decreases \citep{lyn75,pag97}.
We find that a quick increase of the rate of gas falling into the
galaxy followed by a slower decrease in the infall rate does well at
reproducing the data.  We parametrize the gas infall rate as follows.

\begin{equation}
\dot{\xi}_{j,\rm{in}} = A_{\rm in} \, X_j(t=0) \, \left(\frac{t}{\rm{Gyr}}\right) \, e^{-t/\tau_{\rm in}} \label{eq:infall}
\end{equation}

\noindent
The term $X_j(t=0)$ means that the infalling gas is primordial
(metal-free).  The variables $A_{\rm in}$ and $\tau_{\rm in}$ are free
parameters in the model.

\subsection{Supernova Winds}
\label{sec:winds}

The MDFs of dSphs require gas outflow.  If that were not the case, the
metallicities would approach the supernova yields, which are much
larger than observed in even the most metal-rich star in any dSph.
Gas may be lost through supernova winds, stellar winds, or gas
stripping from an external source.  All of these sources undoubtedly
occur over a dSph's lifetime, but supernova winds are the most
straightforward to include in a chemical evolution model.  We ignore
other sources of gas loss.

Our computation of gas loss is fairly simple.  The galaxy loses a
fixed amount of gas for every supernova that explodes.  The blown-out
gas mass does not vary with SN type because the explosion energies for
Types~II and Ia SNe are similar.  See \citet{rec01}, \citet{rom06},
and \citet{mar08} for examples of chemical evolution models that
treated the energy input from the two SNe types differently.  The rate
of gas loss is

\begin{equation}
\dot{\xi}_{j,\rm{out}} = A_{\rm out} \, X_j \, (\dot{N}_{\rm{II}} + \dot{N}_{\rm{Ia}}) \label{eq:winds}
\end{equation}

\noindent
The parameter $A_{\rm out}$ is a free parameter in the model.  An
energy argument shows that the ejected gas mass is of the order of
$10^{4}~M_{\sun}~{\rm SN}^{-1}$.  One supernova explodes with a
typical energy of $10^{51}$~erg \citep{woo95}.  In the late stages of
expansion, the kinetic energy of the ejecta is $E_{\rm ej} \sim 8.5
\times 10^{49}$~erg \citep{tho98}.  A typical line-of-sight velocity
dispersion for a dwarf galaxy is $\sigma_{\rm los} \sim 10~{\rm
  km~s}^{-1}$.  Given the virial theorem ($GM/R = 3\sigma_{\rm
  los}^2$) and the escape velocity ($v_{\rm esc}^2 = 2GM/R$), then the
gas mass ejected as a result of SN blowout is $M_{\rm ej} = E_{\rm
  ej}/v_{\rm esc}^2 = E_{\rm ej}/(6\sigma_{\rm los}^2) \sim 7 \times
10^3~M_{\sun}~{\rm SN}^{-1}$.

A metal-enhanced wind can prevent the galaxy from becoming too
metal-rich without such a large gas loss \citep{vad86}.  For
simplicity, we assume that the SN winds have the same chemical content
as the gas remaining in the galaxy.  See Sec.~\ref{sec:Zwind} for a
further discussion of including metal-enhanced winds in the model.

\subsection{Complete Chemical Evolution Equation}

The complete equation that describes the chemical evolution of the
galaxy's gas is

\begin{eqnarray}
\xi_j(t) &=& M_{\rm gas}(0) + \int_0^{t} (-\dot{\xi}_{j,*} + \dot{\xi}_{j,{\rm II}} + \dot{\xi}_{j,{\rm Ia}} + \label{eq:gce} \\
 & & \dot{\xi}_{j,{\rm AGB}} + \dot{\xi}_{j,{\rm in}} - \dot{\xi}_{j,{\rm out}})\, dt \nonumber
\end{eqnarray}

\noindent
The initial gas mass, $M_{\rm gas}(0)$, is a free parameter.  A
non-zero initial gas mass may seem inconsistent with Eq.~\ref{eq:sfr}
because the gas should form stars as it falls into the galaxy.
However, the galaxy could acquire gas available for star
formation---via gravitational collapse or cooling, for example---on a
timescale faster than the star formation timescale.  We will show that
the non-zero initial gas mass is more important for the more luminous
dSphs.

\subsection{Shortcomings of the Model}
\label{sec:shortcomings}

Our model incorporates realistic conditions in dwarf galaxies.  We
model chemical evolution using an observed Type~Ia SN DTD
\citep{mao10}.  We also take into account the lifetimes of Type~II SN
progenitors, rather than assuming instantaneous recycling.  The delay
helps to shape the metal-poor abundance distributions because it
affects the rapid rise in metallicity after the onset of star
formation.

However, our model is not as sophisticated as some other chemical
evolution models of dwarf galaxies \citep[e.g.,][]{mar08,rev09,saw10}.
In the next section, we show the best model fits to eight different MW
satellite galaxies.  The simplicity of our model reduces the
computational demand of finding the best solution.  Nonetheless, we
enumerate some shortcomings which affect the interpretation of the
abundance distributions.

\begin{enumerate}

\item The turn-on time for Type~Ia SNe is poorly constrained.
  \citet{mao10} showed that it is almost certainly 0.1~Gyr or less (at
  least in the Magellanic Clouds and higher redshift elliptical
  galaxies), but the DTD slope ($t_{\rm delay}^{-1.1}$) is divergent
  as $t_{\rm delay}$ approaches zero.  Therefore, the number of
  Type~Ia SN that explode shortly after their progenitors form depends
  sensitively on the turn-on time.  The uncertainty in the turn-on
  time translates to a large uncertainty in the Fe abundance
  distribution.  With all other model parameters held fixed, an
  earlier turn-on time would cause the metallicity of the MDF peak to
  increase and [$\alpha$/Fe] at a given metallicity to decrease.  See
  \citet{mat09} for a detailed discussion of the effect of adjusting
  the ratio of prompt to delayed Type~Ia SNe.

\item The SN yields are imperfect.  As we mention in
  Sec.~\ref{sec:dsphs}, we needed to increase the [Mg/H] output of the
  model by 0.2~dex \citep[see][]{fra04}.  Furthermore, Ti is severely
  underproduced in our model.  Therefore, we do not consider Ti
  abundances at all.  %The Type~II, rather than Type~Ia, SN yields
  %exert most of the influence over these two elements.

\item Our model assumes instantaneous mixing.  Relaxing this
  approximation would require multiple zones, which we do not consider
  for the sake of computational simplicity.  See \citet{mor02},
  \citet{mar06,mar08}, \citet{rev09}, and \citet{saw10} for
  three-dimensional chemical models of dwarf galaxies.

\item We also assume instantaneous gas cooling.  The cooling time for
  gas to become available for star formation (after accretion or
  ejection from SNe and AGB stars) may be longer than the model time
  step, $\Delta t = 1~{\rm Myr}$.  A more proper treatment of the
  cooling time, such as in a hydrodynamical model, might result in
  slightly longer SF durations that we derive with instantaneous
  cooling.

\item On a related note, we also ignore dynamical processes.  Our
  adoption of a single value of $A_{\rm out}$, the gas ejected from
  the galaxy in the wind of one supernova, implicitly assumes that the
  potential of the galaxy is homogeneous and static.  This assumption
  is inconsistent with our allowance of gas to flow into the galaxy.
  Although dark matter dominates the dynamical mass of dSphs, they
  undoubtedly change their dark matter masses during their star
  formation lifetimes \citep{rob05,bul05,joh08}.  Furthermore,
  baryonic (adiabatic) contraction can affect star formation and
  feedback in the dense centers of the dSphs \citep{nap10}.

\item We consider only one parametrization of the gas infall rate.
  Because the star formation rate is proportional to the gas mass, the
  gas infall rate essentially shapes the differential MDF.
  Differently shaped gas infall histories might better reproduce the
  dSph MDFs.  External influences on the gas flow (or alternatively,
  availability of gas cool enough to form stars) that we do not
  consider include reionization \citep{bul00} and tidal and ram
  pressure stripping \citep{lin83}.

\item We model only one episode of star formation.  CMDs have revealed
  extended and possibly bursty SFHs in several dSphs in our sample
  (Fornax and Leo~I and II).  These bursts will not be included in our
  model.  In these cases, we defer to the photometrically derived
  SFHs.  In fact, we suggest for future study a more sophisticated
  analysis that models both the CMD and abundance distributions.

\item The infalling gas is assumed to be metal-free at all times.  In
  reality, the metallicity may have increased over time because the
  source of the new gas may have been blowout from prior SF episodes
  in the galaxy in question or other galaxies.  This gas would have
  been enriched by SNe and other nucleosynthetic sources.

\item The modeling result for a given galaxy represents only part of
  that galaxy's stellar population.  Our spectroscopic samples were
  centrally concentrated to maximize the number of member stars on a
  DEIMOS slitmask, but most dSphs have radial population gradients
  \citep[e.g., Sculptor,][]{bat08}.  As a result, we preferentially
  probe the younger, more metal-rich populations.  MW satellite
  galaxies also shed stars as they interact with the Galaxy.
  \citet{maj00b} identified stars from the Carina dSph beyond Carina's
  tidal radius.  \citet{maj02} and \citet{mun06} discussed the
  implications for Carina's present stellar population.  In
  particular, the remaining stars are on average younger and more
  metal-rich than the lost stars.  Consequently, the spectroscopic
  sample favors the younger, more metal-rich stars.

\end{enumerate}

Some of these shortcomings are observational or theoretical
uncertainties (1--2), which can only be resolved with a more thorough
investigation of SN rates or yields.  Others are simplifications
(3--8), which can be resolved with more sophisticated models.  The
last shortcoming (9) could be resolved by an intensive, wide-field
campaign with the intent to recover spectra for a magnitude-limited
sample of red giants in a dSph.  This project would require a great
deal of telescope time, but it could be accomplished in principle for
one or two dSphs.  Foreground contamination could be minimized by
selecting a dSph at high Galactic latitude or photometrically
pre-selecting likely members \citep[e.g.,][]{maj00a}.

%%%%%%%%%%%%%%%%%%%%%%%%%%%%%%%%%
%%%%%%%%%   SECTION 3   %%%%%%%%%
%%%%%%%%%%%%%%%%%%%%%%%%%%%%%%%%%

\section{Gas Flow and Star Formation Histories}
\label{sec:dsphs}

We apply our chemical evolution model to eight dSphs: Fornax, Leo~I,
Sculptor, Leo~II, Sextans, Draco, Canes Venatici~I, and Ursa Minor.
We use the abundance measurements from \citeauthor*{kir10b}.  For each
galaxy, we attempt to match simultaneously the distribution of [Fe/H]
and the trends of [Mg/Fe], [Si/Fe], and [Ca/Fe] with [Fe/H] by
adjusting the six free parameters listed at the bottom of
Table~\ref{tab:gcevars}.

Unfortunately, some elemental abundances could not be matched for any
combination of parameter values.  In particular, the model
underpredicts [Mg/H] and [Ti/H].  \citet{fra04} constructed a chemical
evolution model for the Milky Way and also encountered trouble in
reproducing the yields.  They concluded that the SN yields should be
modified.  They specifically singled out Mg for being underproduced by
both Type~Ia SNe and low-mass Type~II SNe.  We feel comfortable
modifying the model results for [Mg/H] because chemical evolution
models by different authors over a wide range of galaxy masses and
ages indicate that such modification is necessary.  We add 0.2~dex to
[Mg/H] to bring the model into better agreement with the data.
However, the \citet{nom06} Type~II SN yield for [Ti/Fe] is about
$-0.1$~dex, which is far below the value observed for metal-poor stars
in dSphs or in the MW halo.  Rather than attempting to correct such a
large deficit, we ignore the model result for Ti.
\citeauthor{nom06}\ also ignore their Ti yields in their own chemical
evolution model of the solar neighborhood.

In \citeauthor*{kir10a}, we found the best-fit analytical chemical
evolution models for the same eight dSphs based on their MDFs alone.
We repeat the process here for our more sophisticated model.  As in
\citeauthor*{kir10a}, we use maximum likelihood estimation to find the
best-fit model parameters.

The likelihood that a particular model matches the data is the product
of probability distributions.  Each star is represented by a
probability distribution in a four-dimensional space.  The four
dimensions are [Fe/H], [Mg/Fe], [Si/Fe], and [Ca/Fe].  We denote these
quantities as $\epsilon_{i,j}$, where $i$ represents the $i^{\rm th}$
star and $j$ identifies one of the four element ratios.  The Gaussian
is centered on the star's observed values.  The width in each axis is
the estimate of measurement uncertainty ($\delta \epsilon_{i,j}$) in
that quantity.  Stars with larger uncertainties have less weight in
the likelihood calculation than stars with smaller uncertainties.
(Although Figs.~2--9 show only stars with uncertainties less than
0.3~dex, there is no error cut in the likelihood calculation.
Instead, we downweight stars with large uncertainties.)  The chemical
evolution model traces a path $\epsilon_j(t)$ in the four-dimensional
space. The probability that a star formed at a point $t$ on the path
is $dP/dt = \dot{M}_*(t)/M_*$, where $M_*$ is the galaxy's final
stellar mass.  The likelihood that one star conforms to the model is
the line integral of $dP/dt$ along the path $\epsilon_j(t)$.  The
total likelihood $L$ is the product of the individual likelihoods of
the $N$ stars:

\begin{eqnarray}
L &=& \prod_{i=1}^{N} \int_0^t \left(\prod_j \frac{1}{\sqrt{2\pi}\,\delta\epsilon_{i,j}} \exp \frac{-(\epsilon_{i,j}-\epsilon_j(t))^2}{2(\delta\epsilon_{i,j})^2}\right) \frac{\dot{M}_*(t)}{M_*} \, dt \nonumber \\
 & & \times \; \bigg(\frac{1}{\sqrt{2\pi}\,\delta M_{*,{\rm obs}}} \exp \frac{-(M_{*,{\rm obs}}-M_{*,{\rm model}})^2}{2(\delta M_{*,{\rm obs}})^2} \nonumber \\
 & & \times \; \frac{1}{\sqrt{2\pi}\,\delta M_{\rm{gas},{\rm obs}}} \exp \frac{-(M_{{\rm gas},{\rm obs}})^2}{2(\delta M_{{\rm gas},{\rm obs}})^2}\bigg)^{0.1N} \label{eq:lprod}
\end{eqnarray}

\noindent
The second line of the equation requires that the final stellar mass
of the model ($M_{*,{\rm model}}$) matches the observed stellar mass
($M_{*,{\rm obs}}$) within the observational uncertainties.  We adopt
the stellar masses of \citet{woo08}.  They did not study Canes
Venatici~I.  We assume that galaxy has about the same stellar mass as
Ursa Minor because it has the same luminosity within the observational
uncertainties.  The third line of the equation assures that the dSph
ends up gas free.  We fairly arbitrarily assume an uncertainty of
$\delta M_{{\rm gas},{\rm obs}} = 10^3~M_{\sun}$ because even lower
values of $\delta M_{{\rm gas},{\rm obs}}$ cause the chemical
evolution model to converge on spurious solutions.  The exponent
$0.1N$ sets the relative influence of the final stellar and gas mass
compared to the abundance distributions.  This value was chosen so
that these quantities did not dominate the likelihood but also so that
the modeled galaxies ended up gas-free and with about the correct
stellar mass.

\begin{deluxetable*}{lccccccccccc}
\tabletypesize{\scriptsize}
\tablewidth{0pt}
\tablecolumns{12}
\tablecaption{Galaxy Properties and Chemical Evolution Model Parameters\label{tab:gcepars}}
\tablehead{\colhead{dSph} & \colhead{$L$} & \colhead{$M_*$} & \colhead{$\sigma_{\rm los}$} & \colhead{$R_e$} & \colhead{$D_{\rm GC}$} & \colhead{$A_*$} & \colhead{$\alpha$} & \colhead{$A_{\rm in}$} & \colhead{$\tau_{\rm in}$} & \colhead{$A_{\rm out}$} & \colhead{$M_{\rm gas}(0)$} \\
 & ($10^5~L_{\sun}$) & ($10^5~M_{\sun}$) & (km~s$^{-1}$) & (pc) & (kpc) & $\left(\frac{10^6~M_{\sun}}{\rm Gyr}\right)$ &  & $\left(\frac{10^9~M_{\sun}}{\rm Gyr}\right)$ & (Gyr) & $\left(\frac{10^3~M_{\sun}}{\rm SN}\right)$ & ($10^6~M_{\sun}$)}
\startdata
Fornax           & $180 \pm  50$ & $190 \pm  50$ &      $10.7 \pm 0.2$ & $714 \pm  40$ &      $141 \pm  12$ & $5.02^{+2.21}_{-1.00}$ & $0.98^{+0.15}_{-0.04}$ & $2.46^{+0.70}_{-0.17}$ & $0.31^{+0.01}_{-0.04}$ & \phn $ 1.51^{+0.03}_{-0.06}$ &      $14.58^{+1.77}_{-2.46}$ \\
Leo~I            & $ 56 \pm  16$ & $ 45 \pm  13$ & \phn $ 9.0 \pm 0.4$ & $295 \pm  49$ &      $257 \pm  76$ & $0.92^{+0.67}_{-0.05}$ & $0.71^{+0.01}_{-0.17}$ & $1.17^{+0.05}_{-0.10}$ & $0.35^{+0.02}_{-0.01}$ & \phn $ 3.89^{+0.16}_{-0.08}$ & \phn $ 7.32^{+0.27}_{-0.21}$ \\
Sculptor         & $ 22 \pm  10$ & $ 12 \pm   5$ & \phn $ 9.0 \pm 0.2$ & $282 \pm  41$ & \phn $ 85 \pm  23$ & $0.47^{+0.09}_{-0.12}$ & $0.83^{+0.14}_{-0.08}$ & $0.70^{+0.12}_{-0.08}$ & $0.27^{+0.02}_{-0.02}$ & \phn $ 5.36^{+0.16}_{-0.17}$ & \phn $ 0.50^{+0.62}_{-0.25}$ \\
Leo~II           & $6.6 \pm 1.9$ & $ 14 \pm   4$ & \phn $ 6.6 \pm 0.5$ & $177 \pm  13$ &      $221 \pm  50$ & $0.43^{+0.93}_{-0.10}$ & $0.66^{+0.17}_{-0.40}$ & $0.48^{+0.23}_{-0.07}$ & $0.42^{+0.05}_{-0.09}$ & \phn $ 6.59^{+0.26}_{-0.31}$ & \phn $ 0.05^{+3.00}_{-0.04}$ \\
Sextans          & $4.1 \pm 1.2$ & $8.5 \pm 2.4$ & \phn $ 7.1 \pm 0.3$ & $768 \pm  47$ & \phn $ 98 \pm  13$ & $0.52^{+0.45}_{-0.18}$ & $0.50^{+0.20}_{-0.25}$ & $1.15^{+0.51}_{-0.20}$ & $0.22^{+0.03}_{-0.04}$ & \phn $ 9.60^{+0.86}_{-0.72}$ & \phn $ 1.55^{+2.12}_{-1.20}$ \\
Draco            & $2.7 \pm 0.4$ & $9.1 \pm 1.4$ &      $10.1 \pm 0.5$ & $220 \pm  11$ & \phn $ 92 \pm  29$ & $0.88^{+0.30}_{-0.28}$ & $0.34^{+0.16}_{-0.14}$ & $1.27^{+0.25}_{-0.18}$ & $0.22^{+0.02}_{-0.02}$ & \phn $ 9.51^{+0.43}_{-0.52}$ & \phn $ 2.32^{+1.06}_{-1.20}$ \\
Can.\ Ven.~I     & $2.3 \pm 0.4$ & $  6 \pm 2  $ & \phn $ 7.6 \pm 0.5$ & $546 \pm  36$ &      $210 \pm  29$ & $0.46^{+0.41}_{-0.26}$ & $0.36^{+0.37}_{-0.32}$ & $0.86^{+0.64}_{-0.22}$ & $0.21^{+0.04}_{-0.06}$ & \phn $ 8.83^{+0.90}_{-0.70}$ & \phn $ 0.27^{+0.81}_{-0.26}$ \\
Ursa Minor       & $2.2 \pm 0.7$ & $5.6 \pm 1.7$ &      $11.5 \pm 0.6$ & $445 \pm  44$ & \phn $ 70 \pm  19$ & $1.21^{+0.53}_{-0.11}$ & $0.26^{+0.07}_{-0.12}$ & $1.47^{+0.64}_{-0.13}$ & $0.17^{+0.02}_{-0.03}$ &      $11.04^{+0.71}_{-0.65}$ & \phn $ 0.54^{+0.71}_{-0.17}$ \\
\enddata
\tablerefs{$L$ (luminosity): \citet{martin08a} for Canes Venatici~I, \citet{irw95} for the others.  $M_*$ (stellar mass): \citet{woo08}, except that we have assumed that Canes Venatici~I has about the same $M_*$ as Ursa Minor.  $\sigma_{\rm los}$ (line-of-sight velocity dispersion) and $R_e$ (2-D projected half-light radius): \citet{wol10} and references therein.  $D_{\rm GC}$ (Galactocentric distance): Coordinates from \citet{mat98}.  See \citeauthor*{kir10b}, Table~1, for the sources of the heliocentric distances.}
\end{deluxetable*}

\begin{deluxetable}{lcccc}
\tablecolumns{5}
\tablewidth{0pt}
\tablecaption{Star Formation Durations\label{tab:duration}}
\tablehead{\colhead{dSph} & \colhead{Duration\tablenotemark{a}} & \colhead{$f_{10G}$ (D05)\tablenotemark{b}} & \colhead{$f_{10G}$ (O08)\tablenotemark{c}} & \colhead{$\tau$ (O08)\tablenotemark{d}} \\
 & (Gyr) &  &  & (Gyr)}
\startdata
Fornax           &  1.3 &     0.73 &                                             0.73 & \phn 7.4 \\
Leo~I            &  1.4 &     0.75 &                                             0.76 & \phn 6.4 \\
Sculptor         &  1.1 &     0.05 &                                             0.14 &     12.6 \\
Leo~II           &  1.6 &     0.56 &                                             0.70 & \phn 8.8 \\
Sextans          &  0.8 &     0.00 & \phantom{\tablenotemark{e}}0.00\tablenotemark{e} &     12.0 \\
Draco            &  0.7 &     0.06 &                                             0.49 &     10.9 \\
Can.\ Ven.~I     &  0.9 &  \nodata &                                          \nodata &  \nodata \\
Ursa Minor       &  0.4 &     0.00 & \phantom{\tablenotemark{e}}0.00\tablenotemark{e} &     12.0 \\
\enddata
\tablecomments{Our star formation durations for Fornax and Leo~I and II are almost certainly too short because our chemical evolution model does not permit multiple SF bursts.}
\tablenotetext{a}{Star formation duration derived from our model, based on spectroscopic, multi-element abundances.}
\tablenotetext{b}{Fraction of stars formed more recently than 10~Gyr ago, based on an analysis of HST photometry \protect \citep{dol05}.}
\tablenotetext{c}{Fraction of stars formed more recently than 10~Gyr ago, based on a different analysis of HST photometry \protect \citep{orb08}.}
\tablenotetext{d}{Stellar mass-weighted mean age, based on {\it Hubble Space Telescope} photometry \protect \citep{orb08}.}
\tablenotetext{e}{\protect \citet{orb08} did not measure these values but took them from \protect \citet{dol05}.}
\end{deluxetable}

For computational simplicity, we minimize the quantity ${\hat L} =
-\ln L$:

\begin{eqnarray}
{\hat L} &=& -\sum_{i=1}^{N} \ln \int_0^t \left(\prod_j \frac{1}{\sqrt{2\pi}\,\delta\epsilon_{i,j}^2} \exp \frac{-(\epsilon_{i,j}-\epsilon_j(t))^2}{2\delta\epsilon_{i,j}^2}\right) \frac{\dot{M}_*(t)}{M_*} \, dt \nonumber \\
 & & + \; 0.1N \bigg(\frac{(M_{*,{\rm obs}}-M_{*,{\rm model}})^2}{2(\delta M_{*,{\rm obs}})^2} \nonumber \\
 & & + \; \frac{(M_{{\rm gas},{\rm obs}})^2}{2(\delta M_{{\rm gas},{\rm obs}})^2} + \ln (2\pi) + \ln (\delta M_{*,{\rm obs}}) + \ln (\delta M_{\rm{gas},{\rm obs}})\bigg) \label{eq:lsum}
\end{eqnarray}

\noindent
We find the values of the six parameters that minimize ${\hat L}$
using Powell's method.  We calculate uncertainties on the model
parameters via a Monte Carlo Markov chain.  We perform at least
$10^4$~trials for each dSph after a burn-in period of $10^3$~trials.
The dSphs with shorter SF durations require less computation time, and
we were able to perform up to $5 \times 10^4$~trials for some of the
dSphs.  As in \citeauthor*{kir10a}, the model uncertainties are the
two-sided 68.3\% confidence intervals.  These uncertainties
incorporate only observational uncertainty and not systematic model
errors.  Table~\ref{tab:gcepars} lists the solutions for each dSph in
order of decreasing luminosity.

Table~\ref{tab:duration} lists the total star formation durations for
the most likely models.  The duration is not a free parameter but a
result of the model.  The table also lists some timescales derived
from HST CMDs \citep{dol05,orb08}.  It is not possible to measure
photometrically the total star formation duration for predominantly
ancient stellar populations because 10~Gyr isochrones are extremely
similar to 13~Gyr isochrones.  Therefore, we have quoted $f_{10G}$,
the fraction of stars formed more recently than 10~Gyr.  For small or
zero values of $f_{10G}$, the CMD shows that the population is
ancient, but there is no time resolution.  We also show the stellar
mass-weighted mean age $\tau$ \citep{orb08}.  For the three dSphs with
intermediate-aged populations (Fornax and Leo~I and II), $\tau$
combined with $f_{10G}$ gives some idea of the star formation
duration.  For example, Fornax formed $1 - f_{10G} = 27\%$ of its
stars beyond 10~Gyr ago, but the mean age is just 7.4~Gyr.  Half of
Fornax's stars formed over at least 2.6~Gyr, and the other half formed
even more recently.  Our abundance-derived duration of \agefor~Gyr is
inconsistent with this photometric star formation duration.  For
Fornax and Leo~I and II, we defer to the photometrically derived SFHs
(see item~7 of Sec.~\ref{sec:shortcomings}).  They are more realistic
because they permit an arbitrary number of SF episodes.  For the
galaxies whose CMDs identify them to be ancient, our abundance
distributions are far more sensitive probes of the SF duration than
the CMD.

In the following sections, we discuss the derived star formation and
gas flow histories for each dSph and compare them to previous
photometrically and spectroscopically derived SFHs.

\subsection{Fornax}

\begin{figure*}[t!]
\centering
 \columnwidth=.5\columnwidth
 \includegraphics[width=0.495\textwidth]{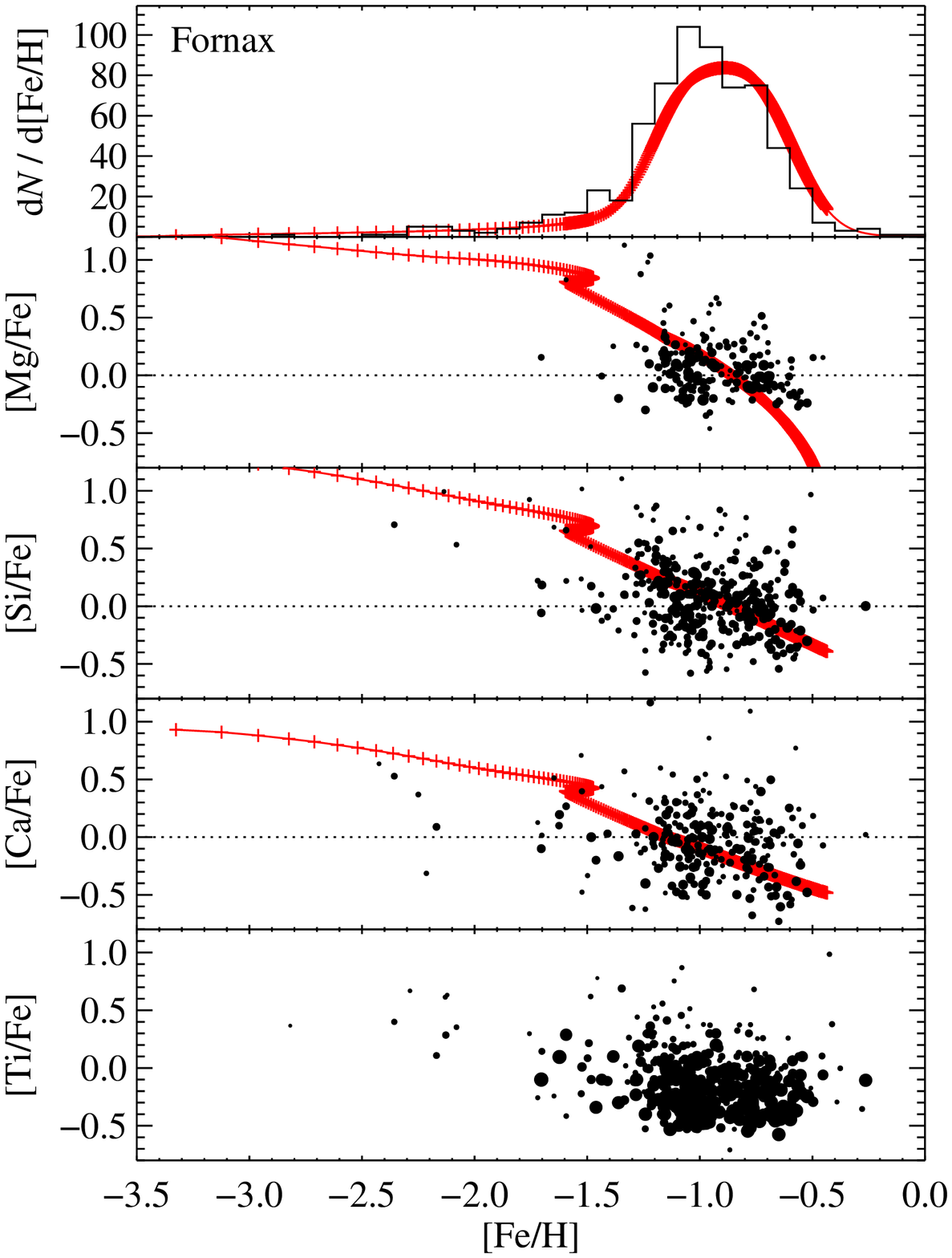}
 \hfil
 \includegraphics[width=0.495\textwidth]{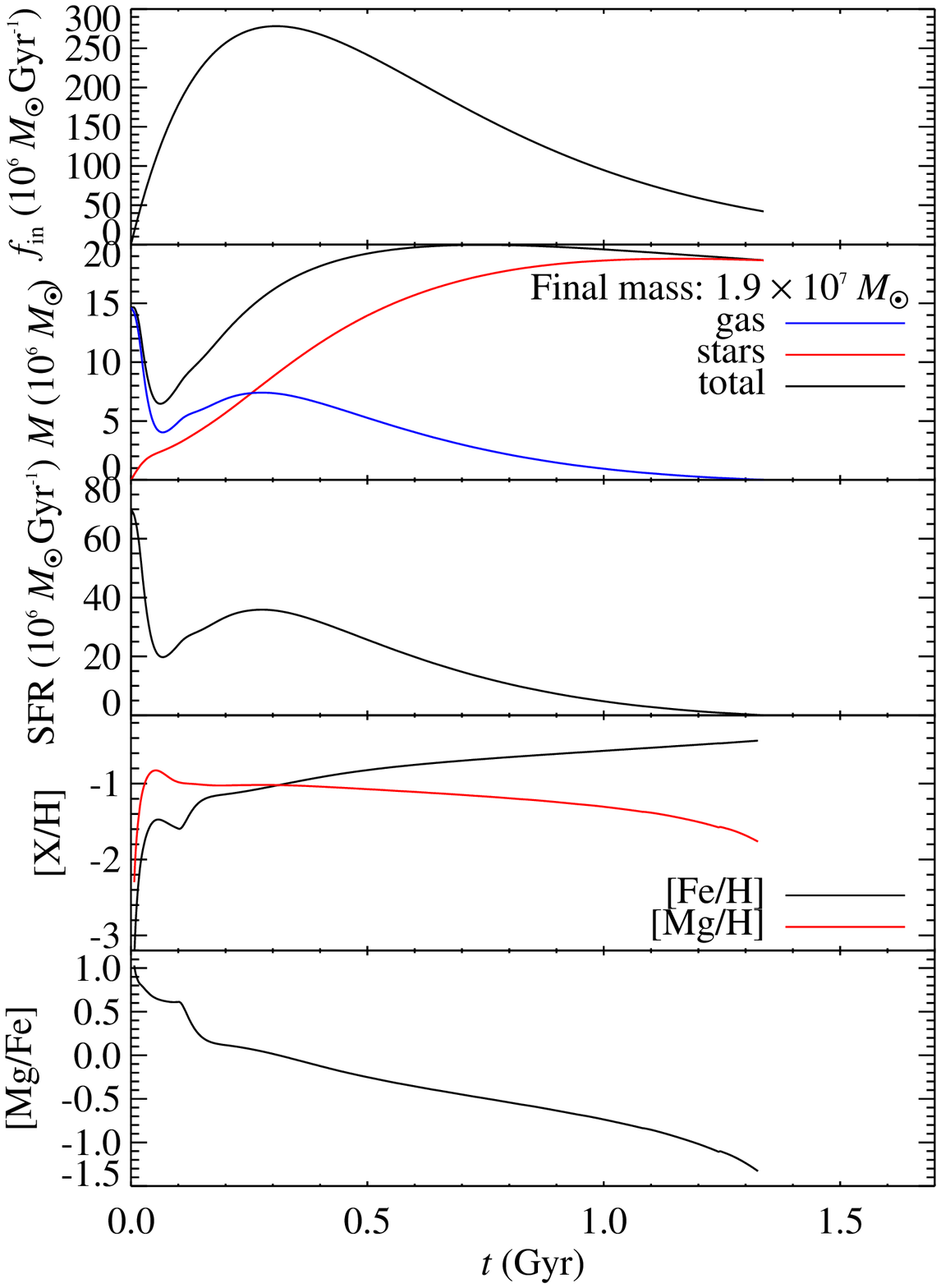}
 \caption{The observed abundance ratios and the best-fit gas flow and
   star formation history model for Fornax.  {\it Left:} The top panel
   shows the observed MDF as the black histogram and the modeled MDF
   in red.  The model is convolved with an uncertainty function to
   mimic the broadening of the histogram induced by observational
   error.  A cross marks each 1~Myr time step, but these are too
   closely spaced to discern for most of the metallicity range.  Very
   few stars are expected to have formed at the low metallicities
   where the crosses are distinguishable.  The other panels show the
   observed [Mg/Fe], [Si/Fe], [Ca/Fe], and [Ti/Fe] ratios as black
   points whose sizes are inversely proportional to measurement
   uncertainties.  Only points with uncertainties less than 0.3~dex
   are shown.  The red lines show the abundance ratios of the stars
   and gas at each time step.  We do not show the model results for
   [Ti/Fe] because the SN yields are inaccurate.  {\it Right:} The gas
   flow and star formation history for the best-fit model.  From top
   to bottom, the panels show the gas inflow rate; the stellar,
   gas-phase, and total baryonic mass; the star formation rate; the
   iron and magnesium abundances; and the [Mg/Fe] ratio, all as a
   function of time.  The second panel also gives the final stellar
   mass in the model.\label{fig:for}}
\end{figure*}

We begin our discussion with the most luminous of the mostly intact MW
dSph satellites, Fornax.  Its [$\alpha$/Fe] distribution
(Fig.~\ref{fig:for}) shows the least evidence of correlation with
[Fe/H] of all eight dSphs studied here.  In the range $-1.3 \la
\mathfeh \la -0.5$, the four [$\alpha$/Fe] element ratios span almost
1~dex at a fixed metallicity with no evidence of a slope with [Fe/H].
The rarer stars more metal poor than $\mathfeh \approx -1.3$ have
higher average [$\alpha$/Fe].

The large range of [$\alpha$/Fe] and the lack of correlation with
[Fe/H] each suggest bursty or inhomogeneous star formation.  A bursty
SFH would cause spikes and depressions in [$\alpha$/Fe] as [Fe/H]
increases monotonically \citep[e.g.,][]{gil91}, even if the star
formation were well-mixed over the whole galaxy at all times.
Measurement uncertainties might blur the division between the
[$\alpha$/Fe] spikes in different bursts.  Alternatively, if the SN
nucleosynthetic products were not well-mixed, the [$\alpha$/Fe] value
of a star would reflect the particular SFH of its birth site rather
than the galaxy as a whole.  Consequently, the abundance distribution
would be a composite of several different SFHs.  Coupled with
measurement uncertainties, the composite distribution may look like an
uncorrelated scatter of points, such as the distribution in
Fig.~\ref{fig:for}.  Burstiness and inhomogeneity are not mutually
exclusive.  Both processes might have affected Fornax's SFH.

Based on HST/Wide Field Planetary Camera~2 (WFPC2) photometry,
\citet{buo99} surmised that the field (not globular cluster)
population of Fornax endured three major bursts of star formation
separated by about 3~Gyr.  \citet{sav00}, \citet{bat06},
\citet{gul07}, and \citet{col08} provided additional photometric and
spectroscopic evidence of multiple discrete populations, including a
burst 4~Gyr ago.  \citet{gre99}, \citeauthor{bat06}, and
\citeauthor{col08}\ additionally showed that the younger, more
metal-rich populations are more centrally concentrated.  Thus, it
seems that star formation in Fornax was both bursty and inhomogeneous.

Our chemical evolution model is incompatible with Fornax's complex
SFH.  First, we model the SFR as a smooth function, not a bursty one.
Second, the model has only one zone and does not account for spatially
segregated star formation.  Consequently, the SFH derived from our
model should be viewed with skepticism.  Most notably, we derive a
total star formation duration of \agefor~Gyr (the time at which star
formation and SN winds exhausted the gas supply, thereby truncating
star formation), whereas every photometric study shows that star
formation in Fornax lasted for most of the age of the Universe.  In
addition, the model does not match the observed flatness of the
[$\alpha$/Fe] distribution for the bulk of the stars.  However, the
model does share one important quality with photometrically derived
SFHs: The initial metal enrichment is very rapid.  The metallicity in
our model reaches $\mathfeh = -1$ at 0.3~Gyr after the commencement of
star formation.  \citet{pon04} deduced that Fornax reached $\mathfeh =
-1$ within a few Gyr.  One advantage of a spectroscopically derived
SFH is that it is sensitive to relative ages, whereas a
photometrically derived SFH is sensitive to absolute ages but has poor
age resolution for old populations.  %Our abundance data may be giving
%a more precise timescale than given by photometry.

\citet{let10} measured multi-element abundances from higher resolution
spectra of 81 Fornax members.  We showed in \citeauthor*{kir10b} that
our abundance measurements match theirs very well.  They pointed out
that centrally selected stars in Fornax will preferentially sample the
young, metal-rich component.  In fact, the most metal-poor star known
in Fornax \citep[$\mathfeh = -3.66$,][]{taf10} is very far ($43'$)
from the center of the dSph.  The discovery emphasizes that selecting
stars in the center of the dSph biases the age and metallicity
distribution.

\subsection{Leo~I}
%%% TONY %%%

\begin{figure*}[t!]
\centering
 \columnwidth=.5\columnwidth
 \includegraphics[width=0.495\textwidth]{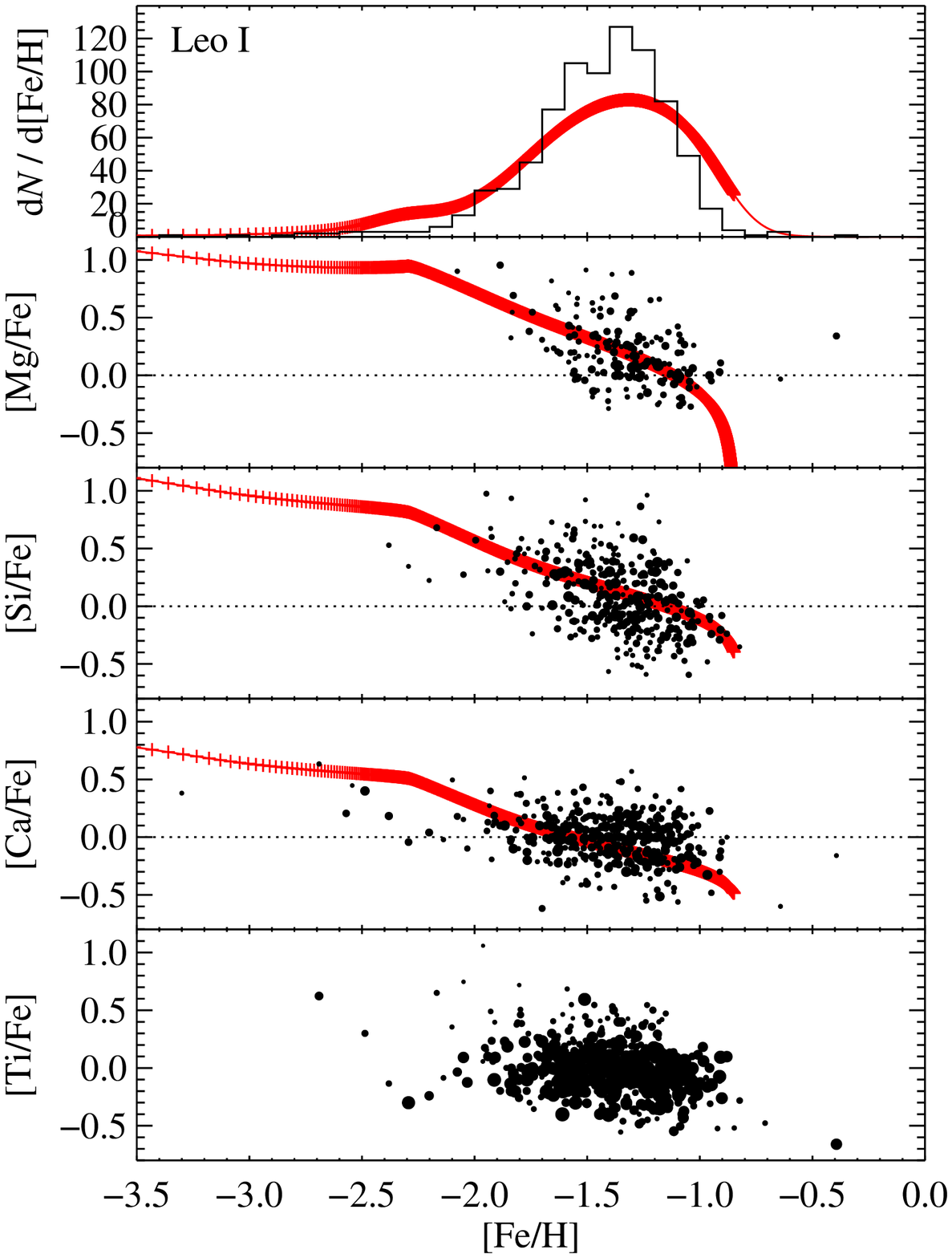}
 \hfil
 \includegraphics[width=0.495\textwidth]{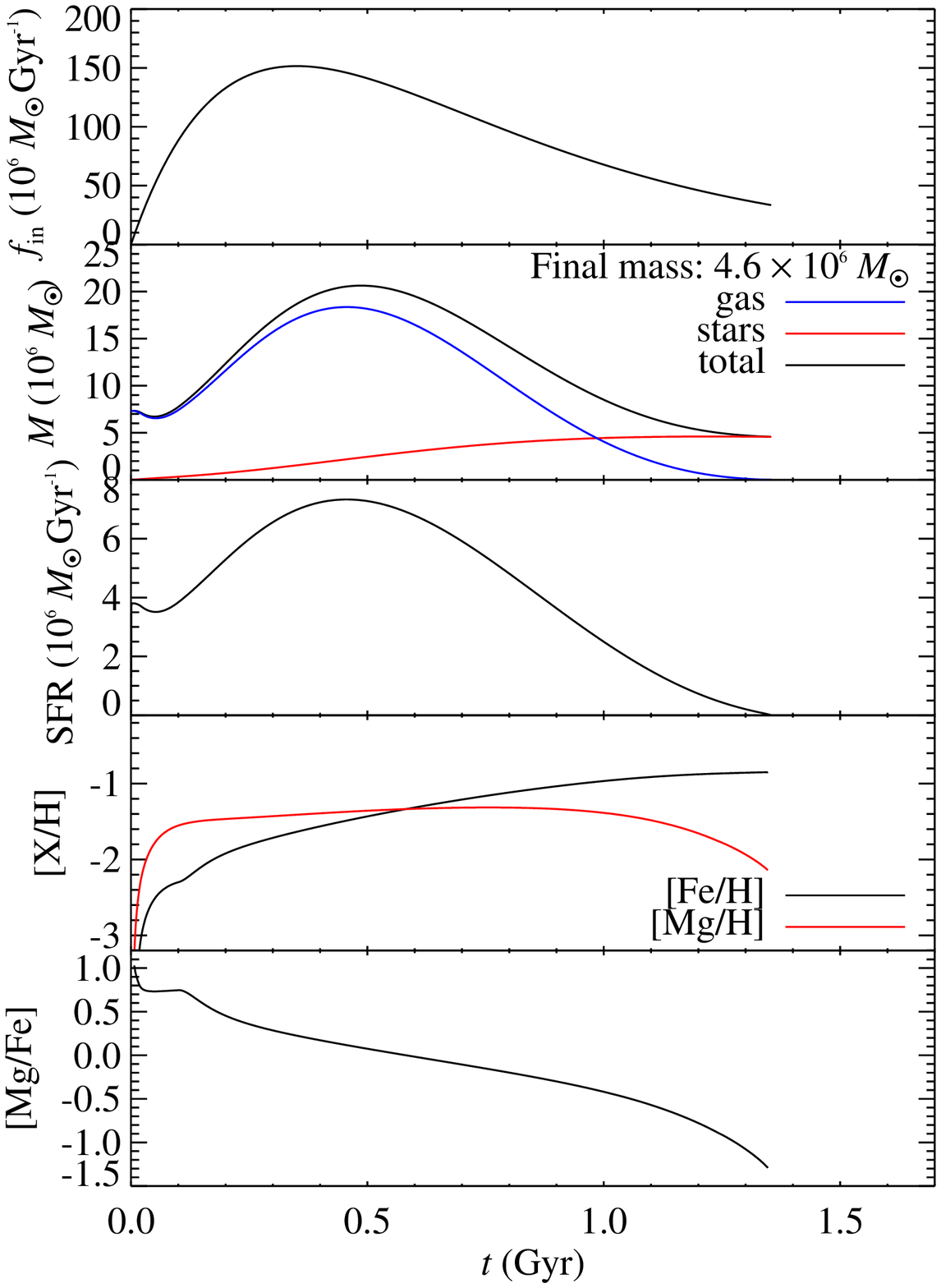}
 \caption{The observed abundance ratios and the best-fit gas flow and
   star formation history model for Leo~I.  See Fig.~\ref{fig:for} for
   a detailed explanation.\label{fig:leoi}}
\end{figure*}

Leo~I is the second most massive dwarf galaxy in our sample.  The
[$\alpha$/Fe] distribution of Leo~I shows a moderate correlation with
[Fe/H].  In particular, the lower metallicity stars ([Fe/H]$ < -1.5$)
show on average higher [$\alpha$/Fe] (except for Ti) than the more
metal-rich stars.

\citet{lee93} obtained the first CCD-based CMD of Leo~I, and they
found hints of a young (3~Gyr) population.  \citet{cap99} and
\citet{gal99a} conducted the first comprehensive studies of Leo~I's
SFH using CMDs obtained with HST/WFPC2.  Because these CMDs reached
the main-sequence turnoff of the oldest ($>10$~Gyr) populations, they
were able to study the multiple stellar populations and complex SFH.
Leo~I was thought to be unique among the MW satellite dSphs for
lacking a conspicuous horizontal branch (HB) until a $12\arcmin \times
12\arcmin$ ground-based survey of Leo~I by \citet{hel00} revealed a HB
structure in its CMD.  The existence of both an extended blue HB and
RR Lyrae stars \citep{hel01} suggested that Leo~I is in fact similar
to other local dSph galaxies in having a $> 10$~Gyr population, but
the majority of stars were still believed to have formed later than
7~Gyr ago.  However, a recent CMD obtained with HST/Advanced Camera
for Surveys/Wide Field Camera \citep{sme09} reached far deeper than
the earlier ones and showed that at least half of the stars were in
fact formed more than 9~Gyr ago, which is consistent with the abundant
RR Lyrae stars found by \citet{hel01}.  In addition,
\citeauthor{sme09}\ combined their CMD with the spectroscopic MDF of
\citet{bos07} to find that Leo~I experienced two episodes of star
formation around 2 and 5~Gyr ago.

Because our chemical evolution models halt when the gas mass drops to
zero, we are unable to recover the later phases of SFH (i.e., the two
bursts at 2 and 5~Gyr ago).  Nonetheless, our model provides insights
into the early phase with better time resolution.  Overall, our model
matches the observed trend of [$\alpha$/Fe] with [Fe/H] fairly well,
but the model MDF slightly overpredicts the frequency of metal-rich
stars.  The observed MDF also shows a more pronounced peak at [Fe/H]
$= -1.4$ than the model.  The initial starburst that likely led to the
formation of Leo~I lasted for about \ageleoi~Gyr.  This is much
shorter than the star formation duration of $\sim 5$~Gyr derived by
photometric studies.  As with other galaxies in our sample, adding
burstiness to our model would help resolve these discrepancies.
\citet{lan10} suggested that Leo~I is characterized by a low SFR and
intense galactic wind.  The main difference between their model and
ours is that we start with a much higher gas mass (by a factor of
$\sim 400$).  Also, our model requires a highly efficient SFR to match
the observed MDF.  The discrepancies with \citeauthor{lan10}\ partly
result from our choice to use unenhanced galactic winds.
Metal-enhanced winds would reduce the amount of gas required to be
blown out.  As for Fornax, our model is qualitatively consistent with
previously derived SFHs in the sense that the overall metallicity
increases quickly at early times.

Leo~I's orbital dynamics, as studied by \citet{soh07} and
\citet{mateo08}, indicate close passes to the center of the MW.  The
dSph almost certainly lost stars in tidal interactions near its
perigalacticon.  The prevalence of an intermediate-aged (rather than
old) population in Leo~I may be a consequence of this tidal stripping.
Because the stripped stars do not fall in our spectroscopic sample,
our model does not represent some stars that formed early in Leo~I's
history (see Sec.~\ref{sec:shortcomings}, item~9).
%Tidal mass loss may be particularly important for Leo~I because it
%already has a large intermediate-aged population.

%\citet{cap99}: HST SFH
%\citet{gal99a}: HST SFH
%\citet{hel00}: discovery of old population
%\citet{bos07}: CaT
%\citet{soh07}: tidal disruption
%\citet{lan10}: chemical evolution model
%\citet{hel10}: RGB and AGB stars, radial age gradient
%Tammy Smecker-Hane has unpublished ACS observations.

\subsection{Sculptor}

\begin{figure*}[t!]
\centering
 \columnwidth=.5\columnwidth
 \includegraphics[width=0.495\textwidth]{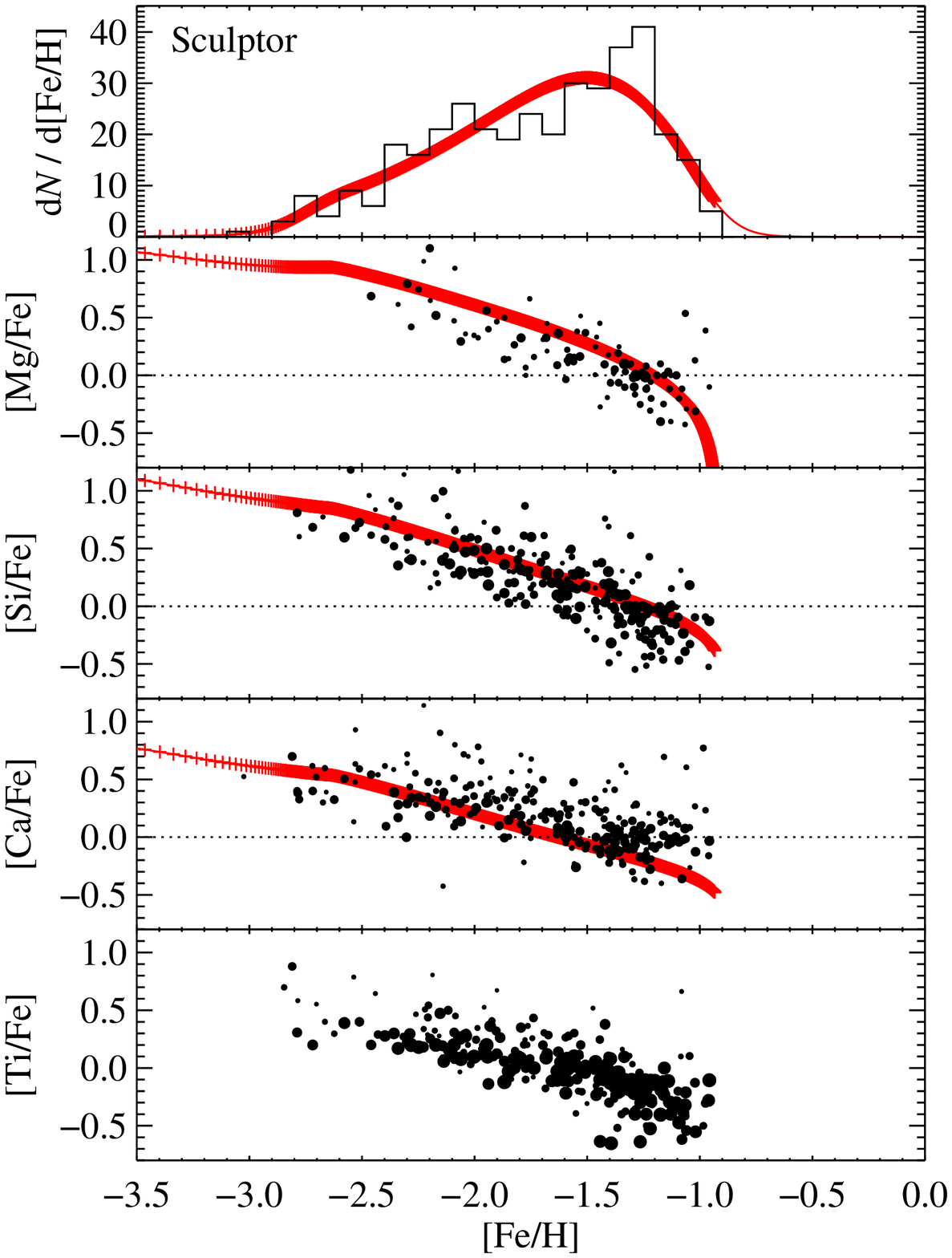}
 \hfil
 \includegraphics[width=0.495\textwidth]{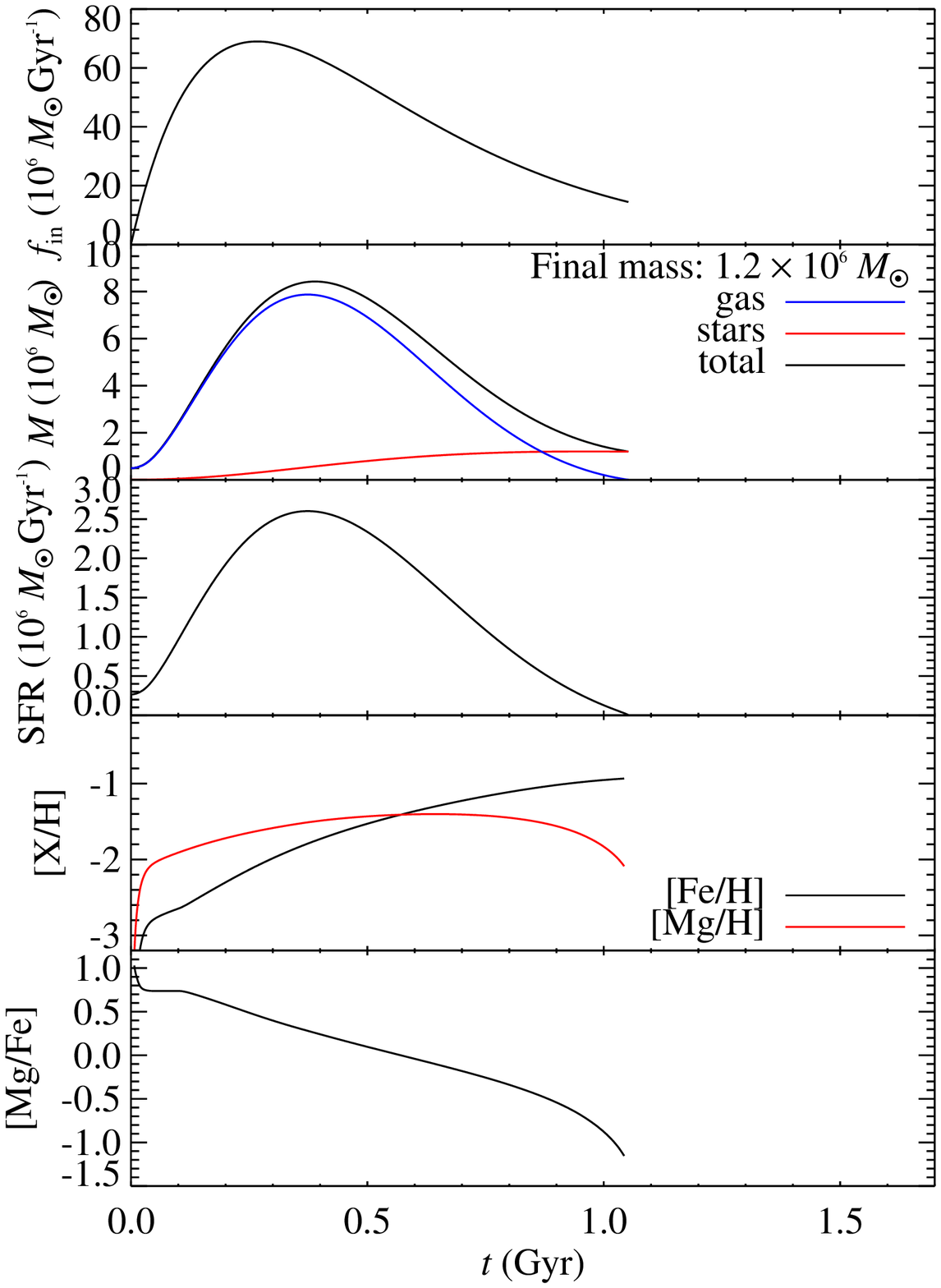}
 \caption{The observed abundance ratios and the best-fit gas flow and
   star formation history model for Sculptor.  See Fig.~\ref{fig:for}
   for a detailed explanation.\label{fig:scl}}
\end{figure*}

Our chemical evolution model for Sculptor produces one of the best
fits to the abundance distributions (Fig.~\ref{fig:scl}) out of all of
the dSphs, particularly for the asymmetrical MDF.  In \citeauthor*{kir10a},
we could not reproduce the width of Sculptor's MDF with an analytical
model of chemical evolution.  Our more sophisticated model, which more
properly treats Fe as a secondary nucleosynthetic product with
multiple origins (Types~II and Ia SNe), yields a broad, well-matched
MDF for the appropriate choice of parameters.  The combination of a
low SFR normalization ($A_*$) and low initial gas mass maintains a
lower rate of star formation than Fornax or Leo~I.  Consequently, the
metal enrichment is less rapid and the SN-induced gas blowout is less
severe.  The resulting MDF has both metal-poor and metal-rich stars
and is less-peaked than for the more luminous dSphs.

%GHS
\citet{nor78} first drew attention to the possibility that Sculptor
was chemically inhomogeneous.
%This has subsequently been studied both photometrically and
%spectroscopically
%\citep{smi83,gei94,gre94,tol01,tol04,win03,she03,bab05,cle05,gei05}.
%GHS
\citet{dac84} found that the bulk of Sculptor's stars are slightly
younger than the oldest globular clusters (GCs) but older than Fornax.
With HST/WFPC2 photometry, \citet{mon99} found that Sculptor is just
as old as the GCs.  Neither study could determine whether the bluer
stars were a younger population or blue stragglers from the older
population.  \citet{map09} presented evidence that the blue stars are
true blue stragglers, meaning that Sculptor has only an old
population.  However, old does not necessarily mean single-aged.  In
fact, \citet{maj99} found that Sculptor undoubtedly contains multiple
stellar populations based on its HB and red giant branch (RGB)
morphologies.  The existence of a metallicity spread, the depression
of [$\alpha$/Fe] with increasing metallicity, and the radial change in
HB morphology means that star formation lasted for at least as long as
the lifetime of a Type~Ia SN and possibly for a few Gyr
\citep{tol03,bab05}.

Our chemical evolution model conforms to the photometric description
of Sculptor's SFH.  According to our model, Sculptor formed stars for
\agescl~Gyr.  In fact, one of the major advantages of an
abundance-derived SFH is that it can resolve ages of old populations
much more finely than a photometrically-derived SFH.  As a result, we
believe our estimate of the star formation duration to be the most
precise presently available for Sculptor.

\citet{lan04} also found a chemical evolution model to match the five
stars with then-available multi-element abundance measurements
\citep{she03}.  Their model showed a sharp kink or knee at the time
when Type~Ia SNe ejecta began to dilute the [$\alpha$/Fe] ratio with
large amounts of Fe.  Our model shows a less pronounced knee that
occurs at lower [Fe/H] and higher [$\alpha$/Fe] primarily due to our
different treatments of the Type~Ia SN DTD\@.  \citet{rev09} modeled
unpublished abundance measurements by the Dwarf Abundances and Radial
Velocities Team (DART) for Sculptor with a sophisticated
hydrodynamical model.  They found that nearly all of the stars formed
between 10 and 14~Gyr ago, with nearly half of the stars forming at
least 13~Gyr ago.  The model supposed that the stars formed in about
five bursts.  It is possible that adding burstiness to our model would
help to reconcile the model with the observed data, such as the peak
in the MDF at $\mathfeh = -1.3$ and the discrepancy in [Ca/Fe] at high
metallicity.  However, \citeauthor{rev09}'s model predicted many more
stars at $\mathfeh < -3$ than we or DART (who sample a wider area)
observe.  A less intense initial burst (crudely approximated by the
0.3~Gyr SFR rise time in Fig.~\ref{fig:scl}) better matches the
low-metallicity MDF.  Finally, in constructing a chemical evolution
model of Sculptor, \citet{fen06} found that neutron-capture elements
contribute significantly to the ability to discriminate between
different models of star formation.  Large, high-resolution surveys
will add these elements to the dSphs' repertoire of abundance
measurements.

Like Fornax, the central regions of Sculptor are dominated by a more
metal-rich population than the outer regions \citep{bat08}.  Our
sample is centrally concentrated in order to maximize the sample size.
The selection results in a bias toward metal-rich, presumably younger
stars, possibly shortening the derived the SF duration compared to
what we would deduce from a more radially extended sample.

We also presented Sculptor's abundance distributions in
\citeauthor*{kir09} \citep{kir09}.  Minor modifications to the
abundance measurements (\citeauthor*{kir10b}) and the restriction of
the plot to points with measurement uncertainties less than 0.3~dex in
either axis cause Fig.~\ref{fig:scl} to appear slightly different from
Figs.~10--12 in \citeauthor*{kir09}.  The differences do not affect
any of the conclusions of \citeauthor*{kir09}.

%\citet{dac84}: older than Fornax
%\citet{mon99}: WFPC2, age
%\citet{hur00}: old population, low-level SF until 2~Gyr ago
%\citet{win03}: CaT MDFs
%\citet{lan04}: chemical evolution model, predicted MDF and AMR
%\citet{bab05}: width of RGB
%\citet{cle05}: MDF
%\citet{fen06}: chemical evolution model, survival after reionization
%\citet{bat08}: centrally concentrated metal-rich stars
%\citet{map09}: blue stragglers
%\citet{rev09}: hydrodynamical simulation
%\citet{kir09}

\subsection{Leo~II}

\begin{figure*}[t!]
\centering
 \columnwidth=.5\columnwidth
 \includegraphics[width=0.495\textwidth]{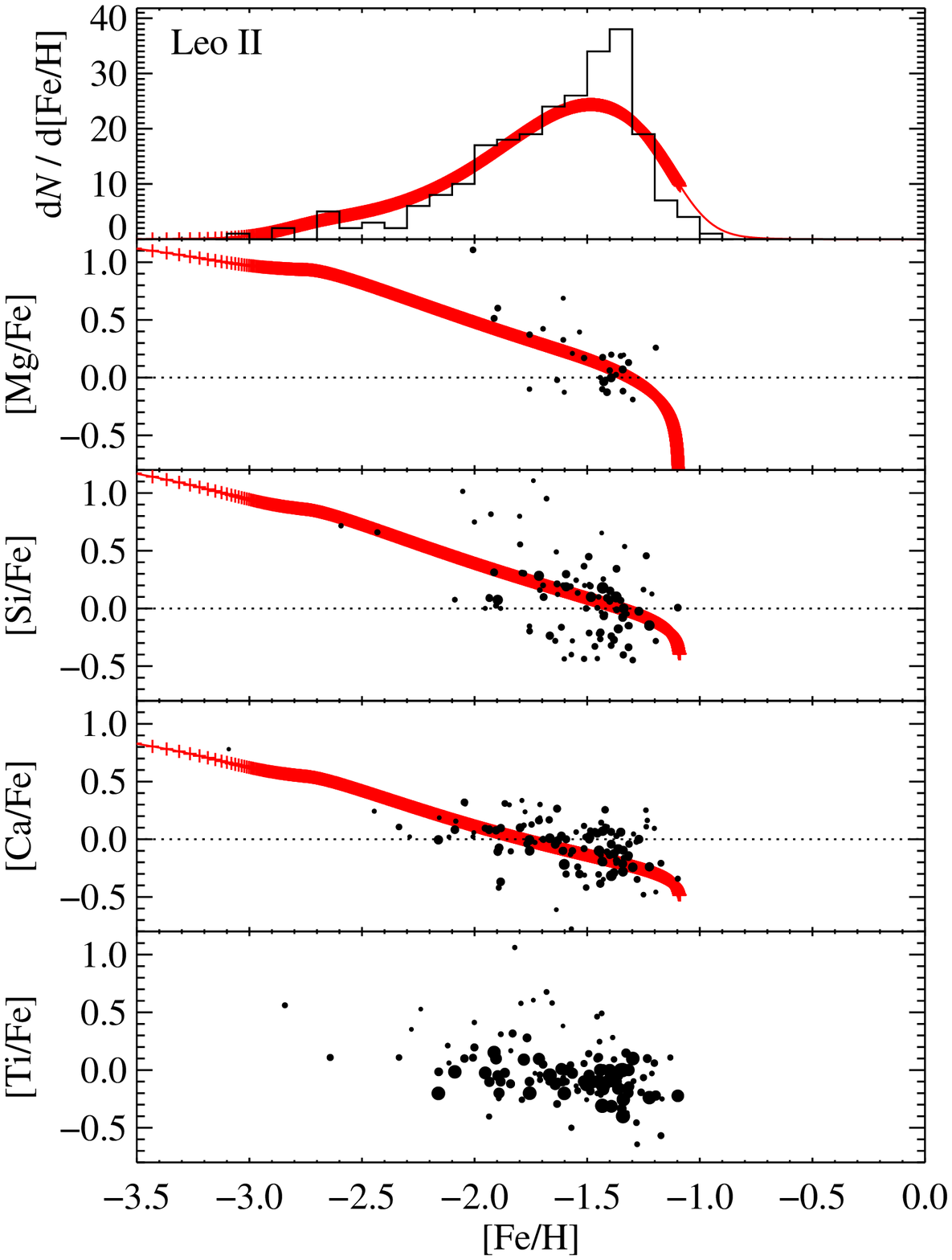}
 \hfil
 \includegraphics[width=0.495\textwidth]{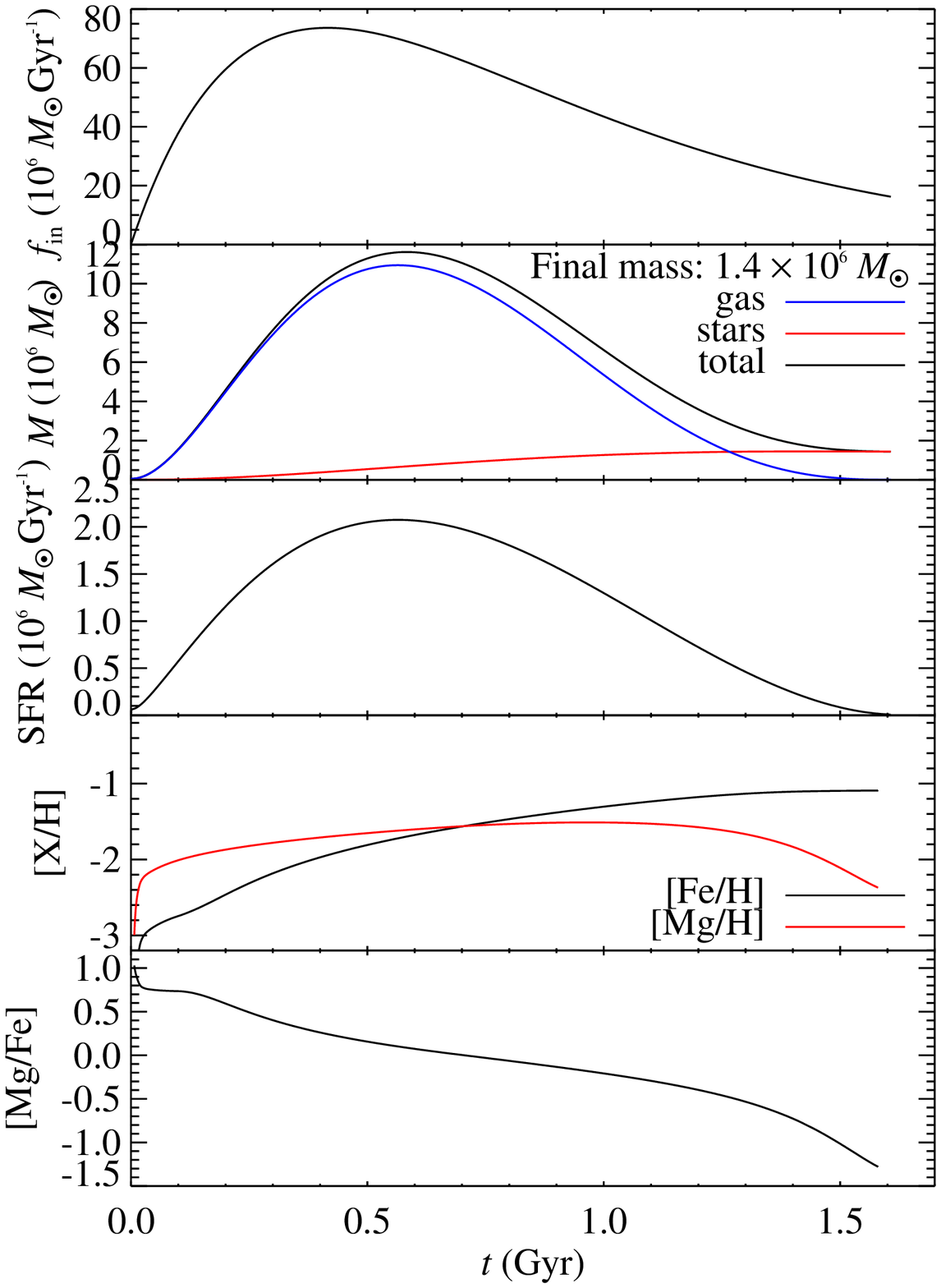}
 \caption{The observed abundance ratios and the best-fit gas flow and
   star formation history model for Leo~II.  See Fig.~\ref{fig:for}
   for a detailed explanation.\label{fig:leoii}}
\end{figure*}

The abundance distributions for Leo~II resemble Sculptor in many ways.
The MDF slowly rises to a peak followed by a sharp cut-off, and
[$\alpha$/Fe] declines smoothly with increasing [Fe/H].  The best-fit
SFH model shows a great deal of gas loss, like Sculptor.
\citet{bos07} also suggested that Leo~II may have experienced more
intense galactic winds than Leo~I due to a lower peak in the MDF.  In
fact, we find that the mass lost per SN ($A_{\rm out}$) is higher in
Leo~II ($\aoutleoii \times 10^3~M_{\sun}~{\rm SN}^{-1}$) than in Leo~I
($\aoutleoi \times 10^3~M_{\sun}~{\rm SN}^{-1}$).

Perhaps by virtue of its large Galactocentric distance (221~kpc),
Leo~II has maintained star formation for longer than Sculptor.
\citet{mig96} found from HST/WFPC2 photometry that the dSph started
forming stars 14~Gyr ago and continued forming stars for about 7~Gyr.
In a reanalysis of the same data, \citet{orb08} determined that 30\%
of Leo~II's stars formed earlier than 10~Gyr ago and 67\% formed
between 5 and 10~Gyr ago.  \citet{she09} resolved the age-metallicity
degeneracy in the CMD by using metallicities based on spectral
synthesis of Keck/LRIS spectra.  They found a significant population
of stars as young as 3~Gyr.  However, they pointed out a number of
caveats that may introduce large errors into their age measurements.

We derive a star formation duration of \ageleoii~Gyr.  Although it is
the longest duration that we measure for the eight dSphs, it does not
approach the photometrically derived durations.  The smoothness of the
modeled SFR may mask the true duration of SF.  The abundance
distributions---particularly [Si/Fe] and [Ti/Fe]---show a smattering
of points beyond the main trend line.  These stars may represent
stellar populations of temporally separated bursts.  \citet{rev09}
showed that a model with about 13 SF episodes matches the dispersion in
[Mg/Fe] at a given [Fe/H] \citep[observations by][]{she09} fairly
well.  Our model for Leo~II, like Sculptor, may benefit by adding
burstiness.

%\citet{mig96}: CMD ages
%\citet{bos07}: CaT
%\citet{rev09}: hydrodynamical simulation
%\citet{she09}: LRIS [$\alpha$/Fe]
%\citet{lan10}: chemical evolution model

\subsection{Sextans}
\label{sec:sex}

\begin{figure*}[t!]
\centering
 \columnwidth=.5\columnwidth
 \includegraphics[width=0.495\textwidth]{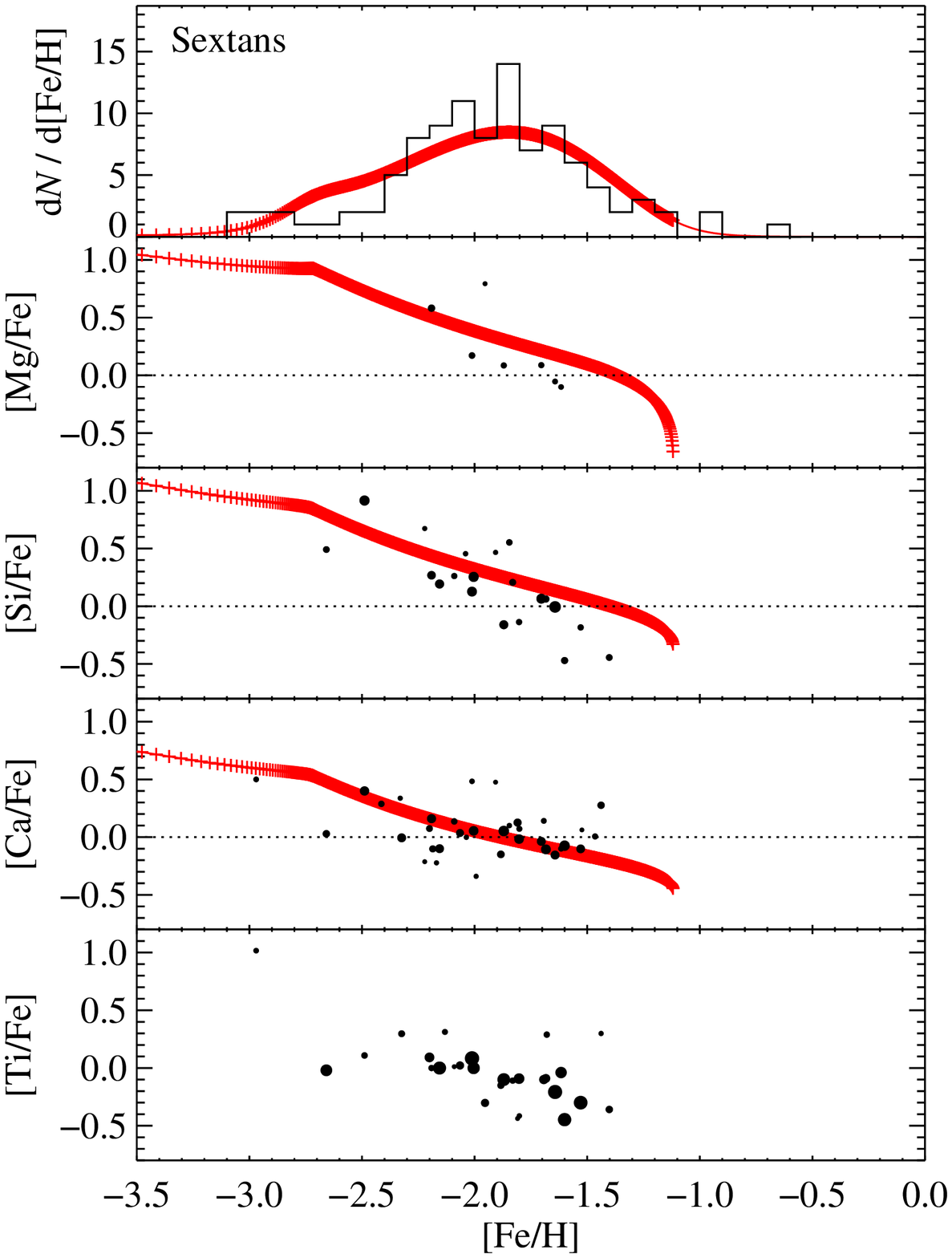}
 \hfil
 \includegraphics[width=0.495\textwidth]{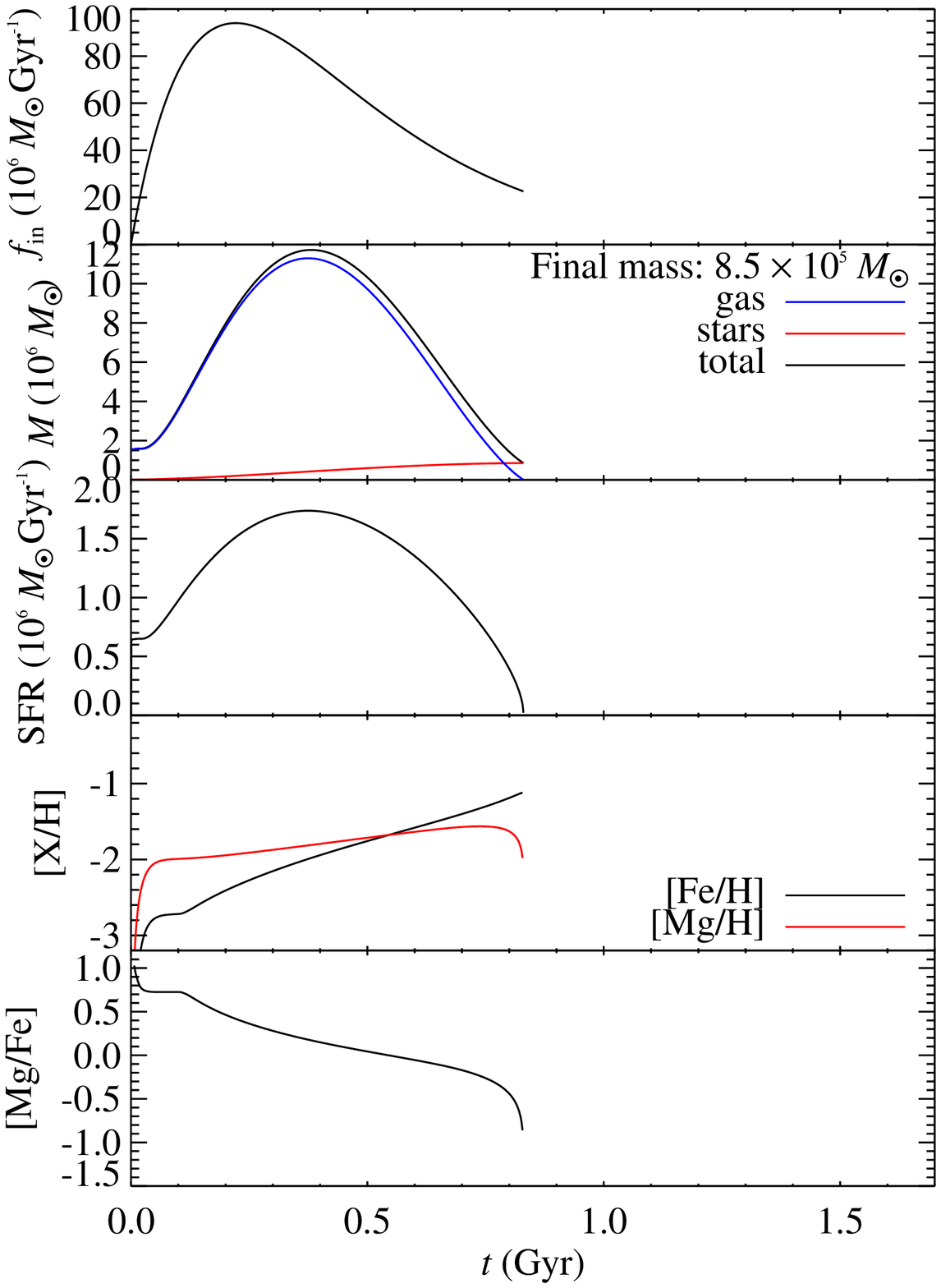}
 \caption{The observed abundance ratios and the best-fit gas flow and
   star formation history model for Sextans.  See Fig.~\ref{fig:for}
   for a detailed explanation.\label{fig:sex}}
\end{figure*}

%Battaglia et al. (2010)

Sextans, Draco, and Ursa Minor form a class of galaxies with similar
abundance distributions and SFH models.  Their MDFs are fairly
symmetric (less so for Ursa Minor) with a clump of stars at $\mathfeh
\sim -3$.  Their [$\alpha$/Fe] ratios decline smoothly with increasing
     [Fe/H].  The dispersion in [$\alpha$/Fe] at a given [Fe/H] is
     fairly small.  Most of the derived star formation parameters are
     similar (infall normalization, $A_{\rm in} \sim 1.1-1.5 \times
     10^9~{\rm Gyr}$; infall timescale, $\tau_{\rm in} \sim 0.2$;
     outflow rate, $A_{\rm out} \sim 10^4~M_{\sun}~{\rm SN}^{-1}$).

The small bump in the MDF at $\mathfeh \sim -3$ deserves some
discussion because it appears in Sextans, Draco, and Ursa Minor.  A
depression in the MDF appears between the bump and the bulk of the
MDF.  This bump might indicate a small, rapid SF burst at early times
followed by an epoch of minimal star formation, possibly because the
SNe from the initial burst blew out the gas.  When the galaxy
reacquired more cool gas, the bulk of SF began.  The few available
[$\alpha$/Fe] measurements in the bump are large, indicating that the
stars in the bump formed before the onset of Type~Ia SNe.  Because our
model does not permit individual bursts, we can not support this
speculation beyond our qualitative argument.

Despite the low metallicity and low luminosity of Sextans,
\citet{bel01} found that the dSph has at least two stellar populations
based on its HB and RGB morphology.  With HST/WFPC2 photometry,
\citet{orb08} found no stars older than 10~Gyr.  \citet{lee09}
measured Sextans's SFH based on wide field photometry coupled with an
algorithm that self-consistently derives the SFH and chemical
evolution of the galaxy.  They deduced that SF in Sextans occurred
mainly between 11 and 15~Gyr ago, but some stars formed as recently as
8~Gyr ago.  However, they assumed that Sextans is a closed box.  In
\citeauthor*{kir10a}, we showed that the MDF is inconsistent with a closed
box.  We allow gas to leave the system, which would bring an earlier
end to SF than in a closed box.  As a result, we find a SF duration of
just \agesex~Gyr.

%\citet{bel01}: multiple populations
%\citet{iku02}: chemical evolution model, extended star formation
%\citet{lan04}: chemical evolution model, predicted MDF and AMR
%\citet{lee09}: SFH as a function of radius
%\citet{rev09}: hydrodynamical simulation

\subsection{Draco}

\begin{figure*}[t!]
\centering
 \columnwidth=.5\columnwidth
 \includegraphics[width=0.495\textwidth]{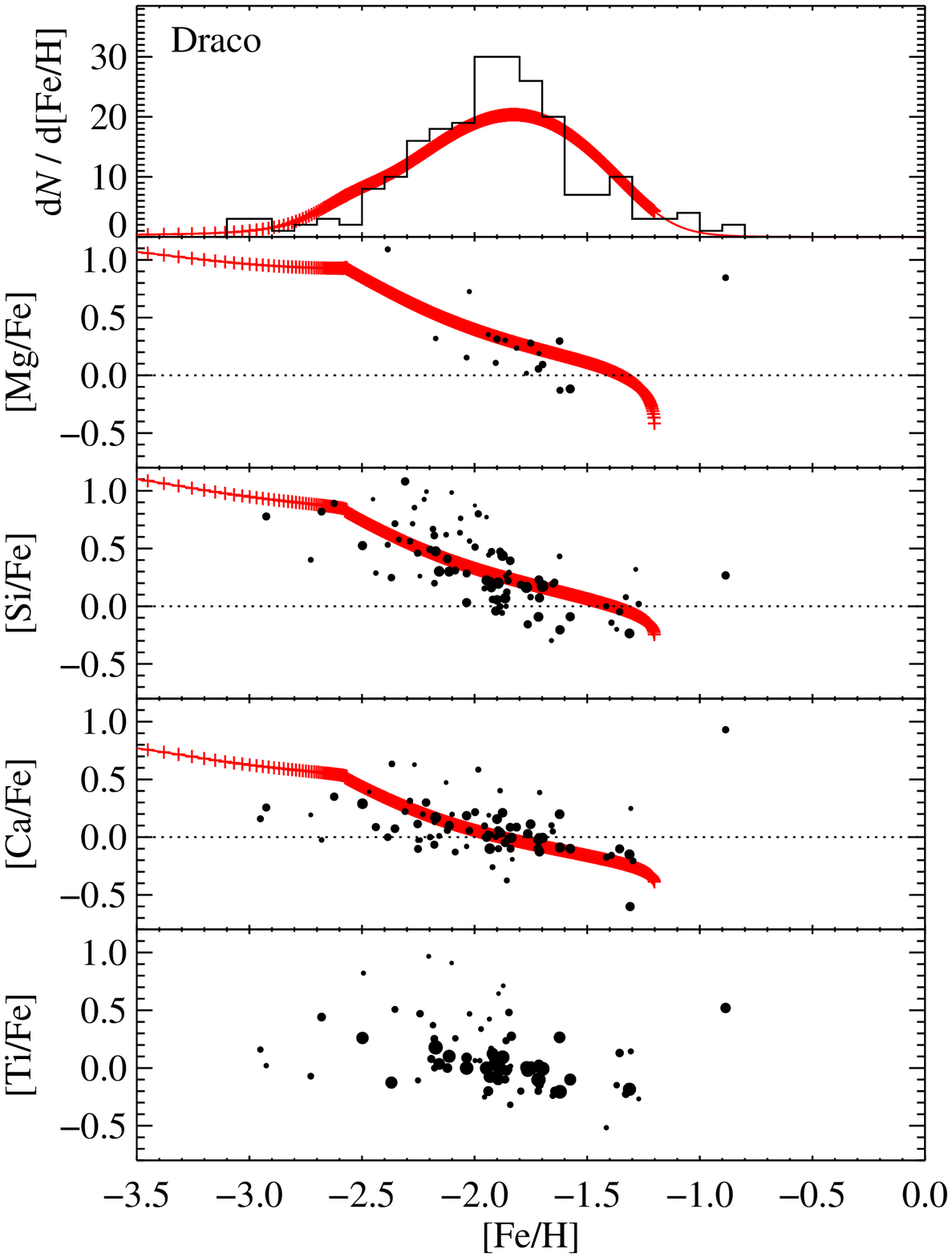}
 \hfil
 \includegraphics[width=0.495\textwidth]{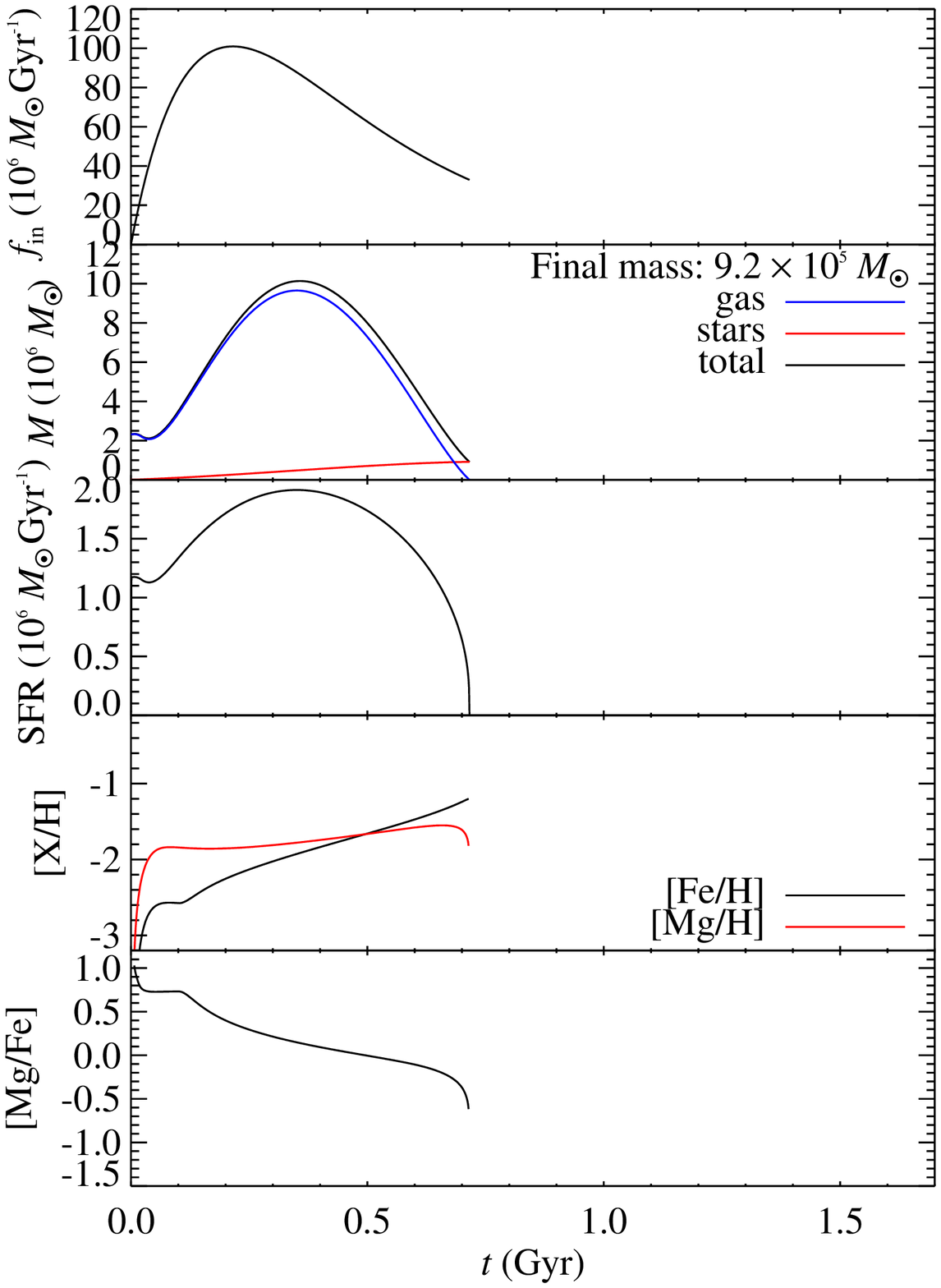}
 \caption{The observed abundance ratios and the best-fit gas flow and
   star formation history model for Draco.  See Fig.~\ref{fig:for} for
   a detailed explanation.\label{fig:dra}}
\end{figure*}

Because we conducted a more intense observational campaign on Draco
than on Sextans, we better sample Draco's abundance space.  The better
sampling does not change our qualitative description of the trio
comprised of Sextans, Draco, and Ursa Minor (see Sec.~\ref{sec:sex}).
The metal-rich side of Draco's MDF seems tiered, with fewer stars than
our model predicts at $\mathfeh = -1.5$ and $-1.2$.  The tiers may
indicate discontinuous periods of SF.

%GHS
As a consequence of its proximity, Draco was one of the first dSphs
subjected to spectroscopic scrutiny.  This system has a stellar mass
comparable to globular clusters, which are homogeneous in iron-peak
elements.  Therefore, the discovery of a metal abundance spread within
this system \citep{kin80,kin81,ste84,smi84,leh92} proved to be a
notable peculiarity.  Furthermore, Draco contains stars more
metal-poor than any globular cluster.  The first attempt to interpret
the metallicity distribution within Draco was that of \citet{zin78}.
He compared metallicities derived for 23 red giants from the Hale 5-m
multichannel scanner to a chemical evolution model that incorporated
gas loss (with a rate proportional to the SFR) but no gas inflow. In
order to account for the low metallicity of Draco, \citet{zin78}
inferred that this system had lost some 90--99\% of its initial gas
mass. Subsequent spectroscopic and photometric work has more
extensively documented the MDF and increased the number of elements
for which abundances have been measured
\citep{she98,she01a,apa01,bel02,win03,smi06,far07,abi08,coh09}.

%Unlike some other dSphs, Draco shows only limited photometric evidence
%of extended star formation \citep{gri98,apa01}.  Conesequently, the
%mass loss inferred by \citet{zin78} would have had to occur relatively
%early.  With the DEIMOS data now available it is possible to revisit
%the chemical evolution history of Draco. One interesting question that
%can be addressed is whether a gas-loss-only model is still viable, or
%whether gas inflow must also be invoked, as with the more-massive dSph
%systems. Other efforts to model the SFH of Draco are
%\citet{iku02,mar06}.
%GHS

HST/WFPC2 photometry \citep{gri98} and wide-field Isaac Newton
Telescope photometry \citep{apa01} showed little evidence for stars
younger than 10~Gyr in Draco.  On the other hand, \citet{iku02}, who
also pointed out the similarities between Sextans, Draco, and Ursa
Minor, found a longer SF duration: between 3.9 and 6.5~Gyr.  However,
\citeauthor{iku02}, like \citet{lee09}, assumed that a closed box was
an adequate description of the galaxy.  In \citeauthor*{kir10a}, we
determined that failing to account for gas outflow overpredicts the
peak metallicity of the MDF and that failing to account for gas infall
results in an MDF shape that does not match the observations.  Our
abundance-based SF duration, relaxing the closed box assumption, is
\agedra~Gyr.  Strangely, based on the same HST/WFPC2 data that
\citeauthor{gri98}\ used, \citet{orb08} determined that half of the
stars in Draco are younger than 10~Gyr.  \citeauthor{orb08}\ derived
SFHs for many dSphs, and they did not mention Draco explicitly in
their text.  As a result, we do not know why their SFH diverged from
that of \citeauthor{gri98}

\citet{coh09} analyzed high-resolution spectroscopic abundances for
eight newly observed stars and six stars from the literature.  They
fit a toy model with low- and high-metallicity plateaus in [X/Fe].
The low-metallicity plateau has a maximum metallicity of $\mathfeh =
-2.9$ for [Mg/Fe] and $-2.4$ for [Si/Fe].  We do not see a
low-metallicity plateau because our sample does not include enough
metal-poor stars.  Instead, we observe a smooth, monotonic decline in
all four [$\alpha$/Fe] ratios as a function of increasing [Fe/H].  The
absence of a low-metallicity plateau for the metallicity range of our
sample suggests that Type~Ia SNe were exploding for nearly the entire
SF lifetime of Draco.
%The absence of a high-metallicity plateau indicates that the SFR never
%achieved a constant rate where Types~II and Ia SNe could achieve an
%equilibrium.

\citet{mar06,mar08} constructed a hydrodynamical model of a Draco-like
dSph.  In order for [$\alpha$/Fe] to drop to 0.2~dex, their modeled
dSph must have evolved for at least 2~Gyr.  However, at small
radius---the location of most spectroscopic surveys, including the
majority of our Draco sample---[$\alpha$/Fe] does drop to lower values
sooner than in the dSph as a whole.  Nonetheless,
\citeauthor{mar08}\ predicted mostly stars with [$\alpha$/Fe] larger
than 0.2~dex with a plateau at low metallicity.  We observe neither of
these qualities.  Nonetheless, their model does qualitatively
reproduce important features of dSph abundance distributions,
including radial gradients in both [Fe/H] and [$\alpha$/Fe], the shape
of the MDF, and an anti-correlation between metallicity and velocity
dispersion.

Finally, we point out that, according to our model, Draco lost an
enormous amount of gas from SN winds during its SF lifetime.
\citet{lan07} used Draco and Ursa Minor as case studies in the
importance of SN winds.  One interesting divergence from our model is
that they found that a wind intensity proportional to the SFR rather
than the SN rate better voided the dSph of gas by the present time, in
agreement with the observed absence of gas.  Our different
prescription for the Type~Ia DTD may mitigate the difference between
the SFR and SN rate.

%\citet{gri98}: single-epoch SF
%\citet{apa01}: up to 90\% of population is older than 10~Gyr
%star formation as recent as 2~Gyr
%\citet{iku02}: chemical evolution model, extended star formation
%\citet{win03}: CaT MDFs
%\citet{lan04}: chemical evolution model, predicted MDF and AMR
%\citet{smi06}: metallicity inhomogeneity
%\citet{mar06,mar08}: hydrodynamical simulation
%\citet{lan07}: chemical evolution model, effects of galactic winds
%\citet{coh09}: high-resolution, multi-element abundances

\subsection{Canes Venatici~I}

\begin{figure*}[t!]
\centering
 \columnwidth=.5\columnwidth
 \includegraphics[width=0.495\textwidth]{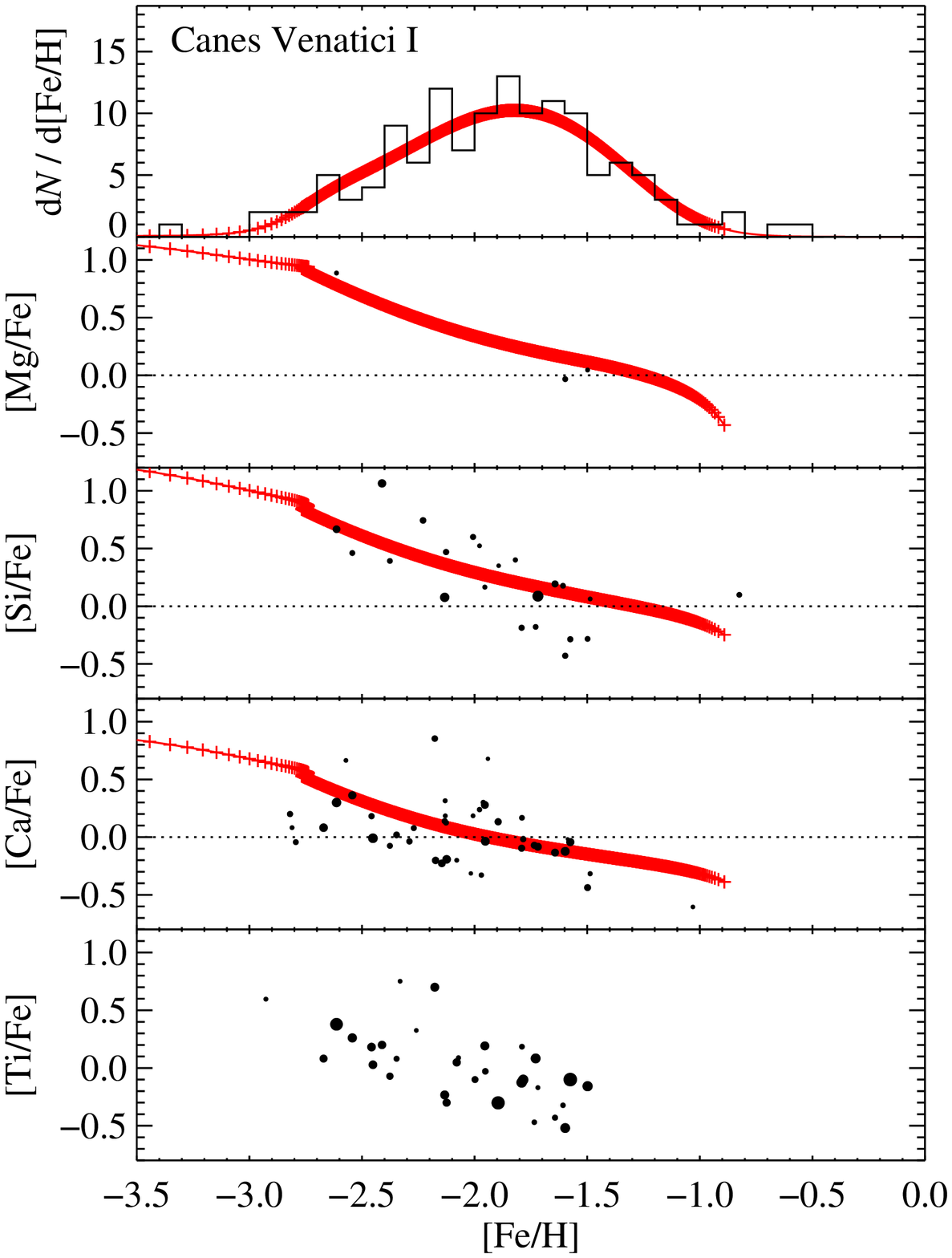}
 \hfil
 \includegraphics[width=0.495\textwidth]{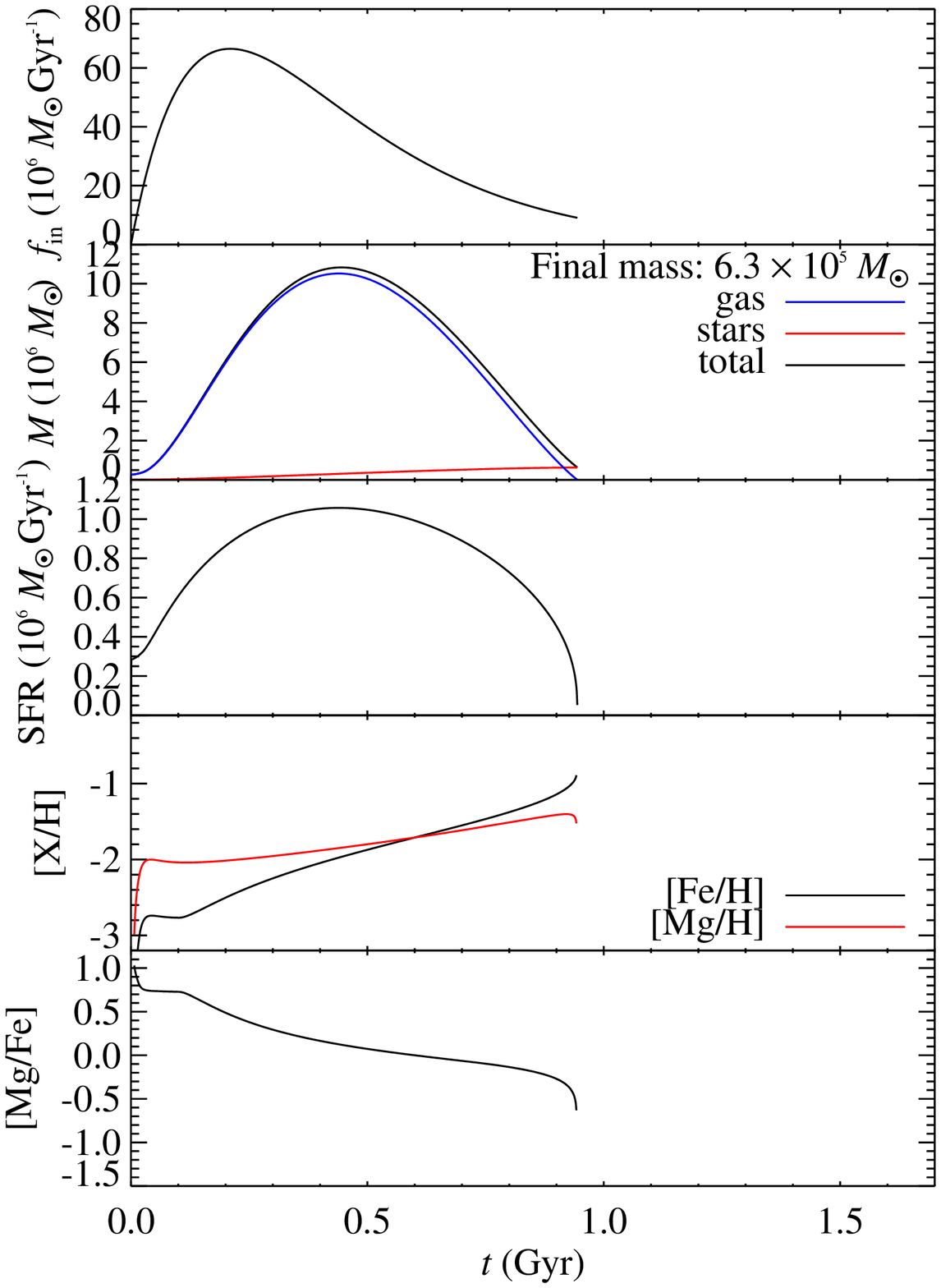}
 \caption{The observed abundance ratios and the best-fit gas flow and
   star formation history model for Canes Venatici~I.  See
   Fig.~\ref{fig:for} for a detailed explanation.\label{fig:cvni}}
\end{figure*}

Of all of our dSph models, that for Canes Venatici~I adheres most
closely to the observed abundance distributions, in part because of
the sparse sampling.  The MDF is a perfect match, and the predicted
[$\alpha$/Fe] line passes through the observed locus of points, except
for veering to slightly high [$\alpha$/Fe] values at high [Fe/H].
Unfortunately, only three stars pass the [Mg/Fe] uncertainty cut of
0.3~dex.  More measurements of [Si/Fe] and [Ca/Fe] help us to
determine a SF duration of \agecvni~Gyr and an unusually low SFR
exponent of $\alpha = \alphacvni$.  The weaker dependence on gas mass
shapes the SFR profile in such a way that produces a more symmetric
MDF while preserving a steadily declining [$\alpha$/Fe] distribution
with increasing [Fe/H].

Because Canes Venatici~I was discovered recently \citep{zuc06}, few
photometric studies exist.  \citet{martin08b} found that the dSph
contains mostly stars older than 10~Gyr, but 5\% of the stars could be
as young as 1.4~Gyr.  \citet{kue08}, with a shallower CMD, found
possible evidence for a population as young as 0.6~Gyr.  They also
found three candidate anomalous Cepheid variables, indicating an
intermediate-age population.  Because the young population is much
smaller than the old population, our chemical evolution model and its
SF duration should be viewed as applicable to the dominant old
population.

%\citet{martin08b}: huge range of ages
%\citet{kue08}: stars ranging in age from 13~Gyr to 0.6~Gyr

\subsection{Ursa Minor}
%%% JUDY %%%

\begin{figure*}[t!]
\centering
 \columnwidth=.5\columnwidth
 \includegraphics[width=0.495\textwidth]{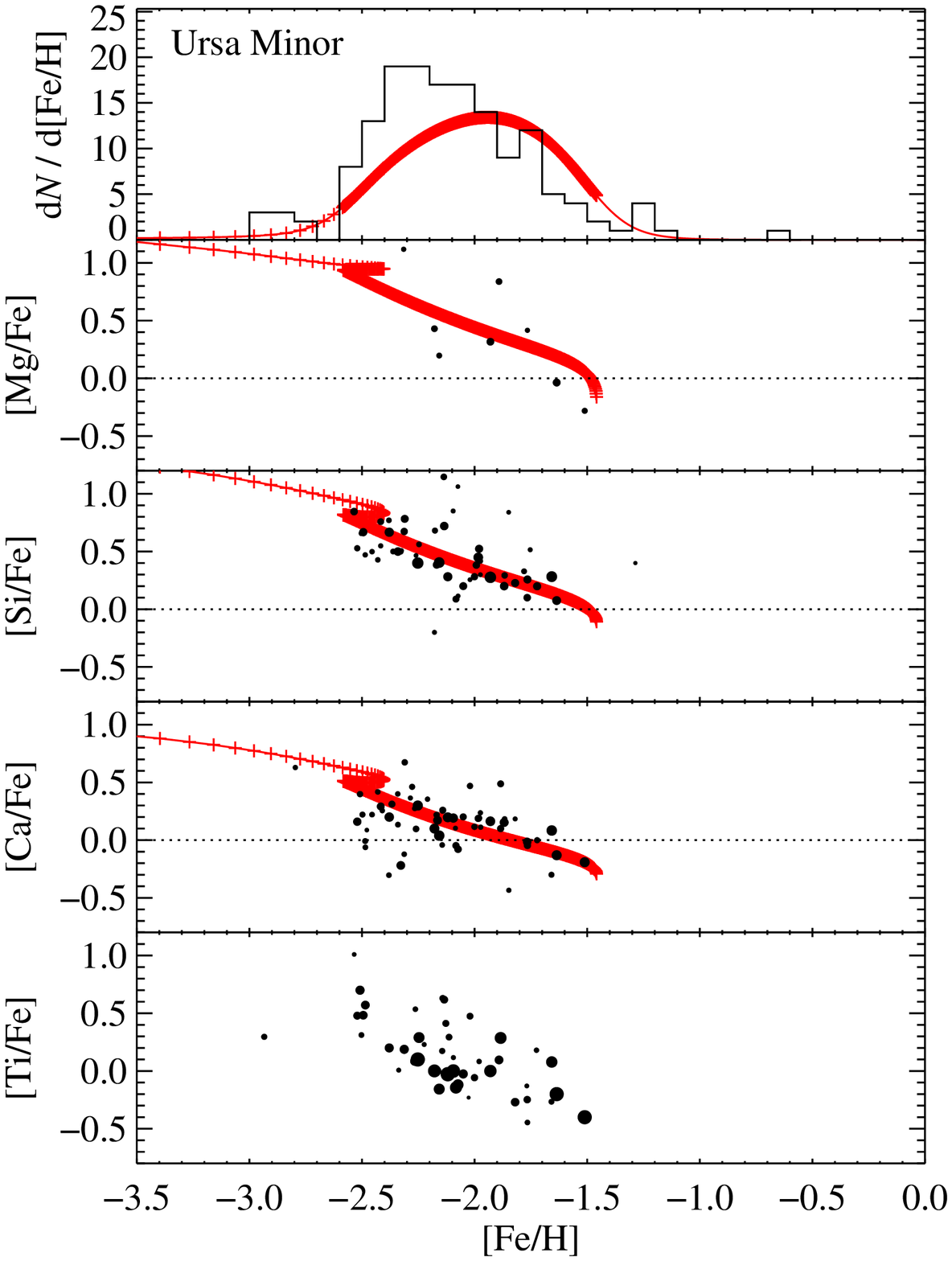}
 \hfil
 \includegraphics[width=0.495\textwidth]{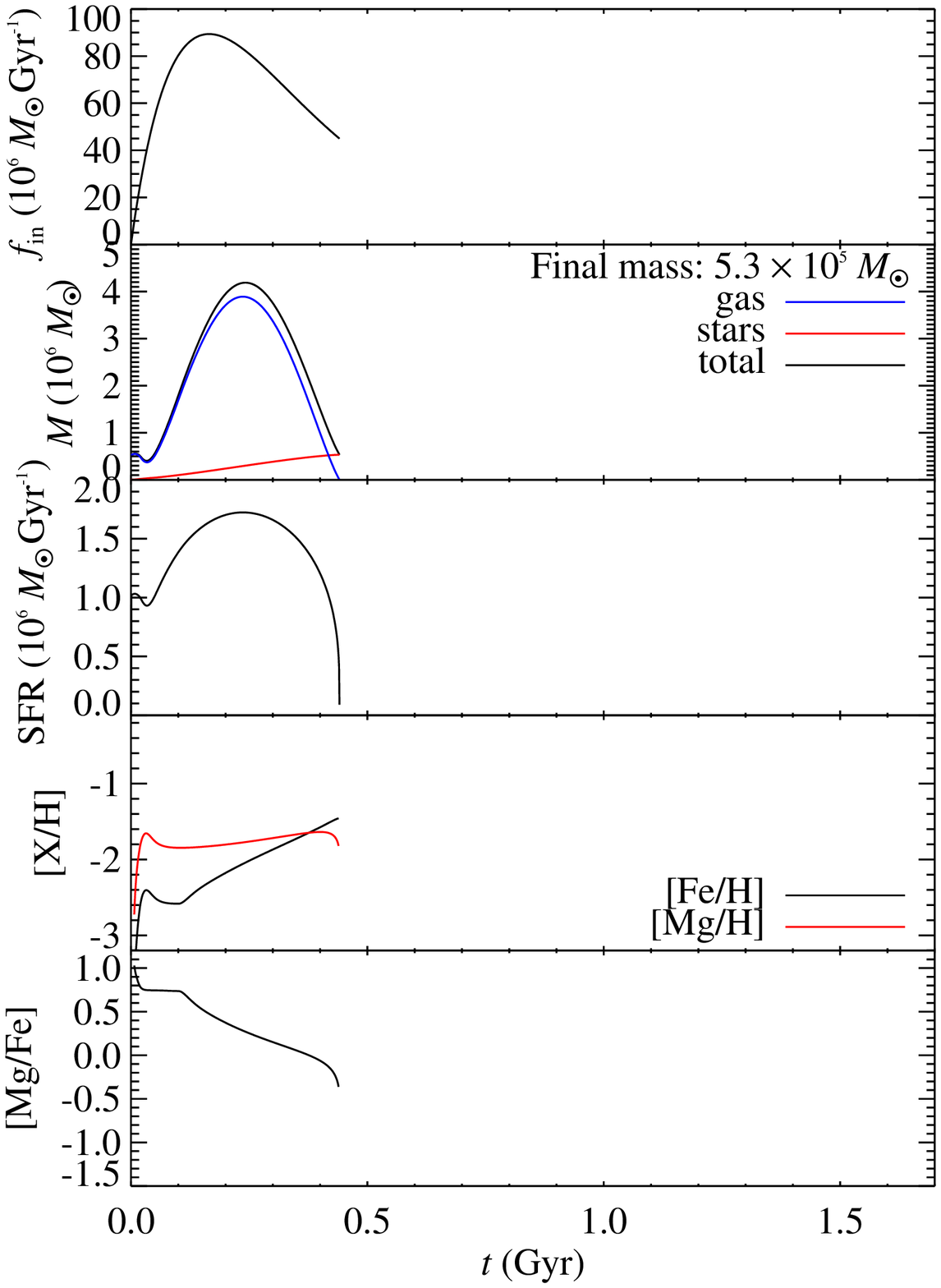}
 \caption{The observed abundance ratios and the best-fit gas flow and
   star formation history model for Ursa Minor.  See
   Fig.~\ref{fig:for} for a detailed explanation.\label{fig:umi}}
\end{figure*}

%GHS
The low-mass Ursa Minor dSph has sometimes been studied in comparison
with the Draco dSph, in regard to both its metallicity inhomogeneity
and stellar population
\citep{zin81,ste84,bel85,she01b,bel02,win03,abi08}.  A relatively
small age spread and an ancient mean age \citep{ols85,mig99,carr02}
also makes it also an interesting contrast to halo globular
clusters. However, spectroscopy has shown that Ursa Minor has a heavy
element abundance spread of more than 1~dex
\citep{zin81,she01a,win03,sad04,coh10} even though its stellar mass is
similar to that of a GC.
%GHS

\citet{cud86} conducted a photometric survey of Ursa Minor down to the
HB.  With $\sim 450$ members, they found that the stellar population
resembles that of an old, metal-poor GC with a steep RGB and a blue
horizontal branch.  The HST/WFPC2 imaging study of \citet{mig99}
confirmed this SFH: a single major burst of star formation about
14~Gyr ago with a duration of less than 2~Gyr.  Our best-fit model
agrees with these earlier results.  From our observed abundance
distributions, we deduce that almost all of the star formation in Ursa
Minor occurred over an interval of only \ageumi~Gyr.  In contrast,
\citet{iku02} derived an extended period of star formation lasting for
about 5~Gyr from their closed-box analysis of the CMD.  In
\citeauthor*{kir10a}, we showed that Ursa Minor's MDF is inconsistent with
a closed box.  \citet{coh10} used metallicities from moderate
resolution spectra combined with ages from isochrones to reaffirm that
most of the stars in Ursa Minor are quite old.

MDFs have been generated from photometric surveys by \citet{bel02} and
from moderate resolution spectroscopy by \citet{win03}.  That of
\citet{bel02} is a good match to our observed MDF given in Fig.~9.
Both show a sharp rise to a peak metallicity of about $-2$~dex with a
more gradual decline towards higher [Fe/H].  The best fit chemical
evolution model for Ursa Minor produces an MDF that fails to match the
rapid rise seen at $\mathfeh \la -2.3$~dex.

\citet{coh10} provided detailed abundance analyses for a sample of 16
RGB stars, 6 of which came from earlier work by \citet*{she01a} or
from \citet{sad04}.  Their trends for [Mg/Fe], [Si/Fe], [Ca/Fe], and
[Ti/Fe] agree qualitatively with those found here, but their sample
has better coverage of the regime $\mathfeh < -2.5$~dex, where they
found a plateau in [$\alpha$/Fe].  At very low metallicity,
[$\alpha$/Fe] in our models reaches highly supersolar ratios, which
are larger than those observed at the metal-poor end of the Ursa Minor
population by \citeauthor{coh10}.

Previous chemical evolution models of Ursa Minor include those of
\citet{lan04}, who found that Ursa Minor has the shortest duration of
star formation of any of the six dSph satellites they studied.  They
deduced that Ursa Minor experienced only a single burst lasting
perhaps 3~Gyr, a moderately high star formation efficiency, and an
intermediate wind efficiency.  In our model the wind efficiency,
$A_{\rm out}$, is the highest of all the dSphs in our sample (see
Tab.~\ref{tab:gcepars}).  \citeauthor{lan04}'s predicted MDF fails at
low [Fe/H], as does ours, by being too extended.  In a later paper,
\citet{lan07} studied the effect of galactic winds.  They concluded
that a strong galactic wind is necessary to reproduce the rather low
[Fe/H] of the peak of the Ursa Minor MDF, but they still failed to
reproduce the sudden scarcity of stars more metal-poor than the MDF
peak.

Both \citet{mar01} and \citet{mun05} have discovered tidal debris
around Ursa Minor.  As we discussed in Sec.~\ref{sec:shortcomings}
(item~9), our observations are centrally concentrated and therefore
biased toward the relatively younger, more metal-rich population that
is still bound to the dSph.  A truly complete analysis of Ursa Minor's
SFH must also include the tidally stripped, unbound stars.

%\citet{mig99}: WFPC2, metal-poor, ancient population
%\citet{carr02}: 90\% of stars are at least 13~Gyr old, but there is a
%blue plume that could be 2~Gyr-old main sequence
%\citet{iku02}: chemical evolution model, extended star formation
%\citet{win03}: CaT MDFs
%\citet{lan04}: chemical evolution model, predicted MDF and AMR
%\citet{lan07}: chemical evolution model, effects of galactic winds
%\citet{coh10}: high-resolution, multi-element abundances
%%\citet{bon10}

%%%%%%%%%%%%%%%%%%%%%%%%%%%%%%%%%
%%%%%%%%%   SECTION 4   %%%%%%%%%
%%%%%%%%%%%%%%%%%%%%%%%%%%%%%%%%%

\section{Further Exploration of the Chemical Evolution Model}
\label{sec:exploration}

In this section, we explore the parameters of the chemical evolution
model that were previously not allowed to vary.  Namely, we examine
the dependence of the outcome of the model on the Type~Ia SN delay
time distribution, the hypernova fraction, and the metal enhancement
of supernova winds.  We have chosen Sculptor as a case study.  In each
of the following three sections, we alter one aspect of the chemical
evolution model for Sculptor.  Then, we use Powell's method to find
the combination of the six free parameters that maximizes the
likelihood estimator, as before.  A Monte Carlo Markov Chain of at
least $10^4$ trials provides the two-sided 68.3\% confidence intervals
for the first two altered models.  Table~\ref{tab:systematics}
compares the results of the new models with the original model.

\begin{deluxetable*}{lcccc}
\tablecolumns{5}
\tablewidth{0pt}
\tablecaption{Sensitivity of Sculptor Model Parameters to Assumptions\label{tab:systematics}}
\tablehead{\colhead{Parameter} & \colhead{Baseline} & \colhead{${\rm{min}}(t_{\rm{delay}}) = 0.1~{\rm{Gyr}}$} & \colhead{$\epsilon_{\rm{HN}} = 0.5$} & \colhead{Metal-Enhanced Wind}} \\
\startdata
$A_*$ ($10^6~M_{\sun}~{\rm Gyr}^{-1}$)        & $0.47^{+0.09}_{-0.12}$ & $0.12^{+0.03}_{-0.06}$ & $0.68^{+0.10}_{-0.18}$ & $0.85$ \\
$\alpha$                                      & $0.83^{+0.14}_{-0.08}$ & $0.93^{+0.30}_{-0.09}$ & $0.84^{+0.16}_{-0.06}$ & $0.93$ \\
$A_{\rm in}$ ($10^9~M_{\sun}~{\rm Gyr}^{-1}$) & $0.70^{+0.12}_{-0.08}$ & $0.07^{+0.01}_{-0.01}$ & $1.02^{+0.15}_{-0.14}$ & $0.29$ \\
$\tau_{\rm in}$ (Gyr)                         & $0.27^{+0.02}_{-0.02}$ & $0.84^{+0.06}_{-0.06}$ & $0.21^{+0.02}_{-0.01}$ & $0.13$ \\
$A_{\rm out}$ ($10^3~M_{\sun}~SN^{-1}$)       & $5.36^{+0.16}_{-0.17}$ & $5.38^{+0.18}_{-0.21}$ & $5.14^{+0.13}_{-0.17}$ & $0.53$ \\
$M_{\rm gas}(0)$ ($10^6~M_{\sun}$)            & $0.50^{+0.62}_{-0.25}$ & $0.60^{+1.01}_{-0.32}$ & $0.00^{+0.17}_{-0.00}$ & $0.33$ \\
SF duration (Gyr)                             & 1.05 & 3.66 & 0.82 & 1.26 \\
\enddata
\tablecomments{We were unable to compute uncertainties for the Metal-Enhanced Wind model because the model is numerically unstable to small perturbations.  The SF duration is a derived value, not a free parameter, and we did not calculate its uncertainty.}
\end{deluxetable*}

\subsection{Type Ia Delay Time Distribution}
\label{sec:tIa3}

\begin{figure*}[t!]
\centering
 \columnwidth=.5\columnwidth
 \includegraphics[width=0.495\textwidth]{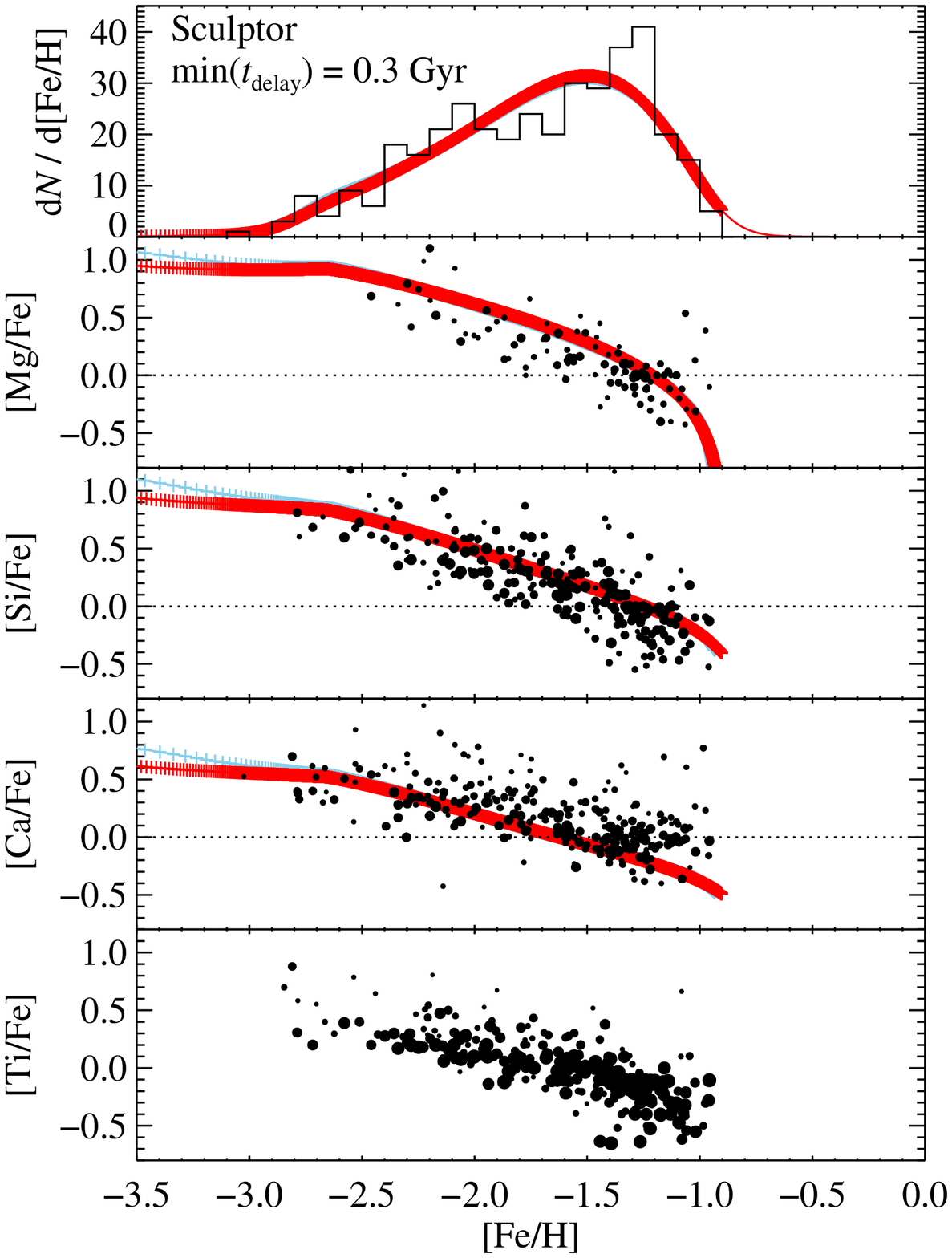}
 \hfil
 \includegraphics[width=0.495\textwidth]{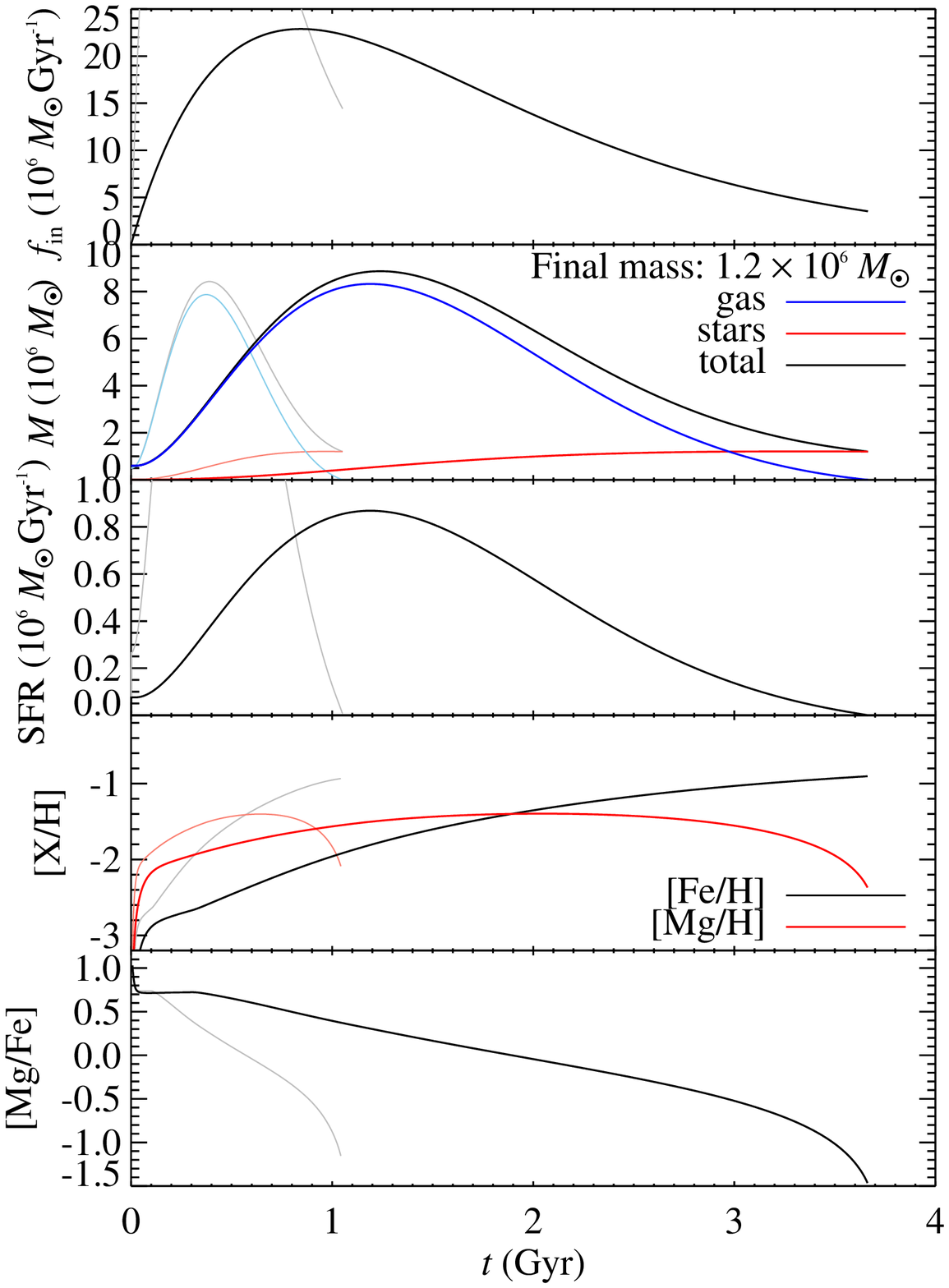}
 \caption{The observed abundance ratios and the best-fit gas flow and
   star formation history model for Sculptor.  The dark red ({\it
     left}) or heavier ({\it right}) lines show the model with the
   longer minimum Type~Ia SN delay time of 0.3~Gyr
   (Sec.~\ref{sec:tIa3}).  The light blue ({\it left}) or faded ({\it
     right}) lines show the original value of 0.1~Gyr, as in
   Fig.~\ref{fig:scl}.\label{fig:tIa3}}
\end{figure*}

We have adopted the Type~Ia DTD of \citet{mao10}.  The model is very
sensitive to the delay time of the {\it first} Type~Ia SN to explode
after the onset of star formation.  Unfortunately, this quantity is
poorly measured.  We have chosen 0.1~Gyr because that is the maximum
value that \citeauthor{mao10}'s DTD seems to allow.  However, the DTD
was measured in a range of galaxies with widely varying star formation
environments.  The details of the combined DTD (Fig.~\ref{fig:dtd})
may not be appropriate for dSphs.  For example, \citet{kob98} and
\citet{kob09} suggested that single-degenerate Type~Ia SNe will be
inhibited at low metallicity ($\mathfeh \la -1$).  Nonetheless, the
decline of [$\alpha$/Fe] with increasing [Fe/H] in
Figs.~\ref{fig:leoi}--\ref{fig:umi} demands that some kind of Type~Ia
SN explode.  Thus, the low-metallicity Type~Ia SNe in dSphs may be
mergers of double-degenerate binaries only.  The removal of the
single-degenerate channel could affect the DTD.

In order to explore the impact of changing the DTD on the chemical
evolution model, we have recomputed the most-likely model parameters
for Sculptor with a minimum Type~Ia delay time of 0.3~Gyr instead of
0.1~Gyr.  We did not change the DTD normalization.
Figure~\ref{fig:tIa3} shows the result compared to the original model
(Fig.~\ref{fig:scl}).  The abundance distribution is identical except
for the low-metallicity [$\alpha$/Fe] plateau, which is flatter for
the longer delay time because the mass dependence of the Type~II SN
yields is muted.  However, the right panel of Fig.~\ref{fig:tIa3}
shows that the SFH has changed dramatically.  In particular, the
timescale of SF has been expanded.  In fact, the differences in the
SFHs can be explained by multiplying the time variable in the original
model by about 3.5.  The result is less intense star formation over a
longer time.  In the end, just as many stars are formed and just as
much gas is blown out as in the original model.

We conclude that the Type~Ia SN DTD is a major uncertainty in our
model.  The abundance data alone does not help to determine the
minimum delay time.  The timescales in our models can be multiplied by
a factor constrained only by the poorly known minimum Type~Ia SN delay
time.

\subsection{Hypernova Fraction}
\label{sec:epshn05}

\begin{figure*}[t!]
\centering
 \columnwidth=.5\columnwidth
 \includegraphics[width=0.495\textwidth]{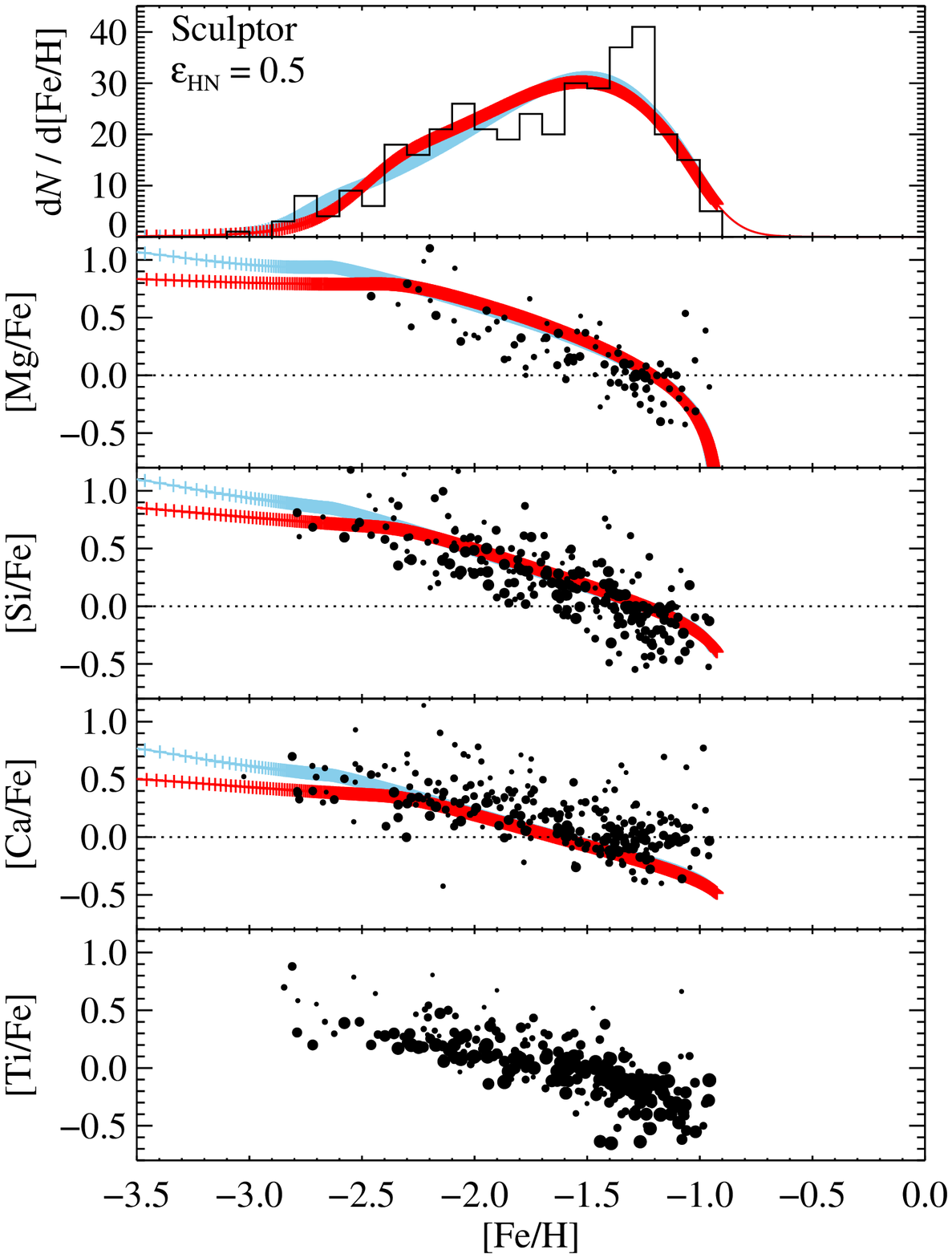}
 \hfil
 \includegraphics[width=0.495\textwidth]{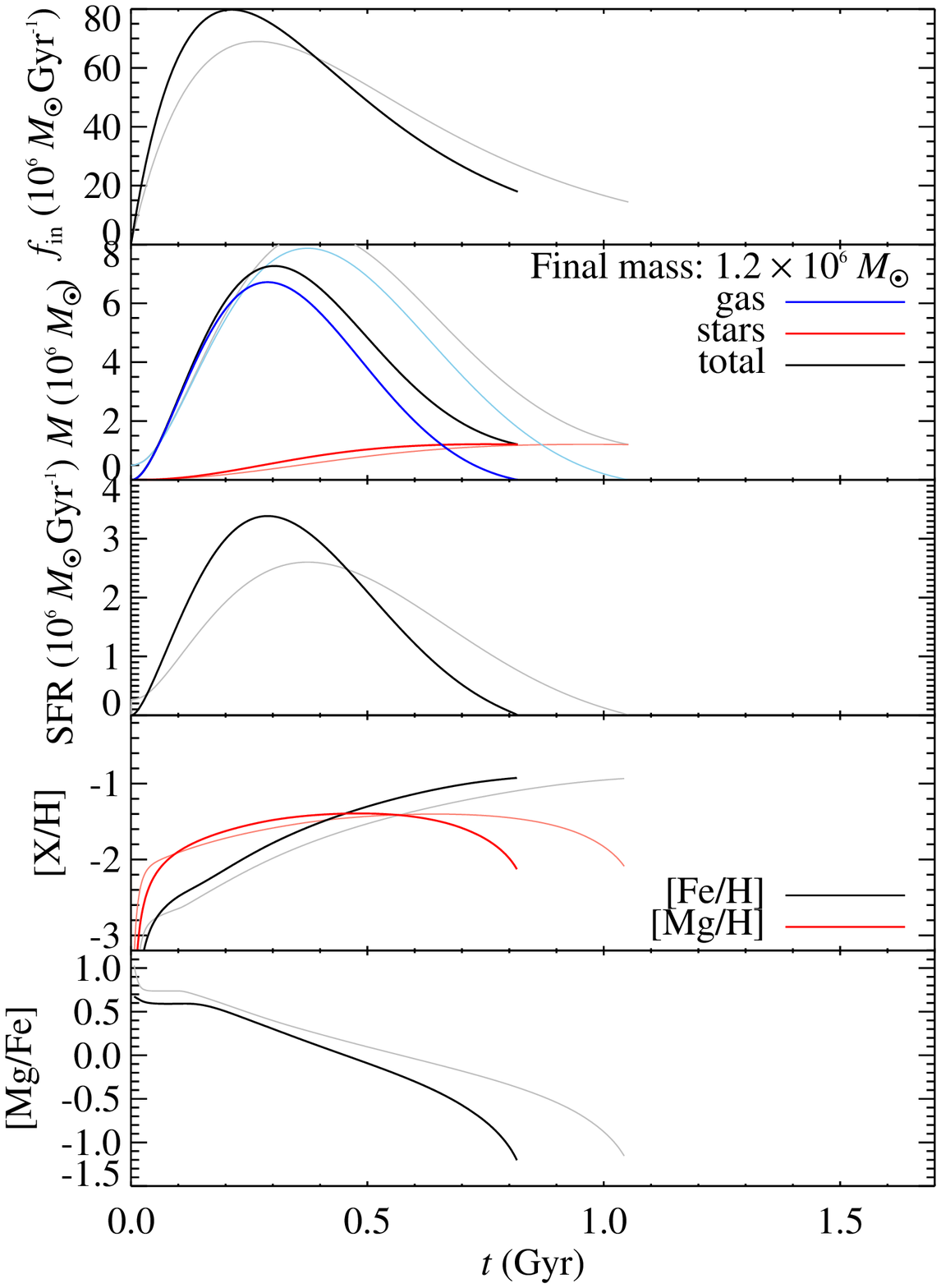}
 \caption{The observed abundance ratios and the best-fit gas flow and
   star formation history model for Sculptor.  The dark red ({\it
     left}) or heavier ({\it right}) lines show the model with a
   hypernova fraction of $\epsilon_{\rm HN} = 0.5$
   (Sec.~\ref{sec:epshn05}).  The light blue ({\it left}) or faded
   ({\it right}) lines show the original value of $\epsilon_{\rm HN} =
   0$, as in Fig.~\ref{fig:scl}.\label{fig:epshn05}}
\end{figure*}

SN~1998bw was immediately identified to be unusual because of its
association with a gamma ray burst and a light curve that suggested
relativistically expanding gas \citep{gal98}.  \citet{iwa98}
determined that the explosion energy for SN~1998bw was about 30 times
larger than the average SN.  The energy of the explosion has
consequences for the nucleosynthesis.  \citet{nom06} calculated
nucleosynthetic yields for SNe at a variety of explosion energies.

One of the fixed parameters in our model is the fraction of stars that
explode as very energetic hypernovae ($\epsilon_{\rm HN}$).  We
initially chose $\epsilon_{\rm HN} = 0$ (no HNe) because it seemed to
better match the abundance patterns at the lowest metallicities (e.g.,
[Ca/Fe] in Sculptor).  In order to explore the effect of HNe, we have
also found the most likely model for Sculptor with $\epsilon_{\rm HN}
= 0.5$.  This is the value that \citeauthor{nom06}\ chose for their
own chemical evolution model of the solar neighborhood.  \citet{rom10}
further explored the effect of changing $\epsilon_{\rm HN}$.

Figure~\ref{fig:epshn05} compares the result of the model with
$\epsilon_{\rm HN} = 0.5$ with the original model ($\epsilon_{\rm HN}
= 0$).  The abundance distributions are nearly identical except at
$\mathfeh < -2.3$.  The model with larger $\epsilon_{\rm HN}$ reaches
higher [Fe/H] before Type~Ia SNe turn-on.  This ensures that the
lowest metallicity stars are not polluted by Type~Ia SNe ejecta.  The
result is a plateau in [$\alpha$/Fe] at low [Fe/H].  We further
discuss the presence of such a plateau in the [Ca/Fe] ratio of
Sculptor and the absence of plateaus in other dSphs in
Sec.~\ref{sec:universalpattern}.

The effect on the SFH is more noticeable than on the abundance
distributions.  The total star formation duration shortens to
\epshnagescl~Gyr from \agescl~Gyr.  The HN model also requires no
initial gas, though the original model for Sculptor already did not
require very much gas.  Less gas is lost to supernova winds in the HN
model.

In conclusion, the inclusion of HNe has a minor effect on the
abundance distributions and SFH.  The most notable result is that very
metal-poor stars ($\mathfeh < -2.3$) in the HN model have
[$\alpha$/Fe] ratios that are inconsistent with any amount Type~Ia SN
ejecta.  Instead, these stars incorporate the ejecta of only Type~II
SNe or HNe.

\subsection{Metal-Enhanced Supernova Winds}
\label{sec:Zwind}

\begin{figure*}[t!]
\centering
 \columnwidth=.5\columnwidth
 \includegraphics[width=0.495\textwidth]{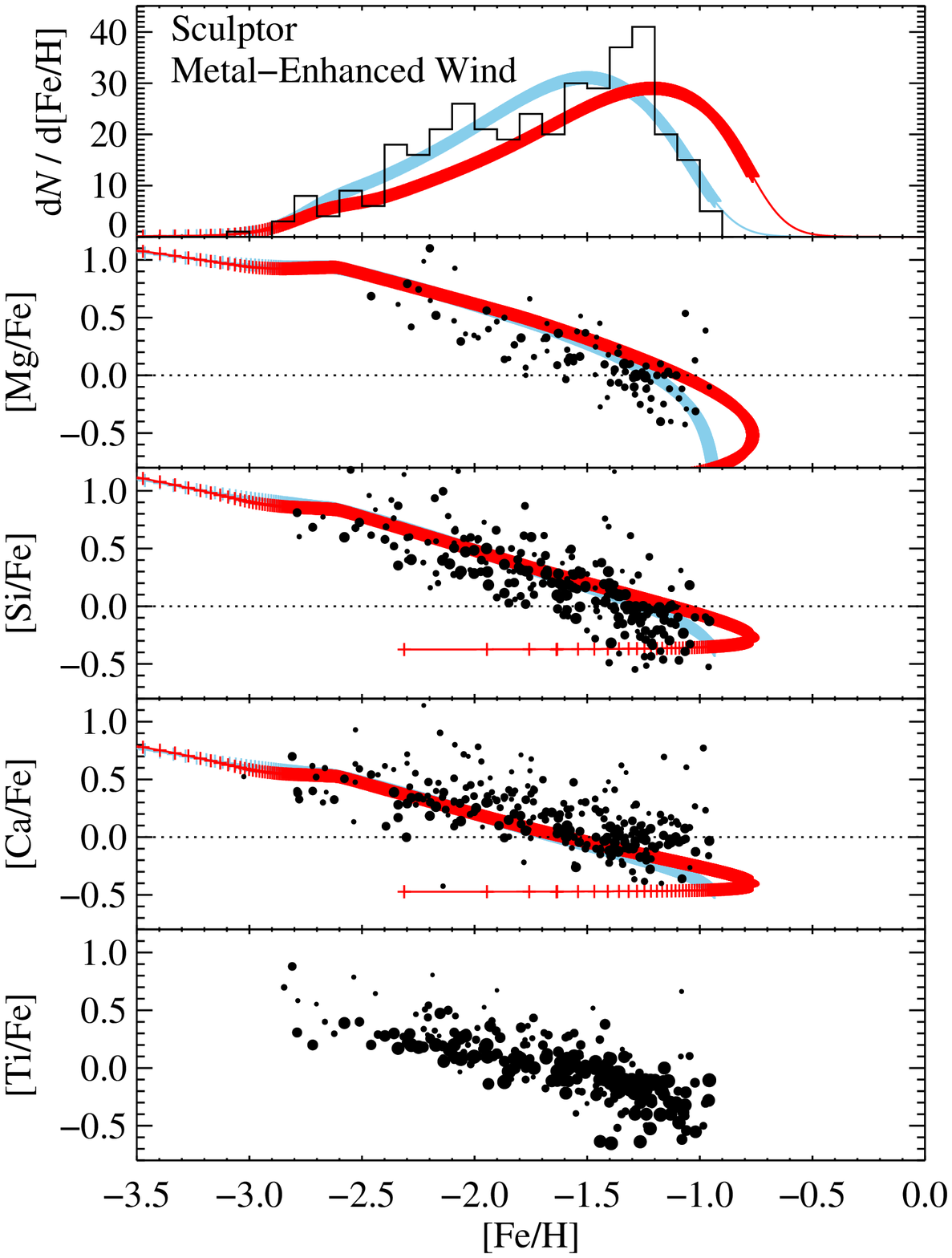}
 \hfil
 \includegraphics[width=0.495\textwidth]{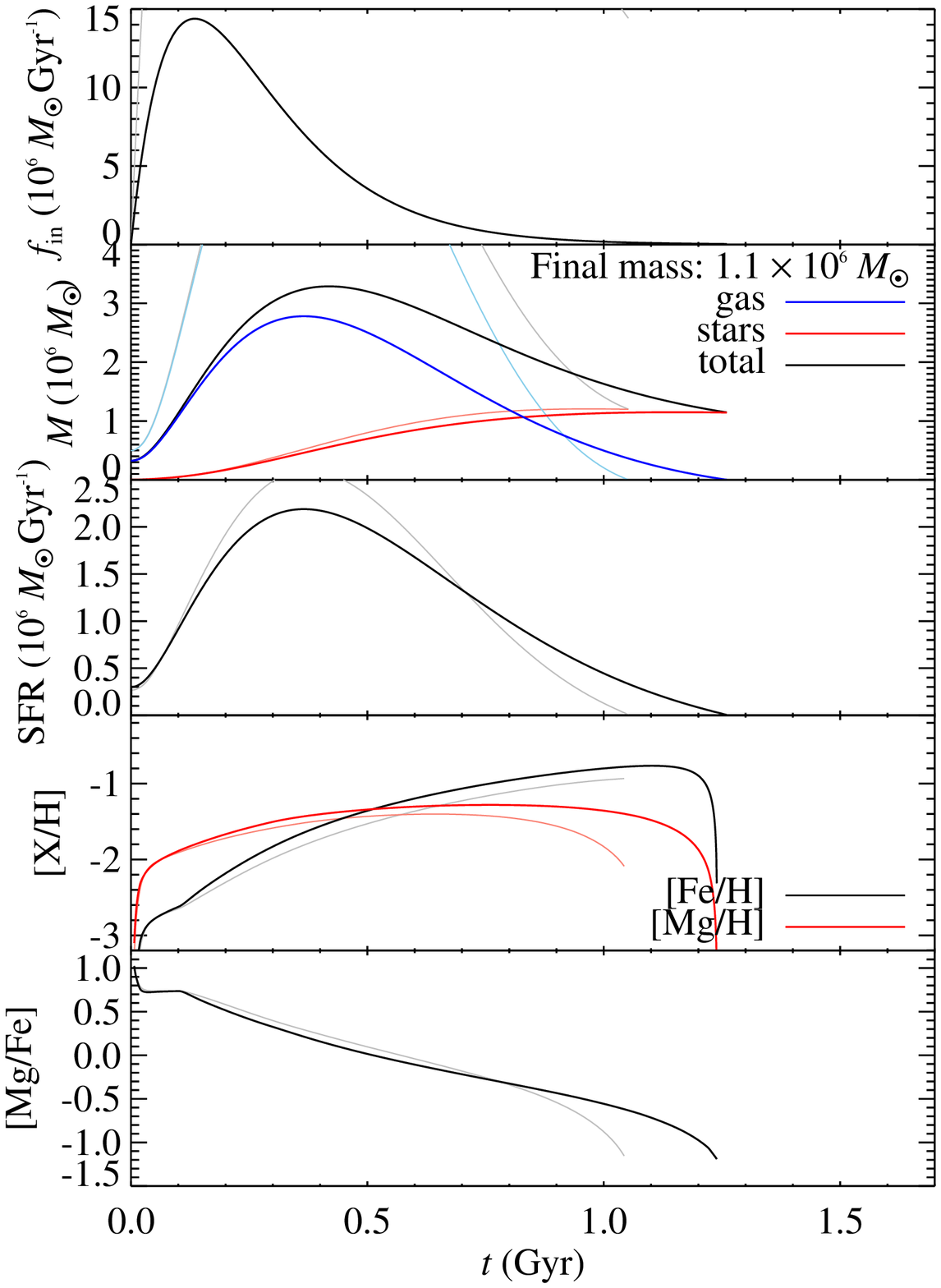}
 \caption{The observed abundance ratios and the best-fit gas flow and
   star formation history model for Sculptor.  The dark red ({\it
     left}) or heavier ({\it right}) lines show the model with a
   metal-enhanced wind (Sec.~\ref{sec:Zwind}).  The light blue ({\it
     left}) or faded ({\it right}) lines show the original model with
   an unenhanced wind, as in Fig.~\ref{fig:scl}.\label{fig:Zwind}}
\end{figure*}

The SNe in our model expel gas without regard to its composition.
However, SN winds might be expected to be more metal-rich than the
average gas-phase metallicity because metals are more opaque (and
therefore more susceptible to radiation pressure) than hydrogren and
helium and because the same SNe that create the metals could blow them
away \citep{vad86,mac99}.  In this section, we explore the effect of a
metal-enhanced SN wind.  We refer the reader to \citet{rob05} for a
more thorough discussion of a model that included metal-enhanced winds
from dwarf galaxies.

We paramaterize the metallicity dependence of the wind by $f_Z$, which
can vary between 0 and 1.  Thus, we replace Eq.~\ref{eq:winds} with

\begin{eqnarray}
\dot{\xi}_{j,{\rm out}} &=& \left\{\begin{array}{lcr}
     A_{\rm out} \, X_j \, (\dot{N}_{\rm{II}} + \dot{N}_{\rm{Ia}}) (1-f_Z) &~~~& j = {\rm H, He} \\
     A_{\rm out} \, X_j \, (\dot{N}_{\rm{II}} + \dot{N}_{\rm{Ia}}) \left[f_Z \left(\frac{1}{Z} - 1 \right) + 1 \right] &~~~& {\rm otherwise} \\
                              \end{array} \right. \label{eq:Zwind}
\end{eqnarray}

\noindent
If $f_Z = 0$, then the wind is unenhanced.  If $f_Z = 1$, then the
winds expel only metals and no hydrogen or helium.  For this
experiment, we fix $f_Z$ at 0.01.  Although that value seems small,
the effect on the SFH is dramatic.

The modeled metallicity distribution (Fig.~\ref{fig:Zwind}) does not
fit the observed distribution as well as for the original model.
Instead, there is an overabundance of metal-rich stars.  The
metal-rich discrepancy could be mitigated by increasing $A_{\rm out}$
(the total amount of gas lost per SN) at the cost of worsening the
match at intermediate metallicities.  The predicted [$\alpha$/Fe]
distributions change only at $\mathfeh \ga -1.2$.  Metal-enhanced gas
loss causes the hook back toward lower [Fe/H] in the [$\alpha$/Fe]
diagrams.  Because the SFR is very low by the time [Fe/H] begins to
decrease, very few stars are formed during this time.

The most dramatic effect on the SFH is that much less gas is lost over
the lifetime of SF in the metal-enhanced wind model than in the
original model.  With an unenhanced wind, Sculptor ejects
\sclgaslostorig\ of the gas that it starts with or accretes.  With a
metal-enhanced wind, that number decreases to \sclgaslostZwind.  In
both models, Sculptor forms about $1.2 \times 10^6~M_{\sun}$ of stars.
The implications for galaxy evolution are dramatic.  In the first
case, over $10^8~M_{\sun}$ of gas is required to catalyze star
formation in Sculptor.  Nearly all of this gas is returned to the ISM.
In the metal-enhanced wind case, star formation in Sculptor requires a
gas mass of only a few times its final stellar mass.  The mass of
metals returned to the intergalactic medium in both cases is the same,
but in the metal-enhanced wind model, the metals in the ejected gas
are much more concentrated.  Changes to other aspects of the SFH are
subtle.

We conclude that the amount of metal enhancement in the SN blowout
dramatically affects the gas dynamics of the dSph.  Even a 1\% metal
enhancement reduces the total amount of gas required for star
formation by a factor of 40.  However, a model with $f_Z = 0.01$
results in a worse match to the observed metallicity distribution than
the original model with an unenhanced wind.  A lower, non-zero value
of $f_Z$ might produce better agreement with the observed abundance
data while reducing the amount of gas infall required from the
unenhanced wind scenario.  The literature on galactic chemical
evolution contains a diversity of SN feedback treatments.  We refer
the reader to the articles we have already mentioned
\citep[e.g.,][]{rec01,lan04,rob05,rom06,mar08} for more thorough
treatments.

%%%%%%%%%%%%%%%%%%%%%%%%%%%%%%%%%
%%%%%%%%%   SECTION 5   %%%%%%%%%
%%%%%%%%%%%%%%%%%%%%%%%%%%%%%%%%%

\section{Trends with Galaxy Properties}
\label{sec:trends}

\begin{figure}[t!]
\includegraphics[width=\linewidth]{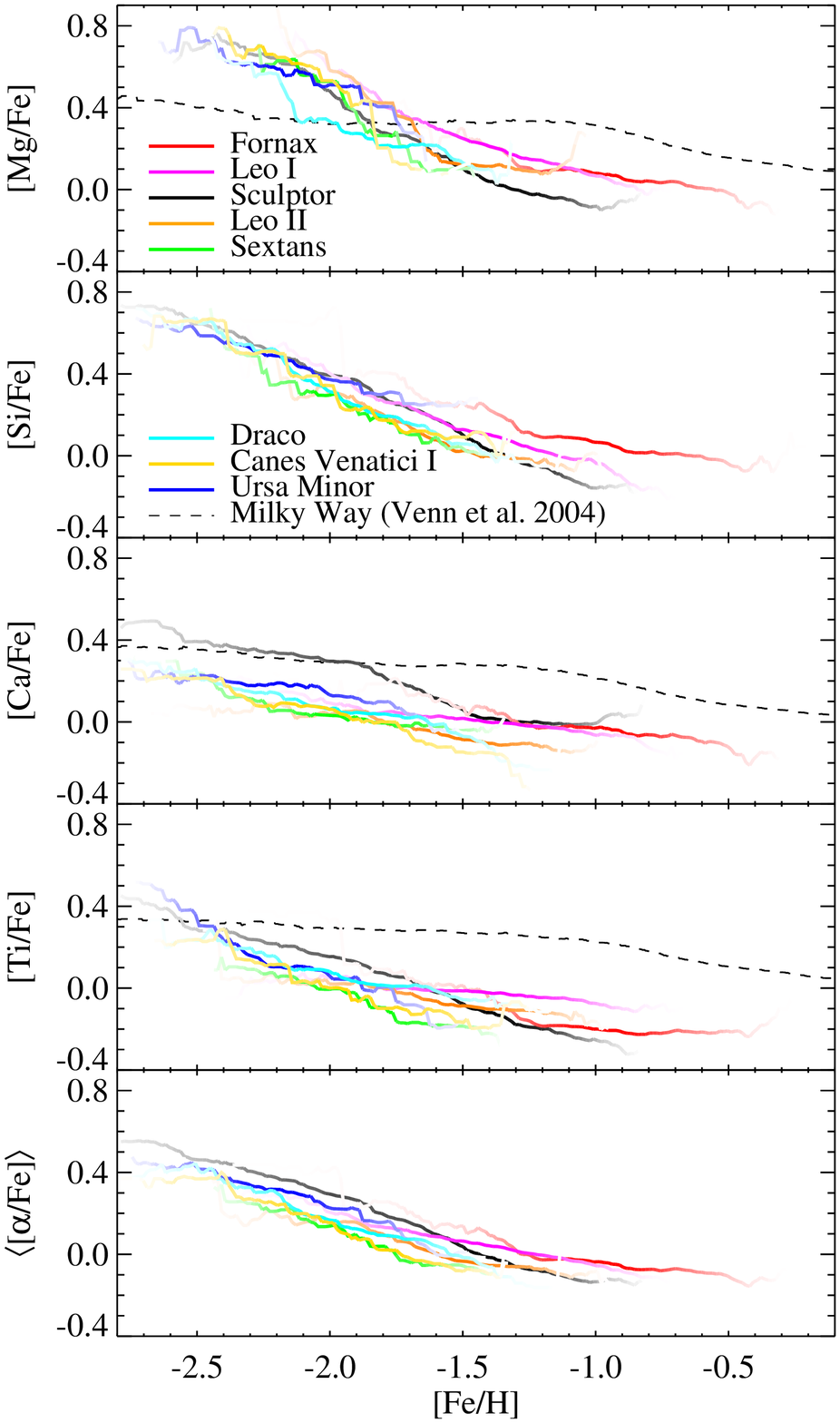}
\caption{The moving averages, inversely weighted by measurement
  uncertainty, of abundance ratios for the eight dSphs and for the
  Milky Way \citep[][who compiled data from the references given in
    footnote~7]{ven04}.  The bottom panel shows
  $\langle[\alpha/\rm{Fe}]\rangle$, the average of the top four
  panels.  The line weight is proportional to the number of stars
  contributing to the average.  The legend lists the dSphs in
  decreasing order of luminosity.  Except for [Ca/Fe] in Sculptor, the
  abundance ratios do not show a low-metallicity plateau, which
  indicates that Type~Ia SNe explode for nearly the entire duration of
  star formation.  Our data are sparse at $\mathfeh < -2.5$, and
  Type~Ia SNe need not explode at times corresponding to those low
  metallicities.  Only the galaxies luminous enough to reach $\mathfeh
  \ga -1$ eventually achieve an equilibrium between Types~II and Ia
  SNe and therefore a plateau at high
  metallicity.\label{fig:alphatrends}}
\end{figure}

We now discuss trends of the abundance distributions and derived SFH
parameters with observed galaxy properties, such as luminosity,
velocity dispersion, half-light radius, and Galactocentric distance.
We show that luminosity is the only galaxy property that shows any
convincing correlation with the properties of the abundance
distributions.

\subsection{General [$\alpha$/Fe] Trends}
\label{sec:gentrends}

Figure~\ref{fig:alphatrends} shows the trend lines of the different
element ratios with [Fe/H].  The trend line is defined by the average
of the element ratio, weighted by the inverse square of the
measurement uncertainties, in a moving window of 0.5~dex in [Fe/H].
The moving averages relax the uncertainty cut of 0.3~dex (used for
Figs.~\ref{fig:for}--\ref{fig:umi}) to 1~dex, meaning that all of the
measurements from the catalog (\citeauthor*{kir10b}) are included.
The bottom panel shows the average of four element ratios, which is
called $\langle[\alpha/\rm{Fe}]\rangle$.  The weight of the line fades
as fewer stars contribute to the average near the ends of the MDF.
The figure legend lists the dSphs in order of decreasing luminosity.
For comparison, some panels of the figure also display the trends for
the Milky Way halo and disk for available element ratios
\citep{ven04}\footnote{The data from \protect \citet{ven04} is a
  compilation of data from the following sources: \citet{ben03},
  \citet{bur00}, \citet{edv93}, \citet{ful00,ful02},
  \citet{gra88,gra91,gra94}, \citet{han98}, \citet{iva03},
  \citet{joh02}, \citet{mcw95}, \citet{mcw98}, \citet{nis97},
  \citet{pro00}, \citet{red03}, \citet{rya96}, and \citet{ste02}.}.

Fig.~\ref{fig:alphatrends} presents the broad trends of the evolution
of [$\alpha$/Fe] with increasing [Fe/H].  It does not convey the width
of the dispersion of the [$\alpha$/Fe] distributions at a given
metallicity, nor does it show the details at the margins of the MDF.
The extremely metal-poor stars, which represent some of the oldest
known stars, are not shown in Fig.~\ref{fig:alphatrends}.

\subsubsection{Universal Abundance Pattern in dSphs}
\label{sec:universalpattern}

The figure does show that the abundance distributions of dSphs evolve
remarkably similarly.  Although the dSphs span different ranges of
[Fe/H], $\langle[\alpha/\rm{Fe}]\rangle$ follows roughly the same
trend line.  This similarity contradicts the reasonable expectation
that different dSphs should show a knee in [$\alpha$/Fe] at different
values of [Fe/H] \citep[e.g.,][]{mat90,gil91,tol09}.  In fact,
\citeauthor{tol09}\ did indeed find a knee in at $\mathfeh = -1.8$ in
DART's preliminary measurements for [Ca/Fe] in Sculptor.  Our
measurements of [Ca/Fe] in Sculptor also show a knee at the same
metallicity and the same [Ca/Fe].  Ursa Minor possibly has a knee in
[Ca/Fe], but with a lower [Ca/Fe] plateau.  In agreement with
\citeauthor{tol09}'s and others' predictions for lower mass systems to
experience less intense SF, Ursa Minor's possible knee occurs at lower
[Fe/H] than Sculptor's knee.  However, the knee is apparent only in
[Ca/Fe] and only in Sculptor and possibly Ursa Minor.  The element
ratios that would better identify the onset of Type~Ia SNe, [Mg/Fe]
and [Si/Fe], do not show a knee for any dSph.

The lack of knees for $\mathfeh > -2.5$ and the lack of
low-metallicity plateaus in the [$\alpha$/Fe] distributions implies
that Type~Ia SNe exploded throughout almost all of the SFHs of all
dSphs.  Of course, the very first stars, which have yet to be found,
must be free of all SN ejecta.  The stars to form immediately after
the first SNe must incorporate only Type~II SN ejecta.  The very
lowest metallicity stars in dSphs likely represent this population.
Stars with $\mathfeh \ga -2.5$ formed after the Type~Ia SN-induced
depression of [$\alpha$/Fe].  We have already explored the possibility
of low-metallicity plateaus in [Ca/Fe], but we discount the absence of
Type~Ia SN products as the cause because [Ca/Fe] is the only element
ratio to show the plateau.  We speculate instead that
metallicity-dependent Type~Ia nucleosynthesis
\citep[e.g.,][]{tim03,how09} might shape the [Ca/Fe] distribution
differently from the other element ratios.

High-metallicity plateaus can form when the SF achieves a constant
rate for a duration long enough for the ratio between Types~II and Ia
SNe to be constant.  The SFR would achieve an equilibrium between the
production of $\alpha$ elements and Fe.  The value of [$\alpha$/Fe] at
the plateau depends on the IMF and SN delay time distribution.  The
SFR need not be strictly constant.  As \citet{rev09} pointed out, a
bursty SF profile with a high duty cycle can mimic a constant SFR.  In
that case, we would expect a scatter about the mean value of
[$\alpha$/Fe] at a given [Fe/H], but the mean value would not
necessarily evolve with increasing [Fe/H].  We do observe
high-metallicity plateaus, seen in Fig.~\ref{fig:alphatrends}.  The
trends for [Mg/Fe] and [Si/Fe] do not completely flatten, but the
slopes at $\mathfeh > -1$ are less than the slopes at $\mathfeh <
-1.5$.  The trends for [Ca/Fe] and [Ti/Fe] do completely flatten for
some dSphs.  Only the more luminous dSphs, which reached metallicities
of $\mathfeh \ga -1.2$, achieved the high-metallicity plateau.  The
[$\alpha$/Fe] ratios of Sextans, Draco, Canes Venatici~I, and Ursa
Minor do not flatten.  We conclude that dSphs with high enough SFRs to
reach stellar masses of at least $10^6~M_{\sun}$ experienced roughly
constant SF at late times, corresponding to metallicities $\mathfeh
\ga -1.2$.

Beneath the apparently universal path in [$\alpha$/Fe]-[Fe/H] space,
the abundance trends vaguely group by luminosity.  Higher luminosity
dSphs tend to have slightly higher values of [$\alpha$/Fe] at a given
[Fe/H] than lower luminosity dSphs.  The tracks for Sextans, Draco,
and Canes Venatici~I tend to lie below the other dSphs.  Fornax and
Leo~I tend to lie above Sculptor and Leo~II.  These divisions are
reminiscent of the groupings we proposed in \citeauthor*{kir10a} based on
MDF shapes.  We classified Fornax, Leo~I, and Leo~II as
``infall-dominated'' and Sextans, Draco, Canes Venatici~I, and Ursa
Minor as ``outflow-dominated.''  Sculptor sat in its own class.  The
similar groupings based on MDF and [$\alpha$/Fe] unsurprisingly
reaffirm that the SFH shapes both the MDF and the element ratio
distributions.

The MW satellite galaxies more luminous than Fornax sample a regime of
greater integrated star formation and higher metallicity.
\citet{pom08} measured [$\alpha$/Fe] for individual red giants in the
disk of the Large Magellanic Cloud (LMC), and \citet{muc08} measured
the same for red giants in LMC globular clusters.  The stars span the
range $-1.2 \le \mathfeh \le -0.3$ with one additional star at
$\mathfeh = -1.7$.  The [Ca/Fe] ratios of the disk stars decline
slightly with increasing [Fe/H], but the other element ratios are
nearly flat.  In fact, the LMC stars seem to follow the same
[$\alpha$/Fe] trends as Fornax or Leo~I, albeit shifted to higher
[Fe/H], except for [Ti/Fe].  The average [Ti/Fe] in the LMC is about
0.1~dex higher than Leo~I and 0.3~dex higher than Fornax.  The
Sagittarius dSph also shows a higher average [Ti/Fe] than Fornax or
Leo~I \citep{cho10}.  Also, [Ti/Fe] in Sagittarius declines with
increasing [Fe/H] over the entire range that
\citeauthor{cho10}\ sampled ($-1.5 \le \mathfeh \le +0.1$).

The available evidence indicates that the evolution of [$\alpha$/Fe]
with [Fe/H] is nearly universal in MW satellite galaxies except for
[Ti/Fe] at $\mathfeh \ga -1.3$.  The average values of [Ti/Fe] for the
dSphs and the LMC at these metallicities vary from about $-0.3$
(Sculptor) to $0.0$ (LMC and Sagittarius), and the slopes vary from
$\Delta \rm{[Ti/Fe]} / \Delta \rm {[Fe/H]} \approx -0.8$ (Sagittarius)
to $0.0$ (Fornax).  Ti is both an $\alpha$ element and an iron-group
element, and it has an appreciable yield from both Types~II and Ia SNe
\citep{woo95}.  Therefore, [Ti/Fe] responds to changes in the SFR and
the IMF differently from the ``purer'' $\alpha$ elements, like Mg and
Si.  Unfortunately, our chemical evolution model failed to reproduce
realistic values of [Ti/Fe] because the theoretical Type~II SN yields
of Ti were too small.  We suggest that future work explore ratios such
as [Mg/Ti] to better understand why [Ti/Fe] behaves differently in
different dwarf galaxies at high [Fe/H].

\subsubsection{[Mg/Fe]}

Our data set for the first time has enabled the exploration of the
bulk properties of [$\alpha$/Fe] in dSphs that span two orders of
magnitude in luminosity.  In particular, Fig.~\ref{fig:alphatrends}
shows that [Mg/Fe] values higher than in the MW are not unique to the
extremely metal-poor stars in dSphs \citep[e.g.,][]{fre10b} but also
exist in stars of more modest metallicity ($\mathfeh \la -1.8$).

Factors beyond the SFH may affect the absolute value of [Mg/Fe] and
other element ratios at low metallicity.  First, changing the IMF
alters [$\alpha$/Fe] because Type~II SN yields depend on the mass of
the exploding star.  Second, the early gas mass of the dSph might
change the shape of the low-metallicity [$\alpha$/Fe] distribution
also because SN yields depend on mass.  The first SNe in a galaxy can
more efficiently enrich a small gas mass than a large gas mass.
Massive SNe explode before less massive SNe, and massive SNe generally
produce higher [$\alpha$/Fe].  As a result, [$\alpha$/Fe] at low
metallicity could depend on the initial gas mass that was enriched by
the first SNe.  This effect possibly explains the larger [Mg/Fe] in
dSphs than in the MW.  We suggest that the stars at $\mathfeh \sim
-2.5$ in dSphs were enriched by SNe of higher average mass than the
stars at $\mathfeh \sim -2$ in the MW.  Finally, the shape of the
abundance distribution might depend on the early gas mass because SN
yields also depend on metallicity.  In addition to sampling higher
mass SNe, stars at a given [Fe/H] in a lower mass galaxy sample lower
metallicity SNe than stars at the same [Fe/H] in a higher mass galaxy.

\subsubsection{Unexplained Details}

Many details in Figure~\ref{fig:alphatrends} defy obvious
explanations.  For example, the [Ca/Fe] ratio is flatter than the
other element ratios.  Sculptor has a strangely large [Ca/Fe] at low
[Fe/H].  The [Si/Fe] trend for Fornax is above the other dSphs'
trends, but the other element ratios seem consistent.  Similarly, the
[Ti/Fe] ratio---and only [Ti/Fe]---for Leo~I lies above the other
dSphs.  Ursa Minor, despite being the least luminous dSph in the
figure, has the second largest [$\alpha$/Fe] at a given metallicity
for much of the metallicity range.  The slope of [Mg/Fe] flattens for
all of the dSphs at $\mathfeh \ga -1.2$, but the slope of [Si/Fe]
flattens only for Fornax and Leo~II.

We suggest that future work examine the abundance catalog in more
detail.  For example, element ratios with a denominator other than Fe
could constrain the IMF.  The predicted yields of [Mg/Si] decrease
from $+0.2$ for a progenitor mass of $18~M_{\sun}$ to $-0.3$ for a
progenitor mass of $40~M_{\sun}$ \citep{nom06}.  Our data set
possesses the sample size and precision to address such questions.

\subsection{Trends in Chemical Evolution Model Parameters}

\begin{figure*}[t!]
\includegraphics[width=\linewidth]{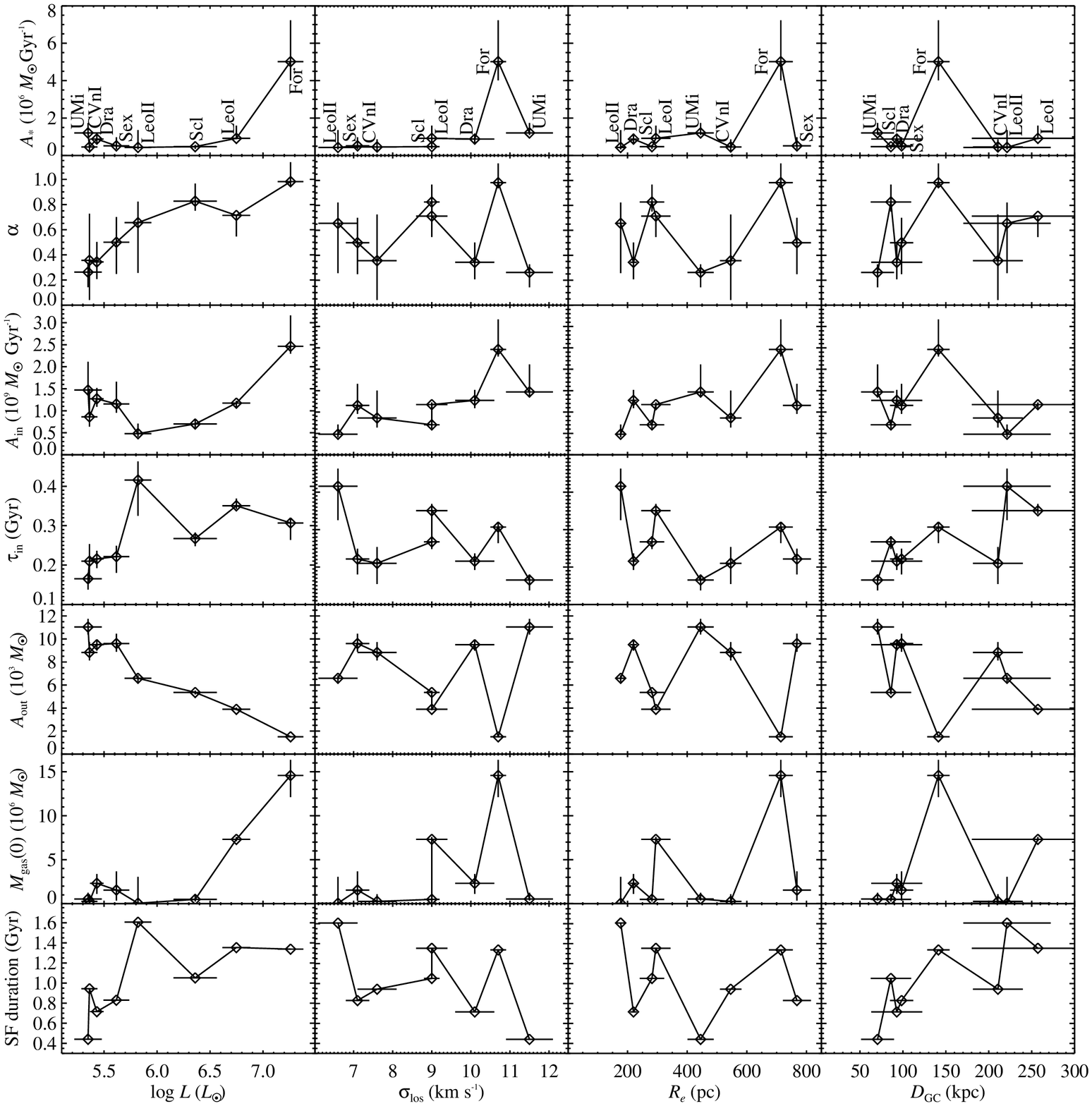}
\caption{The trends of the best-fit chemical evolution models with
  galaxy properties: luminosity \citep{irw95,martin08a}, line-of-sight
  velocity dispersion, two-dimensional projected half-light radius
  \citep[both from][and references therein]{wol10}, and Galactocentric
  distance.  From top to bottom, the parameters are the SFR
  normalization, SFR exponent (Eq.~\ref{eq:sfr}), gas infall rate, gas
  infall timescale (Eq.~\ref{eq:infall}), gas expelled per SN
  (Eq.~\ref{eq:winds}), and initial gas mass (Eq.~\ref{eq:gce}).
  Table~\ref{tab:gcepars} gives the same data.  The bottom row also
  shows the duration of star formation (see Table~\ref{tab:duration}),
  which is a quantity derived from the model, not a free parameter.
  The only galaxy property to show a trend is
  luminosity.\label{fig:gcetrends}}
\end{figure*}

We now invoke the best-fit parameters of the chemical evolution model
in a more quantitative discussion of the correlation between abundance
distributions and galaxy properties.  Figure~\ref{fig:gcetrends}
presents the parameters against luminosity, line-of-sight velocity
dispersion, half-light radius, and Galactocentric distance.  In
addition to the model parameters, the bottom row of the figure shows
the star formation duration, which is a quantity derived from the
best-fit model, not a free parameter.

Luminosity can reasonably be expected to show the best correlation
with quantities related to SF.  Of the four abscissas in
Fig.~\ref{fig:gcetrends}, $L$ is the only one that could be predicted
from our simple chemical evolution model.  Roughly, $L$ is the
integral of past SF, modulated by the reddening and dimming associated
with aging.  Therefore, it is not surprising that the chemical
evolution parameters vary with $L$.  Although we have plotted the SF
parameters against $L$, $L$ is not necessarily the independent
variable.  Luminosity is a present-day quantity, and the stars did not
know the final stellar mass of the galaxy while they were forming.
The SFH determines the present luminosity.

\subsubsection{Star Formation Rate Parameters}

The SFR normalization, $A_*$, is roughly constant at \mbox{$\sim 5
  \times 10^5~M_{\sun}~{\rm Gyr}^{-1}$} for galaxies less luminous
than Leo~I.  The value roughly doubles for Leo~I and increases by an
order of magnitude for Fornax.  The increase in $A_*$ is expected
because a more luminous galaxy must have formed more stars than a less
luminous galaxy.  If the SF timescale does not change much with
luminosity, then the SFR must.  We observe that the SF duration
changes by a factor of about four across the luminosity range.
Therefore, we estimate a range of 40 in luminosity.  The actual $L$
range is 80, but our simple estimate ignored the ages of the stellar
population and the other model parameters which affect the SFR, such
as $\tau_{\rm in}$.

The exponent of the SFR law, $\alpha$, also varies with $L$.  If we
assume that SFR is proportional to gas volume density, then $\alpha$
may indicate the degree to which the gas was concentrated in the
center of the galaxy.  However, we find no correlation between
$\alpha$ and the concentration of the light profiles \citep[][not
  shown in Fig.~\ref{fig:gcetrends}]{irw95}.  Our interpretation of
$\alpha$ is purely speculative because SF is a complex process
affected by many external factors, such as an ionizing radiation
background.  These factors become more difficult to predict for
smaller galaxies \citep[e.g.,][]{gne10}.

\subsubsection{Gas Infall Parameters}

The intensity of infalling gas (or gas cooling to become available for
SF) drives the SFR.  The parameter $A_{\rm in}$ is closely related to
$A_*$.  The dSph cannot maintain a high SFR without the addition of
new gas.  Therefore, a luminous galaxy must have had large values of
both $A_*$ and $A_{\rm in}$.  Alternatively, a luminous galaxy could
have started its life with a large reservoir of gas.  However, in
order to prevent too many metal-poor stars from forming early, new gas
must have been added during the SF lifetime.  The net result is that
$A_*$, $A_{\rm in}$, and $M_{\rm gas}(0)$ are highly covariant.

The most likely timescales for gas infall (or cooling) vary from
\tauinumi\ to \tauinleoii~Gyr.  It may be significant that none of the
timescales exceeds \tauinleoii~Gyr.  We propose three conjectures.
First, $\tau_{\rm in}$ may reflect the time the dSph requires to
accumulate gas.  The central densities of dSphs are similar
\citep{mat98,gil07,str08}.  Therefore, the similar gravitational
potentials of the dSphs themselves might enforce similarly small gas
accretion timescales.

Second, the dSphs' environment may set the $\tau_{\rm in}$ timescale.
Interestingly, $\sim 0.1$~Gyr was the timescale for the Galaxy's
monolithic collapse proposed by \citet*{egg62}.  This collapse time
corresponds to a period when the gas in the vicinity of the MW was
rapidly coalescing into individual structures, such as the
proto-Galaxy and the dSphs.  After $~0.1$~Gyr, gas accretion would
have declined considerably because the MW and its satellites would by
then have accreted the bulk of the surrounding gas.  In the
$\Lambda$CDM paradigm, the formation time for a dSph-sized dark matter
halo is only 0.4~Gyr after the Big Bang \citep{wec02}.  Therefore, our
most likely gas accretion timescales are consistent with both
cosmogonies.

Third, the time from the formation of the first stars to cosmological
reionization is roughly 0.5~Gyr.  \citet{ric05} referred to all eight
of our dSphs as ``true'' or ``polluted fossils,'' meaning that all or
most of their stars formed before reionization.  Our models are
sensitive to the bulk of the population, and not the few younger stars
present in most dSphs.  Therefore, the best-fit values of $\tau_{\rm
  in}$ may be probing the pre-reionization SF timescale.  Fornax must
be a exception because the bulk of its population formed after
reionization.  The majority stellar populations in other dSphs may be
fossils with SF timescales on the order of the reionization time.  The
(small) dispersion among our $\tau_{\rm in}$ values may be a result of
temporally protracted, spatially inhomogeneous reionization
\citep{mir00}.  However, we note that our derived SF durations are
longer than 0.5~Gyr except for Ursa Minor.  To the extent that these
durations are accurate, we surmise that reionization is one of several
mechanisms that inhibited SF in dSphs.

\subsubsection{Supernova Winds}

The role of SN feedback for dSphs has been emphasized repeatedly.
\citet{dek86} posited that SN feedback regulates the SFR for dwarf
galaxies.  It can cause a terminal wind, or it can blow out gas that
is later re-accreted.  For the smallest galaxies, including the dSphs
presented here, radiation feedback also plays a significant role
\citep{dek03}.  The best-fit SN wind intensities, $A_{\rm out}$, also
show a strong correlation with $L$.  More luminous dSphs experienced
more intense winds.  This trend is a direct result of the
metallicity-luminosity relation for dSphs (e.g.,
\citeauthor*{kir10a}).  For reasons discussed in \citeauthor*{kir10a},
more intense gas outflow lowers the effective metal yield.  Therefore,
the less luminous, more metal-poor dSphs naturally show more gas
outflow.  However, we expected that $A_{\rm out}$ also correlate with
the velocity dispersion, a measure of the depth of the potential well.
No such correlation exists.  The lack of correlation is puzzling, but
the gas blowout depends on the unmeasurable mass density profile at
the time of SF and on the locations of the SNe within the
gravitational potential.

\subsubsection{Galaxy Properties Other Than Luminosity}

The model parameters are insensitive to galaxy properties other than
$L$.  The velocity dispersions of dSphs do not span nearly as large a
range as their luminosities, which may partly explain the lack of
dependence on $\sigma_{\rm los}$.  The half-light radius and
luminosity together are related to the galaxy's surface brightness and
stellar density.  It does not seem that the SF parameters in our model
depend significantly on these quantities.  The timescales, $\tau_{\rm
  in}$ and the SF duration, may depend weakly on Galactocentric
distance.  The Pearson linear correlation coefficient between
$\tau_{\rm in}$ and $D_{\rm GC}$ is \tauindgc.  Because $\tau_{\rm
  in}$ basically represents the SF duration (the correlation
coefficient between $\tau_{\rm in}$ and the SF duration is \tauinage),
this relation may indicate that more distant dSphs survive
SF-truncating interactions with the MW longer than closer dSphs.  In
fact, \citet*{sil87} suggested host galaxies competed with their
satellites for gas accretion.  The more distant satellites, such as
dwarf irregulars, successfully accreted more gas to power present star
formation than the closer satellites, such as dSphs.  Orbital history
would be a better indicator of past interaction with the MW.  Orbital
parameters based on proper motions are available for Fornax
\citep{pia07}, Sculptor \citep{pia06}, and Ursa Minor \citep{pia05}.
\citet{soh07} also constrained the orbit of Leo~I based on the shape
and dynamics of tidal debris.  We leave orbital analyses for future
work.

We conclude that luminosity is more directly related to a dSph's SFH
than dynamical or morphological properties.  The present luminosity
can not drive the past star formation, but the luminosity does mirror
a single parameter which determines the SFH.  This conclusion is
similar to the fundamental line for dwarf galaxies defined by
\citet{woo08}.  They also found that stellar mass (closely related to
luminosity) is the best predictor of other dSph properties.  However,
stellar mass loss by tidal stripping may obfuscate the correlation
between present stellar mass and past star formation.

%%%%%%%%%%%%%%%%%%%%%%%%%%%%%%%%%
%%%%%%%%%   SECTION 6   %%%%%%%%%
%%%%%%%%%%%%%%%%%%%%%%%%%%%%%%%%%

\section{Summary and Conclusions}
\label{sec:conclusions}

We have made a first attempt at quantitative chemical evolution models
for the large sample of multi-element abundance measurements for MW
dSphs that we published in \citeauthor*{kir10b}.  Our simple model is
a significant improvement to the analytical models of the metallicity
distributions that we explored in \citeauthor*{kir10a}.  We fit the
MDF and [$\alpha$/Fe] distribution simultaneously to derive the SF and
gas flow histories of each of eight dSphs spanning about two orders of
magnitude in luminosity.  Our model produces reasonable fits to the
abundance distributions of dSphs whose color-magnitude diagrams show
that most or all of their stars are older than 10~Gyr.

We draw the following conclusions from our models and from the general
trends in abundance distributions (Fig.~\ref{fig:alphatrends}):

\begin{enumerate}
\item The [$\alpha$/Fe] ratios evolve with metallicity along nearly
  the same path for all dSphs.  The average value of [Mg/Fe], [Si/Fe],
  [Ca/Fe], and [Ti/Fe] drop from $+0.4$ at $\mathfeh = -2.5$ to $0.0$
  at $\mathfeh \approx -1.2$, where the slope flattens.

\item No low-metallicity plateaus or knees exist in [$\alpha$/Fe]
  vs.\ [Fe/H] space for any dSph at $\mathfeh > -2.5$.  We conclude
  that Type~Ia supernovae contributed to chemical evolution for all
  but the most metal-poor stars.

\item The [Mg/Fe] ratio in dSphs exceeds that of the Milky Way at
  $\mathfeh \la -1.8$.  We suggest that the abundance ratios of stars
  in low-mass systems are more sensitive to the mass and metallicity
  dependence of Type~II supernovae yields than stars at the same
  metallicity in higher-mass systems, such as the progenitors of the
  inner MW halo.

\item The dSphs may be grouped based on their [$\alpha$/Fe]
  distributions into roughly the same groups that we defined based on
  their metallicity distributions (\citeauthor*{kir10a}).  The more
  luminous dSphs have infall-dominated MDFs and slightly higher
  $\langle\rm{[\alpha/Fe]}\rangle$ at a given [Fe/H].  The less
  luminous dSphs have outflow-dominated MDFs and slightly lower
  $\langle\rm{[\alpha/Fe]}\rangle$ at the same [Fe/H].

\item Some SF model parameters correlate with present luminosity, but
  not with velocity dispersion, half-light radius, or Galactocentric
  distance except for a possible correlation between gas infall
  timescale and $D_{\rm GC}$.

\item The gas flow histories for all dSphs except Fornax are
  characterized by large amounts of gas loss, probably driven by
  supernova winds.  Less luminous dSphs experienced more intense gas
  loss.

\item Allowing supernova winds to be metal-enhanced drastically
  reduces the amount of gas infall and outflow required to explain the
  observed abundance distributions.

\item The gas infall timescale does not exceed \tauinleoii~Gyr.  This
  possibly reflects the amount of time ancient stars had to form
  before reionization ended star formation.

\item The derived star formation timescales are extremely sensitive to
  the delay time for the first Type~Ia SN.  Increasing the delay time
  from 0.1~Gyr to 0.3~Gyr results in a star formation duration in
  Sculptor inflated by a factor of 3.5.

\item The presence of bumps in the MDFs and stars with [$\alpha$/Fe]
  ratios far from the average trend lines suggests that the SFHs of
  dSphs were characterized by bursts, which are not included in our
  model.  Bursts are a common feature of more sophisticated models.
\end{enumerate}

Some of our conclusions (5--10) depend on the realism of our chemical
evolution model.  Many more sophisticated models exist, and we
encourage their application to our data set.  \citeauthor*{kir10b} contains
the complete abundance catalog.

The major strength of the present work is that we apply the same model
to a homogeneous data set of hundreds of stars in each of eight dSphs.
The sample size and diversity of galaxies has allowed us to present an
overview of chemical evolution in dwarf galaxies.  We have discovered
patterns not apparent in previous data sets due to small samples or
lack of diversity among the well-sampled galaxies.  In particular, we
have shown that [$\alpha$/Fe] distributions of dSphs do not form a
sequence of knees corresponding to the metallicities at which Type~Ia
supernovae began to explode.  Instead, the [$\alpha$/Fe] patterns of
all dSphs are largely the same, but different dSphs sample different
regions in metallicity.

\acknowledgments We thank John Johnson, Hai Fu, Julianne Dalcanton,
Chris Sneden, and Bob Kraft for insightful discussions.  Support for
this work was provided by NASA through Hubble Fellowship grant
51256.01 awarded to ENK by the Space Telescope Science Institute,
which is operated by the Association of Universities for Research in
Astronomy, Inc., for NASA, under contract NAS 5-26555.  SRM
acknowledges support from NSF grants AST-0307851 and AST-0807945, and
from the SIM Lite key project ``Taking Measure of the Milky Way''
under NASA/JPL contract 1228235.  PG acknowledges NSF grants
AST-0507483, AST-0607852, and AST-0808133.

The authors wish to recognize and acknowledge the very significant
cultural role and reverence that the summit of Mauna Kea has always
had within the indigenous Hawaiian community.  We are most fortunate
to have the opportunity to conduct observations from this mountain.

{\it Facility:} \facility{Keck:II (DEIMOS)}

\end{document}